\documentclass[11pt,a4paper]{article}

\usepackage[a4paper,margin=2.3cm]{geometry}
\usepackage[T1]{fontenc}
\usepackage[utf8]{inputenc}
\usepackage{lmodern}
\usepackage{microtype}

\usepackage{amsmath,amssymb}
\usepackage{graphicx}
\usepackage{caption}
\usepackage{subcaption}
\usepackage{booktabs}
\usepackage{siunitx}
\usepackage{titletoc}
\usepackage[numbers,sort&compress]{natbib}
\usepackage{bibunits}
\defaultbibliographystyle{unsrtnat}
\usepackage[colorlinks=true,linkcolor=blue,citecolor=blue,urlcolor=blue]{hyperref}

\graphicspath{{figures/}}

\usepackage[most]{tcolorbox}
\usepackage{xcolor}

\definecolor{main}{HTML}{1F4E79}
\definecolor{sub}{HTML}{A6C9E2}

\newtcolorbox{boxE}{
    enhanced,
    boxrule=0pt,
    borderline={0.75pt}{0pt}{main},
    borderline={0.75pt}{2pt}{sub}
}

\begin{document}

% =====================================================
% First page: title, authors, affiliations only
% =====================================================
\hypersetup{pageanchor=false}
\begin{titlepage}
\thispagestyle{empty}
\centering
\vspace*{2.0cm}

{\LARGE\bfseries Cavity-enhanced superconducting response in an underdoped cuprate\par}

\vspace{1.5cm}

{\large
Angela Montanaro$^{1,*,\dagger}$, Vadim Plastovets$^{2,*}$, Nitesh Khatiwada$^{1}$, Jacopo Fiore$^{3}$, Giacomo Jarc$^{1}$, Abdullah Alabbadi$^{1,4,5}$, Antonio Mastropasqua$^{1}$, Enrico Maria Rigoni$^{1}$, Shahla Y. Mathengattil$^{6,7}$, Simone Dal Zilio$^{8}$, Francesca Fassioli Olsen$^{1}$, Fabio Novelli$^{9}$, Stephan Winnerl$^{10}$, Michael A. Sentef$^{11,12}$, Dante M. Kennes$^{3,12}$, Andrew J. Millis$^{13,14}$, Francesco Piazza$^{2}$, Daniele Fausti$^{1,6,\dagger}$\par
}

\vspace{0.9cm}

{\normalsize
$^{1}$Department of Physics, Friedrich-Alexander University of Erlangen–Nuremberg, 91058 Erlangen, Germany\\[0.25em]
$^{2}$Theoretical Physics III, Center for Electronic Correlations and Magnetism, Institute of Physics, University of Augsburg, 86135 Augsburg, Germany\\[0.25em]
$^{3}$Institut für Theorie der Statistischen Physik, RWTH Aachen, 52074 Aachen, Germany\\[0.25em]
$^{4}$Max Planck Institute for the Science of Light, 91058 Erlangen, Germany\\[0.25em]
$^{5}$Department of Physics, Faculty of Science, Alexandria University, 21511 Alexandria, Egypt\\[0.25em]
$^{6}$Department of Physics, Università degli Studi di Trieste, 34127 Trieste, Italy\\[0.25em]
$^{7}$Elettra Sincrotrone Trieste, 34149 Basovizza, Italy\\[0.25em]
$^{8}$CNR-Consiglio Nazionale Delle Ricerche, Istituto Officina Dei Materiali, 34149 Basovizza, Italy\\[0.25em]
$^{9}$School of Physics and Astronomy, Faculty of Engineering and Physical Sciences, University of Southampton, SO17 1BJ Southampton, UK\\[0.25em]
$^{10}$Institute of Ion Beam Physics and Materials Research, Helmholtz-Zentrum Dresden-Rossendorf, Dresden, Germany\\[0.25em]
$^{11}$Institute for Theoretical Physics and Bremen Center for Computational Materials Science, University of Bremen, 28359 Bremen, Germany}\\[0.25em]
$^{12}$Max Planck Institute for the Structure and Dynamics of Matter, Center for Free Electron Laser Science, 22761 Hamburg, Germany\\[0.25em]
$^{13}$Center for Computational Quantum Physics, The Flatiron Institute, 162 5th Avenue, New York, New York 10010, USA\\[0.25em]
$^{14}$Department of Physics, Columbia University, 538 West 120th Street, New York, New York 10027, USA\\[1.5em]

$^{*}$These authors contributed equally to this work.\\[0.8em]
$^{\dagger}$To whom correspondence should be addressed: \href{mailto:angela.montanaro@fau.de}{angela.montanaro@fau.de}, \href{mailto:daniele.fausti@fau.de}{daniele.fausti@fau.de}
\par

\vfill

% Optional: uncomment if you want a date on the title page.
% {\normalsize June 2026\par}

\end{titlepage}
\hypersetup{pageanchor=true}

% =====================================================
% Main text, including its own bibliography
% =====================================================
\pagenumbering{arabic}
\setcounter{page}{1}
\begin{bibunit}
% =====================================================
% Main text starts here.
% This file has its own BibTeX bibliography at the end.
% =====================================================

\begin{center}
\begin{minipage}{0.85\textwidth}
\small
\textbf{Superconductors carry electrical current without resistance when
paired electrons condense into a coherent macroscopic quantum state. In
underdoped cuprates, evidence suggests that pairing-related correlations
and superconducting fluctuations can survive above the temperature at
which global coherence is lost, pointing to phase fluctuations as a key
limitation on superconductivity in this regime. Motivated by recent
demonstrations of cavity-modified collective states in quantum
materials, we investigate whether superconducting coherence can be
stabilized by engineering the electromagnetic environment of the
superconductor. We study an underdoped
YBa\textsubscript{2}Cu\textsubscript{3}O\textsubscript{7-$\delta$} thin film in
a tunable terahertz cavity formed with a semi-transparent gold mirror.
From temperature-dependent terahertz transmission measurements, we find
that the cavity enhances the superconducting response below the critical
temperature, with an increase of the inferred superfluid weight. The
effect becomes more pronounced at smaller cavity lengths and is
accompanied by an upward shift of the superconducting onset temperature.
Calculations based on a cavity-coupled model for phase-fluctuating
superconductors capture these trends and support an interpretation in
terms of cavity-enhanced phase stiffness. These results showcase the
potential of cavity engineering for designing emergent functionalities
in correlated systems.}\end{minipage}
\end{center}

\vspace{1.5cm}

The control of quantum materials through cavity electrodynamics has seen
a rapid development in recent years~\cite{Garcia-Vidal2021, Schlawin2022, Bretscher2026}. A growing body of theoretical work has proposed that modifying the electromagnetic environment can reshape collective states of
matter~\cite{Laplace2016, Sentef2018, Schlawin2019, Curtis2019, Gao2020, Ashida2020, Latini2021, Chiocchetta2021, Bloch2022, Dmytruk2022, ViasBostrm2023, Lu2024, Andolina2024, Fassioli2025, Lu2025, Flores-Caldern2025, Islam2025, Kozin2025, Plastovets2026}, while a series of experiments has begun
to show that cavity confinement can alter the properties of correlated
electronic systems~\cite{Thomas2021, Appugliese2022, Jarc2023, Jarc2024, Thomas2025, Kipp2025, Enkner2025, Graziotto2026, Helmrich2026, Keren2026, Xu2026}. These advances have established cavity quantum materials as a promising route to
manipulating many-body phases beyond conventional tuning parameters such
as doping, pressure or external fields.

Superconductivity has only recently entered this experimental landscape.
Initial studies of cavity-embedded superconductors have shown that the
superconducting condensate can be directly affected by confinement of
the electromagnetic field. Strong reduction of the superfluid density
has been reported in both an organic superconductor~\cite{Keren2026}
and in NbN~\cite{Xu2026} coupled to optical cavities. These
results provide clear evidence that cavity electrodynamics can influence
superconducting ground states, but they also suggest that the effect is
not generically favourable to superconductivity. This naturally raises
the question of whether cavity electrodynamics can, under appropriate
conditions, enhance rather than suppress superconductivity.

\begin{figure}[t]
  \centering
  \includegraphics[width=1\textwidth]{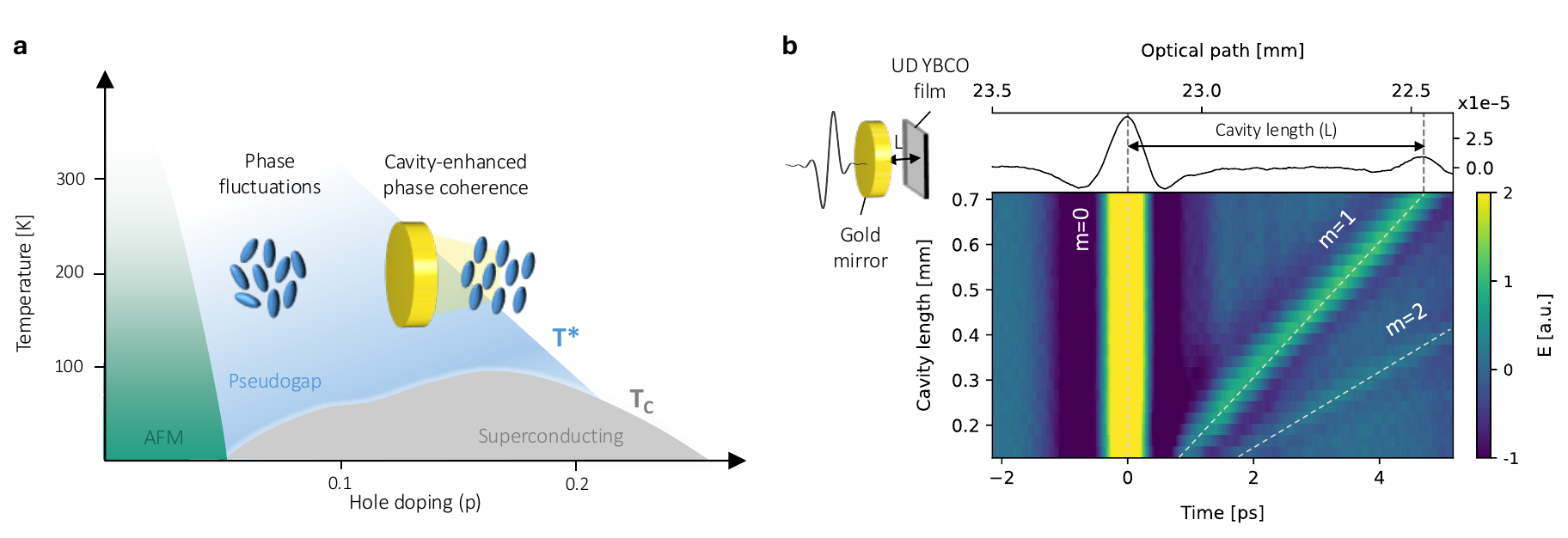}
  \captionsetup{font=footnotesize}
  \caption{\textbf{Cavity control of phase coherence in underdoped YBa\textsubscript{2}Cu\textsubscript{3}O\textsubscript{7-$\delta$}. a,} Phase fluctuations of the order parameter are widely discussed as an important factor limiting superconductivity in high-$T_c$ cuprates. In underdoped (UD) samples, evidence for incoherent local pairing has been reported throughout the pseudogap phase. We investigate whether the presence of a mirror in proximity to the superconductor could effectively act as a cavity, promote phase coherence and enhance the superconducting response. \textbf{b,} The sketch illustrates the cavity geometry. Terahertz (THz) time-domain spectroscopy provides a direct way to track the optical path of photons inside the cavity: the delay between the main (m=0) and first (m=1) transmitted peaks yields the cavity round-trip time and thus a robust calibration of the cavity length. The colormap shows the room-temperature THz transmission as a function of cavity length.}
  \label{fig:main1}
\end{figure}

Underdoped cuprates provide an exceptionally compelling platform in this
respect. In these materials, electron pairing and long-range phase
coherence are widely thought to develop on different temperature scales
rather than emerging simultaneously~\cite{Emery1995, Lee2006, Wang2006}. In
particular, the pseudogap phase appearing above the critical temperature
($T_c$) in underdoped cuprates is widely discussed as a
state in which strong pairing correlations persist up to a higher
temperature $T^*$, even though macroscopic superconducting coherence is
weakened or lost~\cite{Loeser1996, Shi2009, Levchenko2011, Boschini2018, Bastiaans2021, Niu2024} (Figure~\ref{fig:main1}a). From this
perspective, a possible route to enhancing superconductivity would be to
suppress the mechanisms that disrupt phase coherence rather than
strengthening the pairing itself. This possibility is especially
intriguing in light of theoretical proposals that cavity photons can
mediate effective long-range interactions and stabilize collective order
in quantum matter~\cite{Hammer2011, Laplace2016, Gao2020, Chakraborty2021}.

\

Here, we investigate whether a cavity-modified electromagnetic environment can influence superconducting coherence in a phase-fluctuation-dominated cuprate. We study a 100-nm-thick underdoped YBa\textsubscript{2}Cu\textsubscript{3}O\textsubscript{7-$\delta$} (YBCO) film
embedded in a tunable cavity formed by the film itself and a
semi-transparent gold mirror (Figure~\ref{fig:main1}b). The tunable distance between the superconducting film and the mirror sets the fundamental mode of the
cavity and provides a tunable photonic environment for YBCO.

Time-domain terahertz (THz) spectroscopy is a particularly well-suited
technique to this platform for two reasons. First, it provides a direct,
\emph{in situ} and non-invasive measure of the cavity length: the delay
between successive transmitted pulses yields the photon round-trip time
inside the cavity, and therefore the optical path between the film and
the mirror. Second, THz spectroscopy is highly sensitive to the onset of
superconductivity. The low-frequency complex conductivity directly
reveals the inductive response of the condensate through its imaginary
component. For this reason, THz conductivity measurements have long been
used to track the superconducting transition in thin film
cuprates~\cite{Basov2005, Orenstein2006}.

The room-temperature THz electric field transmitted by the cavity
assembly is shown in Figure~\ref{fig:main1}b for different cavity lengths. As the
mirror position is tuned with nanometric precision, the first cavity
reflection (m=1) shifts systematically in time. This delay provides a
robust calibration of the cavity round-trip time and hence of the cavity
length (see Methods for a detailed description of the experimental
setup).

\

\begin{figure}[t]
  \centering
  \includegraphics[width=0.7\textwidth]{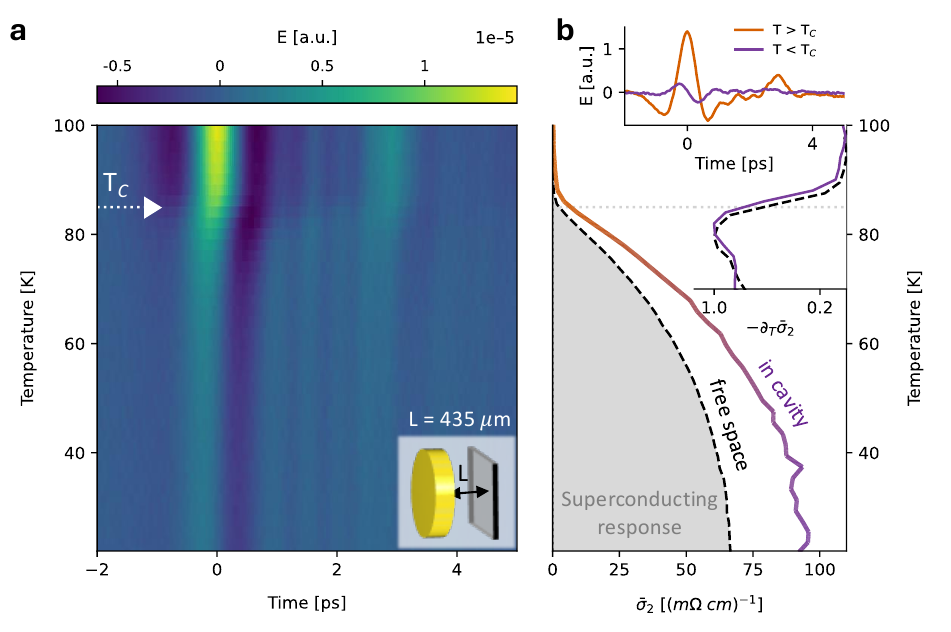}
  \captionsetup{font=footnotesize}
  \caption{\textbf{Enhanced superconducting response in a terahertz cavity. a, }Temperature-dependent transmitted THz electric field for a representative cavity length (L=435 $\mu$m). THz traces recorded at T > $T_c$ (orange) and at T < $T_c$ (purple) are plotted in the inset of panel (b). On cooling below $T_c$, the THz waveform shows both a reduction in transmission and a shift in arrival time of the peak, consistent with the emergence of an inductive superconducting response. \textbf{b,} Frequency-averaged imaginary conductivity, $\bar{\sigma}_2$, extracted from the THz response and plotted as a function of temperature. The frequency averaging range is (0.4 THz – 1 THz). As a measure of the condensate-related superconducting response, $\bar{\sigma}_2$ is systematically larger inside the cavity (solid coloured curve) than outside it (dashed black curve) below $T_c$, indicating a cavity-induced enhancement. Inset, normalized temperature derivatives of the curves in the main panel, shown for the film outside the cavity (dashed line) and inside the cavity (solid line).}
  \label{fig:main2}
\end{figure}

With the cavity geometry calibrated at room temperature, we next
investigate its response across the superconducting transition.
Thermally driven changes do not measurably modify the cavity length over
the temperature range explored here, so the mirror-film separation can
be treated as constant during cooling (Supplementary Information Sec.~\ref{sec:cav_length}). For a fixed mirror position (L = 435 \(\mu\)m), we record the transmitted THz electric field as a function of temperature (Figure~\ref{fig:main2}).

The superconducting transition is immediately visible in the
temperature-dependent map, occurring at approximately 85 K, in agreement
with the sample specifications (Methods). Across the transition, the
transmitted THz field undergoes two pronounced changes: i) its amplitude
is strongly reduced on entering the superconducting state, ii) the
transmitted pulse acquires a clear phase shift. These changes are
illustrated more clearly by representative traces measured above
(orange) and below (purple) the critical temperature (inset in Figure~\ref{fig:main2}b). Below $T_c$, the waveform develops a more
derivative-like shape, reflecting the emergence of a strong inductive
response. In electrodynamics terms, this behaviour arises from the rapid
increase of the imaginary conductivity ($\sigma_{2}$) driven by the
condensate, which enhances reflectivity and shifts the peak of the
transmitted field in time.

Additional structures are visible in the map. In particular, the feature
at \textasciitilde{} 2.9 ps corresponds to the first cavity reflection
(m=1), whose temperature dependence follows that of the main
transmitted pulse. Other weaker features at shorter time delays
originate from the LaAlO\textsubscript{3} (LAO) substrate on which the
YBCO film is grown and are not discussed further here (Supplementary
Information, Sec.~\ref{sec:characterization}).

\

To quantify the cavity-induced change in the superconducting response, we
convert the temperature-dependent THz transmission into the complex
conductivity of the YBCO film using the thin-film Tinkham
formalism~\cite{Tinkham1956}. The analysis is performed with respect to
a reference measurement of the empty cavity having the same length,
consisting of the semi-transparent mirror and the LAO substrate without
the superconducting film. Importantly, the analysis explicitly accounts
for the temperature dependence of all cavity elements as well as for
propagation through the full structure (Supplementary Information, Sec.~\ref{sec:tinkham}). From the extracted conductivity, we evaluate the frequency-averaged imaginary part
\({\bar{\sigma}}_{2} = \frac{1}{\Delta\omega}\int_{\omega_{1}}^{\omega_{2}}{\sigma_{2}(\omega)d\omega}\),
which is proportional to the superfluid weight and therefore provides a
figure of merit of the superconducting condensate response. 

Figure~\ref{fig:main2}b shows this quantity as a function of temperature for the film inside (solid coloured curve) and outside the cavity (dashed black curve). In both cases, \({\bar{\sigma}}_{2}\) rises sharply below $T_c$, as expected for the onset of superconductivity, but its magnitude is larger in the cavity, indicating an enhanced superfluid response. We stress that this conclusion does not rely on a single analysis protocol: independent approaches based on direct time-domain fitting and on the phase shift of the transmitted THz pulse yield
consistent results (Supplementary Information, Sec.~\ref{sec:time_domain}). Taken together, these observations provide evidence that, for this cavity geometry, the superconducting condensate response is enhanced inside the cavity.

A closer inspection of Figure~\ref{fig:main2}b shows that \({\bar{\sigma}}_{2}\)
starts to increase at slightly higher temperature inside the cavity,
with a small but reproducible finite response visible above the critical
temperature of the film without the cavity. The normalized
temperature-gradients of the curves, plotted in the inset of Figure~\ref{fig:main2}b for a temperature range around the transition, highlight this trend, suggesting that the cavity promotes the emergence of the superconducting condensate response at slightly higher temperature. 

We performed several control measurements to rule out trivial origins of this effect, including thermal drifts, differences in thermal contact, direction of
the temperature scan (heating/cooling) and THz-induced heating
(Supplementary Information, Sec.~\ref{sec:control_exp}). In all the tests, the upward shift of the transition temperature is consistently reproduced. Crucially, the shift disappears when the metallic gold layer is removed from the
mirror, and only the mirror dielectric substrate is brought in proximity
to the YBCO film (Supplementary Information, Sec.~\ref{ssec:quartz}).

\begin{figure}[t]
  \centering

  \begin{minipage}[t]{0.62\textwidth}
    \vspace{0pt}
    \centering
    \includegraphics[width=\linewidth]{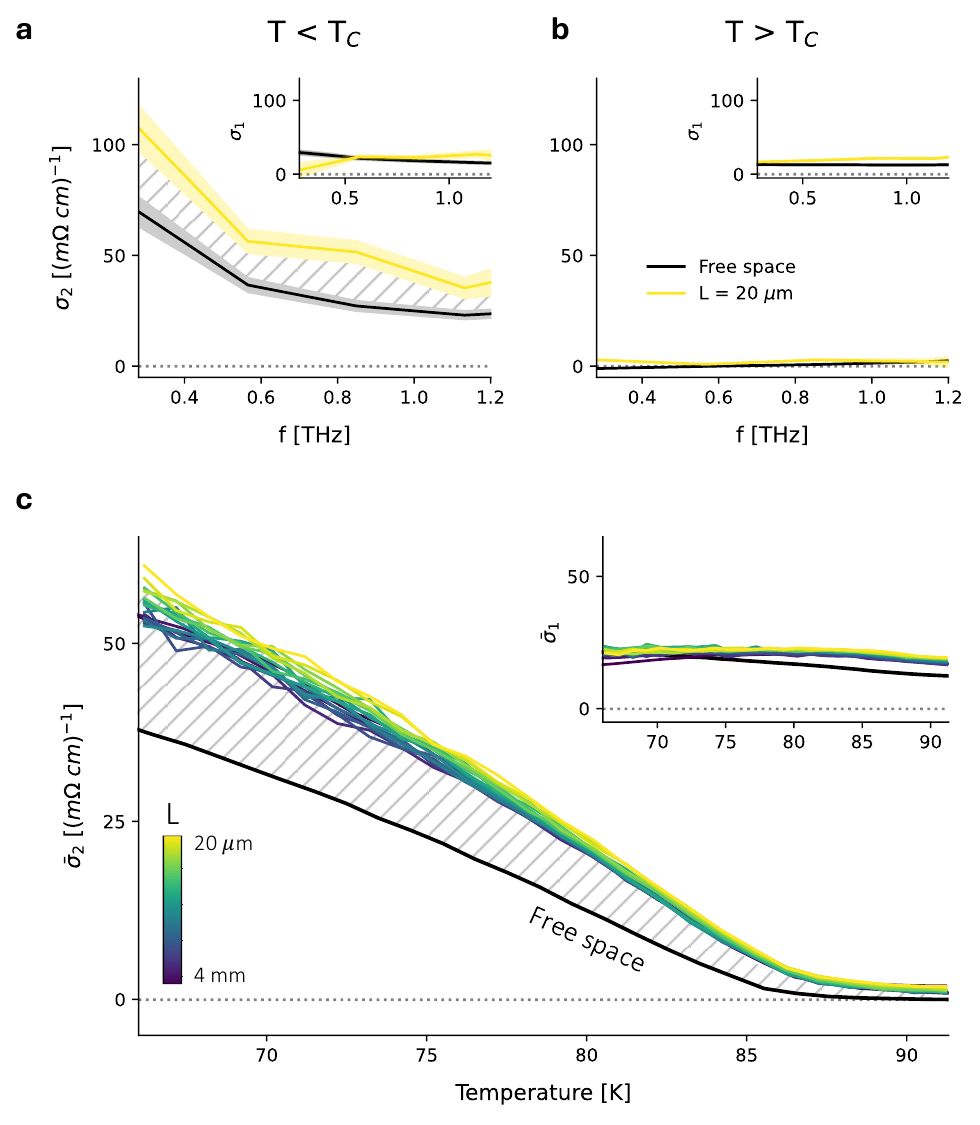}
  \end{minipage}
  \hfill
  \begin{minipage}[t]{0.32\textwidth}
    \captionsetup{font=footnotesize}
    \caption{\textbf{Cavity-length dependence of the enhanced superconducting response.} Frequency-dependent $\sigma_2$ (and $\sigma_1$ in the insets) below (\textbf{a,} T = 66K) and above $T_c$ (\textbf{b,} T = 90K), measured in free space (black curve) and inside the cavity (yellow, L = 20 $\mu$m). The most pronounced cavity-induced change occurs in $\sigma_2$ below the superconducting transition, where the inductive response is enhanced. The shaded areas indicate the propagated standard uncertainties. \textbf{c,} Frequency-averaged $\sigma_2$ (and $\sigma_1$ in the inset) as a function of temperature for different cavity lengths, compared to the one for YBCO in free space (black curve). The enhancement of the superconducting response is larger at shorter cavity lengths.}
    \label{fig:main3}
  \end{minipage}

\end{figure}

\

To resolve how the cavity modifies the electrodynamics across the
superconducting transition, we next vary the mirror-film separation and
focus on the optical conductivity close to $T_c$. For the
shortest cavity length studied (L = 20 \(\mu\)m), Figures~\ref{fig:main3}a and ~\ref{fig:main3}b show the frequency-dependent optical conductivity extracted below and above the transition, respectively. The shaded areas indicate the propagated standard uncertainty in the extraction of the optical conductivity.
Below $T_c$, the imaginary conductivity
\(\sigma_{2}(\omega)\) displays the characteristic 1/\(\omega\)-like
increase towards low frequency expected for the inductive response of
the superconducting condensate. Importantly, this response is larger
inside the cavity (yellow) than outside the cavity (black), as
emphasized by the hatched area between the two curves. By contrast, the
real conductivity \(\sigma_{1}(\omega)\), shown in the insets, exhibits
only minor cavity-dependent variations. Above $T_c$,
\(\sigma_{2}(\omega)\) becomes negligible both inside and outside the
cavity, while \(\sigma_{1}(\omega)\) shows no systematic cavity-induced
change. The most pronounced cavity-induced effect therefore appears in
\(\sigma_{2}\) below the superconducting transition, indicating that the
cavity selectively enhances the inductive, condensate-related response
without significantly affecting the dissipative channel.

This behaviour becomes even clearer when the conductivity is averaged
over frequency for different cavity lengths as a function of temperature
(Figure~\ref{fig:main3}c). The inset shows that the averaged \(\sigma_{1}\) remains essentially independent of cavity length and follows the same
temperature dependence inside and outside the cavity. By contrast, the
averaged \(\sigma_{2}\) in the main panel displays a clear and
systematic cavity-length dependence: below the transition, it is
consistently enhanced in the cavity, and the enhancement grows as the
mirror is brought closer to the film.

\begin{figure}[t]
  \centering
  \includegraphics[width=0.85\textwidth]{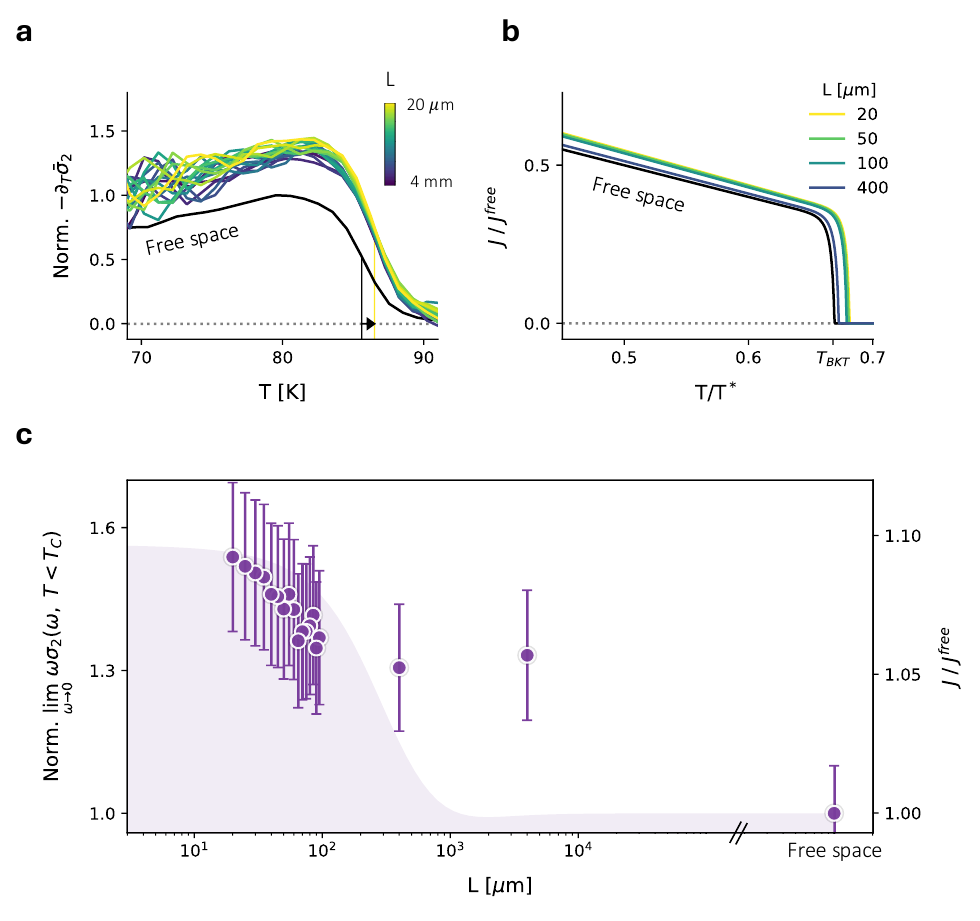}
  \captionsetup{font=footnotesize}
  \caption{\textbf{Signatures of enhanced phase stiffness in the cavity.} \textbf{a,} Normalized temperature derivatives of the data in Figure~\ref{fig:main3}c, highlighting the evolution of the superconducting transition inside the cavity (coloured curves) relative to the film outside the cavity (black curve). The transition shows a small upward shift of the onset temperature, indicated by the horizontal arrow. \textbf{b,} Results of calculations for a cavity-coupled \textit{XY}/BKT model. The model captures the two main experimental trends: an increase in phase stiffness $J$ below the superconducting transition and a cavity-dependent shift of the TBKT. \textbf{c,} Experimental values of $\omega\sigma_2 (\omega\rightarrow0)$ as a function of cavity length (fixed T = 66K). The increase for shorter cavities is in qualitative agreement with the calculated cavity-enhanced phase stiffness (purple shaded area, $T/T^*$=0.55). Error bars represent the propagated standard uncertainties.}
  \label{fig:main4}
\end{figure}

Figure~\ref{fig:main4}a summarizes the temperature evolution of the condensate
response for all cavity lengths. By plotting the normalized
temperature-derivatives of the averaged \(\sigma_{2}\) referenced to the
bare film, two features become apparent. First, the superconducting
response remains systematically larger in the cavity over the full
transition region. Second, the maximum shifts to higher temperature by
up to about 1 K, as highlighted by the horizontal arrow. The black and
yellow vertical lines indicate the transition temperatures in free space
and for L = 20 \(\mu\)m, respectively, estimated from the
second-derivatives of the curves (Supplementary Information, Sec.~\ref{sec:find_tc}).
These trends suggest that cavity confinement affects not only the magnitude of the condensate response in the superconducting
phase, but also its onset.

\

As a minimal framework for interpreting these observations, we use a
phenomenological description of a phase-fluctuating superconductor in
which the superconducting order parameter can be treated, at a
coarse-grained level, as having a nearly fixed amplitude but a
fluctuating phase. In underdoped cuprates, this description is
consistent with the phase diagram in Figure~\ref{fig:main1}, where pairing-related correlations emerge below $T^*$, while long-range superconducting coherence develops only at the lower temperature $T_c$~\cite{Emery1995, Loktev2001, Carlson1999}. For a low-dimensional
layered system such as YBCO in the weak interlayer-coupling limit, this
picture implies a superconducting transition of the
Berezinskii-Kosterlitz-Thouless (BKT) type, driven by the unbinding of
vortex-antivortex pairs~\cite{Berezinskii1971, Kosterlitz1973, Minnhagen1987}. 

It should be stressed that this approach neglects the full three-dimensional character of YBCO and it is not intended here as a microscopic description of the
material. Rather, it provides a natural minimal framework for addressing
how the presence of the cavity may affect phase coherence in
superconductors controlled by phase fluctuations.

Motivated by this picture, we describe the superconducting film within a
long-wavelength approximation to the \textit{XY} model, which is equivalent to
the coarse-grained hydrodynamic London framework~\cite{Benfatto2004}
(see Supplementary Information, Sec.~\ref{sec:model} for details).~This model has a nonlinearity (mode-coupling term) that allows thermal and quantal phase
fluctuations to renormalize the static long wavelength phase stiffness
that is relevant to the BKT transition. 

We consider the film inside a
Fabry-Pérot cavity and couple the space and time derivatives of the
superconductor's phase field to the~electromagnetic field in the usual
way. A crucial point of the analysis is that within the plane of the
film, the vector potential associated with fluctuations of the cavity
modes has a longitudinal component that couples strongly to the
superconducting phase at nonzero frequency and is strongly affected by
the cavity. The latter acts to increase the superfluid stiffness,
reducing the amplitude of the thermal and quantal fluctuations of the
phase field and thus reducing the mode-coupling contribution to the BKT
stiffness. This reduction appears first as an enhancement of the
stiffness \(J\), which in turn suppresses long-wavelength phase
fluctuations below the transition. At the same time, the larger
stiffness raises the energetic cost of vortex excitations, thereby
strengthening vortex-antivortex binding and shifting the BKT transition
to higher temperatures. The model therefore provides a qualitative
framework that is consistent with the experimental observations.

\

The physical understanding of the phase stiffening within this model is
the following. The controllable photon gap created by the cavity mirrors
reduces the hybridization of the vector potential (i.e. the photons
contributing to non-static electric field) with the superconducting
low-energy phase correlations, protecting the latter from electric-field
fluctuations and thus increasing the stiffness compared with free space
(Supplementary Information, Sec.~\ref{ssec:model_J}).

Figure~\ref{fig:main4}b shows the~ cavity-renormalized stiffness \(J\), relative to its free-space value, for a few representative cavity lengths chosen to
match the experimental range. In the calculations, cavity confinement
enhances the phase stiffness below the transition temperature of the
film outside the cavity. When vortex fluctuations are included, this
increase is accompanied by an upward shift of the estimated BKT
transition temperature T\textsubscript{BKT}, consistent with stronger
phase rigidity and delayed vortex unbinding.

Figure~\ref{fig:main4}c compares the measured cavity-length dependence of the
low-frequency limit of \(\omega\sigma_{2}(\omega)\) below
$T_c$ with the corresponding dependence of the calculated
cavity-dependent phase stiffness. In the experiment,
\(\lim_{\omega \rightarrow 0}{\omega\sigma_{2}(\omega)}\), which is
proportional to the superfluid density in the London regime, increases
as the cavity is shortened. Over the same range of cavity lengths, the
calculations follow the same trend.~We note that at the longest cavity
length (L = 4 mm) the measured response remains above the free-space
value, whereas the minimal model would suggest the cavity-induced
enhancement should already be suppressed. The residual response likely
reflects ingredients that are not fully captured by the simplified
phase-only description, including the effective treatment of the
experimental cavity geometry and additional electrodynamic
contributions.

\

In the present configuration, although the mirror positioning can in
principle reach smaller nominal separations, the minimum reproducible
mirror-film distance is ultimately limited by alignment. At sufficiently
small spacing, contact between the mirror and the film can occur,
leading to an apparent shift of the transition temperature caused by
local heating (Supplementary Information, Sec.~\ref{sec:cav_align}). Accessing shorter cavity lengths will therefore require, for instance, the deposition of
cavity-like heterostructures embedding the superconducting layer. Cavity
assemblies in which the layer is positioned closer to the field
antinode, or in general more symmetrically within the cavity, may also
enhance the effective light-matter coupling and increase the magnitude
of the phase-stiffness renormalization.

The same strategy adopted here is likely to be relevant beyond YBCO.
More anisotropic cuprates such as
Bi\textsubscript{2}Sr\textsubscript{2}CaCu\textsubscript{2}O\textsubscript{8+$\delta$} are especially promising, because their stronger two-dimensional and
weaker interlayer coupling make superconducting phase fluctuations more
pronounced~\cite{Corson1999, Yong2012}. In such materials, cavity-induced
modifications of phase coherence may therefore be even more visible,
providing a compelling test of the phase-fluctuation scenario discussed
here. This perspective is also consistent with the recent study of
cavity-altered superconductivity, where pronounced superfluid
suppression was observed both under resonant interfacial coupling in an
organic superconductor~\cite{Keren2026} and non-resonant conditions
in a conventional superconductor~\cite{Xu2026}. Taken together
with our results, this contrast suggests that cavity schemes acting
primarily through dressing of pairing-related modes do not necessarily
reinforce the condensate response, whereas acting on the phase sector
may offer a more favorable route to promoting and sustaining
superconducting phase coherence in fluctuation-dominated cuprates. By
controlling the emergence of long-range coherence, this approach paves
the way to light-matter hybrids exhibiting superconductivity at higher
temperature.

\clearpage

\section*{Methods}
\subsection*{Cavity assembly}
The cavity assembly was designed and realized in-house, inspired by the
setup reported in Ref.~\cite{Jarc2022}. It consists of a
semi-reflecting gold mirror facing a YBCO thin film grown on a
dielectric substrate, forming a planar Fabry-Pérot-like geometry. The
mirror-sample separation is controlled by three piezoelectric actuators
(Physik Instrumente), enabling nanometric tuning of the cavity length.
The entire assembly is integrated into a closed-cycle liquid-helium
cryostat (ARS) with vibration damping.

The cavity mirror is fabricated by evaporating a Ti/Au bilayer (5/10 nm)
onto a 2-mm-thick z-cut quartz substrate. It is mounted on a copper disk
using conductive silver paint and mechanically connected to a copper
plate actuated by the piezo positioners. Thermalization is achieved via
a copper braid connected to the cryostat cold finger. To ensure reliable
operation of the piezo actuators at low temperature, they are thermally
decoupled from the cold stage using a PEEK spacer.

The YBCO sample is a 100-nm-thick film grown on a LaAlO\textsubscript{3} substrate (10 × 5 × 0.5 mm\textsuperscript{3}) with a 10-nm CeO\textsubscript{2} buffer layer (Ceraco). The film has a
critical temperature $T_c$ = 85 K and a specified critical
current density \(J_{c} \simeq 3\ \)MA/cm\textsuperscript{2}. The sample is mounted with
conductive silver paint onto a copper disk featuring a 6 × 3 mm\textsuperscript{2}
aperture for transmission measurements. The disk is housed in a copper
holder allowing both azimuthal and polar alignment, and is in direct
thermal contact with the cryostat cold finger to ensure efficient heat
dissipation.

The temperatures of the mirror and the sample are monitored
independently using silicon diode sensors, while a third diode mounted
on the cold finger is used for PID temperature control. The base
temperature at the cold finger is 9 K. Due to finite thermal
resistances, temperature offsets of approximately 1.5 K at the sample
and up to \textasciitilde20 K at the mirror are observed (Supplementary
Information, Sec.~\ref{sec:characterization}). All the temperatures reported in the main text are the ones measured at the sample's position.

The cavity alignment is performed at room temperature using a visible
laser beam collinear with the THz path. The back-reflected beams from
the mirror and the sample are overlapped in the far field to ensure
parallelism. For measurements without the cavity, the mirror is removed
without disturbing the thermal contacts or the sample mounting. A new
reference measurement is acquired whenever the thermal configuration is
modified.

\subsection*{Terahertz time-domain spectroscopy}

The superconducting response of the YBCO film is probed by
temperature-dependent terahertz time-domain spectroscopy in transmission
geometry. Broadband THz pulses are generated using a large-area GaAs
photoconductive antenna excited by ultrashort optical pulses (800 nm, 20
fs, 12 \(\mu\)J) from a commercial laser system (Carbide + Orpheus-N 2H, Light
Conversion) operating at 40 kHz.

The THz radiation is focused onto the sample using a gold-coated
parabolic mirror, yielding a beam waist of approximately 1 mm at the
sample position. The transmitted THz field is detected via electro-optic
sampling in a 0.5-mm-thick ZnTe crystal using a synchronized probe pulse
(800 nm, 20 fs, \textless50 nJ). The orthogonal polarization components
of the probe beam are measured with a balanced photodetector (Thorlabs),
and the differential signal is recorded using lock-in detection
referenced to the photoconductive antenna bias.

The signal-to-noise ratio of the measurement exceeds
\(2 \times 10^{4}\). The THz generation and detection paths are enclosed
in a nitrogen-purged environment to suppress absorption by atmospheric
water vapor.

\section*{Acknowledgements}
We thank Lara Benfatto and Martin Eckstein for helpful discussions. We
are also grateful to Jürgen Linzmayer and Klaus Wölfel for their
technical support and for the mechanical construction of the cavity
assembly. This work was mainly supported by the Gordon and Betty Moore
Foundation through the Grant CENTQC (No.~GBMF12213). DF acknowledges
support from the European Union's Horizon Europe~research and innovation
programme under the Marie Skłodowska-Curie grant No
HORIZON-MSCA-2023-DN-01 101169225 - SPARKLE. FP acknowledges funding
from the Munich Quantum Valley within the~Hightech Agenda Bayern Plus
supported~by the State Ministry of Science and the Arts. VP was
supported by the Deutsche Forschungsgemeinschaft (DFG, German Research
Foundation) under the Walter Benjamin Programme (No. 566401345). DF,
MAS, JF and DMK acknowledge funding by the Deutsche
Forschungsgemeinschaft (DFG, German Research Foundation) -- 531215165
(Research Unit ``OPTIMAL''). MAS was funded by the European Union (ERC,
CAVMAT, No. 101124492).

% =====================================================
% Main-text references
% =====================================================

\putbib[main_references]
\end{bibunit}

% =====================================================
% Supplementary Materials, starting on a new page, with own bibliography
% =====================================================
\begin{bibunit}
% =====================================================
% Supplementary Materials start here.
% This file has its own BibTeX bibliography at the end.
% =====================================================
\clearpage
\thispagestyle{empty}
% Reset counters and label figures/tables/equations as supplementary.

\setcounter{figure}{0}
\setcounter{table}{0}
\setcounter{equation}{0}
\renewcommand{\thefigure}{S\arabic{figure}}
\renewcommand{\thetable}{S\arabic{table}}
\renewcommand{\theequation}{S\arabic{equation}}

\renewcommand{\thesection}{S\arabic{section}}
\renewcommand{\thesubsection}{S\arabic{section}.\arabic{subsection}}
\setcounter{section}{0}

%\section*{Supplementary Informatopm}
%\addcontentsline{toc}{section}{Supplementary Materials}

\startcontents[sm]

\section*{Supplementary Information}
\printcontents[sm]{}{1}{\setcounter{tocdepth}{2}}

\clearpage
\setcounter{page}{1}
\pagestyle{plain}

\section{Propagation effects}
\begin{figure}
\centering
\includegraphics[width=1.0\linewidth]{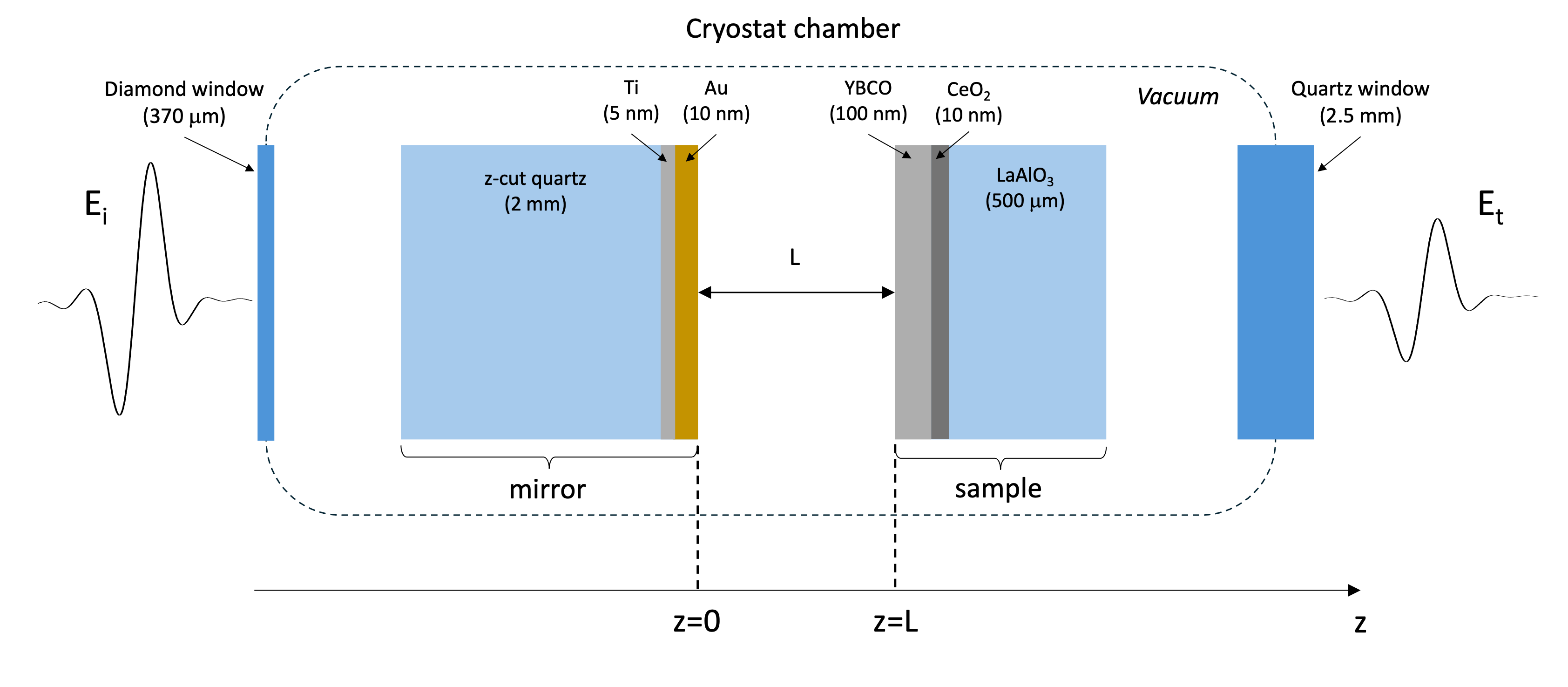}
\caption{\label{fig:cavity_setup}Sketch of the cavity assembly. All the optical elements and their thicknesses are reported. }
\end{figure}

The sample space is schematically shown in Fig.~\ref{fig:cavity_setup}. The measured THz field $E_{t}$ results from the propagation of the incident field through the sequence of optical elements indicated in the sketch. Its amplitude and phase encode the electromagnetic response of the YBCO film, which we extract by carefully modeling the propagation through the entire setup.
 
In principle, a full transfer matrix analysis is required to model the propagation of THz radiation through the whole cavity assembly. We first show that, for the thick dielectric elements in the setup (namely air, diamond, quartz, and LaAlO$_3$), the temporal shape of the THz pulse allows the model to be greatly simplified by time-windowing the signal, reducing the transfer matrix to a product of single-interface transmission coefficients and phase factors (Sec.~\ref{ssec:dielectrics}). The propagation through the cavity region, which involves thin conductive layers, requires a separate treatment and is addressed in Sec.~\ref{ssec:cavity}.

\subsection{Thick dielectric layers \label{ssec:dielectrics}}

The transmitted electric field $E_t(\omega)$ through a slab of material with real refractive index $n$ and thickness $d$, surrounded by vacuum, is given by \cite{born_99}:
\begin{equation}
    E_t(\omega)=\frac{tt^\prime e^{in\omega d/c}}{1-{r^\prime}^2e^{i2n\omega d/c}}
    E_{i}(\omega),
\end{equation}
where $t$, $t^{\prime}$ are the transmission coefficients at the vacuum-medium and
medium-vacuum interfaces, respectively, and $r^\prime$ is the reflection coefficient
at the medium-vacuum interface at normal incidence:
\begin{align}
t&=\frac{2}{n+1}\nonumber\\
t^\prime&=\frac{2n}{n+1}\nonumber\\
r^\prime&=\frac{1-n}{n+1}.
\end{align}

For a wave packet with carrier frequency $\Omega$ and temporal envelope of width $\tau$, e.g.\ a Gaussian pulse $E(t)=E_0e^{-t^2/(2\tau^2)}\sin(\Omega t)$, with $\tau\Omega/(2\pi)\sim1$ as typical for single-cycle THz pulses generated by the photoconductive antenna in our setup, and assuming $n$ to be constant over the pulse bandwidth, the transmitted field in the time domain reads:
\begin{align}
E_t(t)&=\int d\omega\,e^{-i\omega t}tt^\prime e^{in\omega d/c}\sum_{m=0}^{+\infty}
({r^\prime}^2e^{i2n\omega d/c})^mE_i(\omega)\nonumber\\
&\equiv tt^\prime\sum_{m=0}^{+\infty}{r^\prime}^{2m}E_i(t-T-m2T),
\end{align}
where $T=nd/c$ is the single-pass propagation time through the medium. The transmitted field thus consists of a series of time-delayed copies of the input pulse, each separated by the round-trip time $2T$ and weighted by successive powers of ${r^\prime}^2$. When the pulse duration $\tau$ is much shorter than $2T$, consecutive echoes are temporally well separated and the primary transmitted pulse ($m=0$) can be isolated by time-windowing. In this regime, the windowed spectrum reduces to:
\begin{equation}
\label{eq:S_die}
    E_t(\omega)=tt^\prime e^{in\omega d/c}E_i(\omega).
\end{equation}

The THz pulses generated by the photoconductive antenna used in our setup have a central frequency $\Omega/2\pi=0.7$~THz and a duration $\tau\sim0.5$~ps, corresponding to a spectral bandwidth of $\sim1$~THz. As shown in Sec. \ref{sec:characterization}, the round-trip times through the dielectric elements (diamond, quartz, LaAlO$_3$) in the apparatus all significantly exceed this duration.
%\begin{itemize}
 %   \item \textbf{Diamond}: the refractive index is flat and non-absorbing in the few-THz range due to the absence of optically-active phonons. For $n=2.4$ and $d=500~\mu$m, $2T\sim8$~ps \cite{dore_Appl.Opt.AO98}.

 %   \item \textbf{Quartz}: an optically-active phonon at $\sim3.9$~THz lies well outside our spectral range, resulting in negligible dispersion and absorption within the relevant bandwidth. For $n=2.1$ and $d=500~\mu$m, $2T\sim7$~ps \cite{davies_JInfraredMilliTerahzWaves18}.

%    \item \textbf{LaAlO$_3$}: small apparent absorption features near $\sim2$~THz have been attributed to crystal anisotropy rather than physical phonons \cite{zou_AIPAdvances12,lloyd-hughes_Opt.Lett.OL14}. The associated refractive-index dispersion remains at the few-percent level and the extinction coefficient is small throughout the relevant range. For $n=5.1$ and $d=500~\mu$m, $2T=17$~ps.
%\end{itemize}

In all cases, the reshaping of the main peak is well described by the amplitude and phase factors in Eq.~\ref{eq:S_die}, while the delayed echoes visible at later times are consistent with the round-trip time estimates above. Time-windowing can therefore be safely applied to isolate the primary transmitted pulse.

For a stack of $N$ dielectric layers sandwiched between vacuum (indexed $j=0$ and $j=N+1$), the overall transmission function in the windowed approximation is:
\begin{equation}\label{eq:diel_stack}
E_t(\omega)=t_{0,1}\prod_{j=1}^{N}t_{j,j+1}\,e^{in_j\omega d_j/c}\,E_i(\omega),
\end{equation}
where $t_{j,j+1}$ is the transmission coefficient at the interface between layers $j$ and $j+1$, $n_j$ is the refractive index of layer $j$, and $d_j$ its thickness. 

\subsection{\label{ssec:cavity}Cavity transmission}

The two mirrors closing the cavity, the titanium-gold layer and the superconducting YBCO layer, are treated as thin conductive sheets, following the approach of the Tinkham formula \cite{tinkham_Phys.Rev.56}. This formulation allows the complex conductivity to be extracted from the cavity transmissivity by simple algebraic manipulation. For both layers, the condition
\begin{equation}
    d_j\ll\delta_j,\lambda_j
\end{equation}
is satisfied, where
\begin{align}
\delta_j&=\frac{c}{\kappa_j\omega}\nonumber\\
\lambda_j&=\frac{2\pi c}{n_j\omega}
\end{align}
are the penetration depth and the wavelength of THz radiation inside the layer, expressed in terms of the complex refractive index $\tilde{n}_j=n_j+i\kappa_j$. For the metallic layers, this follows from the fact that at THz frequencies both titanium and gold are well within the Hagen-Rubens regime ($\omega\ll\gamma$, with $\gamma$ the scattering rate), where the optical response is governed by the static conductivity and $n=\kappa$ \cite{dressel_02}. The relevant condition then reduces to
\begin{equation}
d\ll\delta\sim\sqrt{\frac{c^2}{2\pi\omega\sigma_0}},
\end{equation}
which holds for both gold ($d\sim10$~nm, $\delta\sim80$~nm) and titanium ($d\sim5$~nm, $\delta\sim330$~nm) at $\omega/(2\pi)\sim1$~THz \cite{ordal_Appl.Opt.AO85,walther_Phys.Rev.B07}. An analogous argument applies to the YBCO layer in both the normal and superconducting states, with the London penetration depth replacing the skin depth in the latter case \cite{pimenov_PhysicaB:CondensedMatter98}.

The complex transmissivity of the cavity is obtained by applying boundary conditions for the electric and magnetic fields at the two thin conductive sheets. Materials with refractive indices $n_l$ and $n_r$ are placed on the left and right of the sheets, respectively, with vacuum inside the cavity. Following the geometry in Figure \ref{fig:cavity_setup}, the electric field is parametrized as
\begin{align}
E(z<0)&=E_+^l(\omega)e^{in_l\omega z/c}+E^l_-(\omega)e^{-in_l\omega z/c}\nonumber\\
E(0<z<L)&=E_+^c(\omega)e^{i\omega z/c}+E^c_-(\omega)e^{-i\omega z/c}\nonumber\\
E(z>L)&=E_+^r(\omega)e^{in_r\omega z/c}.
\end{align}
The transmitted field $E_+^r(\omega)$ is determined by imposing continuity of the electric field at $z=0$ and $z=L$:
\begin{align}
\label{eq:S_ce}
E_+^l+E^l_-&=E_+^c+E^c_-\nonumber\\
E_+^ce^{i\omega L/c}+E^c_-e^{-i\omega L/c}&=E_+^re^{in_r\omega L/c},
\end{align}
and continuity of the magnetic field in the presence of surface currents $K_0$ and $K_L$ at the two interfaces. Using $\mathbf{k}\times\mathbf{E}=-\partial_t\mathbf{B}$ with $B=\mu_0 H$, the magnetic field reads
\begin{align}
H(z<0)&=\frac{n_l}{Z}E_+^l(\omega)e^{in_l\omega z/c}-\frac{n_l}{Z}E^l_-(\omega)
e^{-in_l\omega z/c}\nonumber\\
H(0<z<L)&=\frac{1}{Z}E_+^c(\omega)e^{i\omega z/c}-\frac{1}{Z}E^c_-(\omega)
e^{-i\omega z/c}\nonumber\\
H(z>L)&=\frac{n_r}{Z}E_+^r(\omega)e^{in_r\omega z/c},
\end{align}
where $Z$ is the impedance of vacuum. The boundary conditions for $H$ give:
\begin{align}
\label{eq:S_ch}
n_lE_+^l-n_lE^l_-&=E_+^c-E^c_-+ZK_0\nonumber\\
E_+^ce^{i\omega L/c}-E^c_-e^{-i\omega L/c}&=n_rE_+^re^{in_r\omega L/c}+ZK_L.
\end{align}
The surface currents are related to the local electric fields via the bulk conductivities $\sigma_0$ and $\sigma_L$ of the layers at $z=0$ and $z=L$:
\begin{equation}
K_j=d_j\sigma_jE(z=j),
\end{equation}
where the interface values of $E$ are given by Eq.~\ref{eq:S_ce}. The linear system formed by Eqs.~\ref{eq:S_ce} and \ref{eq:S_ch} is cast as $Ax=y$, with unknown vector $x=\begin{pmatrix}E_-^l& E_+^c& E_-^c& E_+^r\end{pmatrix}$, matrix
\begin{equation}
A=
\begin{pmatrix}
-1&1&1&0\\
0&e^{i\omega L/c}&e^{-i\omega L/c}&-e^{in_r\omega L/c}\\
n_l+Zd_0\sigma_0&1&-1&0\\
0&-e^{i\omega L/c}&e^{-i\omega L/c}&(n_r+Zd_L\sigma_L)e^{in_r\omega L/c}\\
\end{pmatrix}
\end{equation}
and
\begin{equation}
y=E_+^l\begin{pmatrix}
1\\0\\n_l-Zd_0\sigma_0\\0
\end{pmatrix}.
\end{equation}
The solution for the transmitted field $E_t(\omega)=E_+^re^{in_r\omega L/c}$ takes the form of a modified Fabry-Pérot expression \cite{born_99}:
\begin{equation}
\label{eq:S_modtin}
E_t(\omega)=\frac{t_0t_Le^{i\omega L/c}}{1-r_0r_Le^{i2\omega L/c}}E_i(\omega),
\end{equation}
where $E_i(\omega)=E_+^l(\omega)$, and the modified transmission and reflection coefficients, incorporating the surface current contributions, are:
\begin{align}
\label{eq:S_rtc}
t_0&=\frac{2n_l}{n_l+1+Zd_0\sigma_0}\nonumber\\
t_L&=\frac{2}{n_r+1+Zd_L\sigma_L}\nonumber\\
r_0&=\frac{n_l-1+Zd_0\sigma_0}{n_l+1+Zd_0\sigma_0}\nonumber\\
r_L&=\frac{n_r-1+Zd_L\sigma_L}{n_r+1+Zd_L\sigma_L}.
\end{align}
As a consistency check, setting $n_l=1$, $d_0=0$, $L=0$ and $n_r=n_s$ recovers the standard Tinkham formula. In that limit, Eq.~\ref{eq:S_modtin} reduces to
\begin{equation}\label{eq:tinkham}
E_t(\omega)=\frac{2}{n_s+1+Zd\sigma}E_i(\omega),
\end{equation}
while in the absence of the film
\begin{equation}
\tilde{E}_t(\omega)=\frac{2}{n_s+1}E_i(\omega).
\end{equation}
The ratio $\mathcal{T}(\omega)=E_t(\omega)/\tilde{E}_t(\omega)$ then matches exactly the result of Ref.~\cite{tinkham_Phys.Rev.56}, from which the conductivity is retrieved as
\begin{equation}\label{eq:tinkham_first}
\sigma(\omega)=\frac{n_s+1}{Zd}\left(\frac{1}{\mathcal{T}(\omega)}-1\right).
\end{equation}

Combining Eqs.~\ref{eq:diel_stack} and \ref{eq:S_modtin}, the full propagation through the experimental apparatus is given by:
\begin{equation}
\label{eq:S_tot}
E_t(\omega)=\frac{t_{0,1}}{t_{l,l+1}}\prod_{j=1}^{l}t_{j,j+1}e^{in_j\omega d_j/c}
\frac{t_0t_Le^{i\omega L/c}}{1-r_0r_Le^{i2\omega L/c}}
\prod_{j=r}^{N}t_{j,j+1}e^{in_j\omega d_j/c}\,E_i(\omega).
\end{equation}
The first product accounts for propagation through the dielectric stack up to the layer immediately to the left of the first cavity mirror ($j=l$). Note that the factor $t_{0,1}/t_{l,l+1}$ removes the bare Fresnel transmission at the quartz-Ti/Au interface from the dielectric product, as this interface is already accounted for, including the surface current contribution, by $t_0$ in
Eq.~\ref{eq:S_rtc}. The second product covers propagation from the dielectric layer immediately to the right of the second cavity mirror ($j=r$) to the vacuum outside the apparatus. \newline

Equation~\ref{eq:S_tot} is the complete propagation model used to describe the experimental cavity assembly, combining the windowed transmission through the thick dielectric elements with the modified Fabry-Pérot response of the thin conductive cavity mirrors. It provides the basis for the conductivity extraction discussed in the Sec. \ref{sec:tinkham}.

\subsection{Simulations}
To test the validity of our model, we begin by fitting the field transmitted through the apparatus in time domain. For simplicity, we can rewrite Eq.\ \ref{eq:S_tot} as
\begin{equation}
E_t(\omega)=\bar{T}e^{i\omega T_d}\frac{t_0t_Le^{i\omega L/c}}{1-r_0r_Le^{i2\omega L/c}}E_i(\omega),
\end{equation}
where $\bar{T}e^{i\omega T_d}$ encodes all the effects of propagation inside the dielectric layers, which only amounts to a modification of the amplitude (in $\bar{T}$) and a trivial phase shift (in $e^{i\omega T_d}$) of $E_i(\omega)$. Noticing that $\bar{T}_d=T_d+L/c$ actually does not depend on the length of the cavity, since the amount of free space propagation that is gained (lost) by moving the mirror is actually lost (gained) from a shorter (longer) optical path on its left side, we can equivalently redefine 
\begin{equation}\label{eq:cav}
E_t(\omega)=\bar{T}e^{i\omega \bar{T}_d}\frac{t_0t_L}{1-r_0r_Le^{i2\omega L/c}}E_i(\omega).
\end{equation}
Here, the only dependence on cavity length $L$ is encoded in the denominator and produces the echo structure induced by multiple reflections. 

As a first check, we consider the transmitted field in the absence of the YBCO layer at ambient temperature. This geometry corresponds to a \textit{"reference empty cavity"}, in which the cavity mirror is approached to the LAO substrate. Since all the coefficients appearing in the previous expression are now real (see characterization of the optical elements in Sec.~\ref{sec:characterization}), we can directly evaluate the Fourier transform
\begin{equation}\label{eq:td}
E_t(t)=\bar{T}t_0\tilde{t}_L\sum_{m=0}^{+\infty}{(r_0\tilde{r}_L)}^{m}E_i(t-\bar{T}_d-m2T),
\end{equation}
where we set $\tilde{t}_L=t_L(d_L=0)=2/(n_r+1)$ and $\tilde{r}_{L}=r_L(d_L=0)=(n_r-1)/(n_r+1)$. 

\begin{figure}
\includegraphics[width=0.75\textwidth]{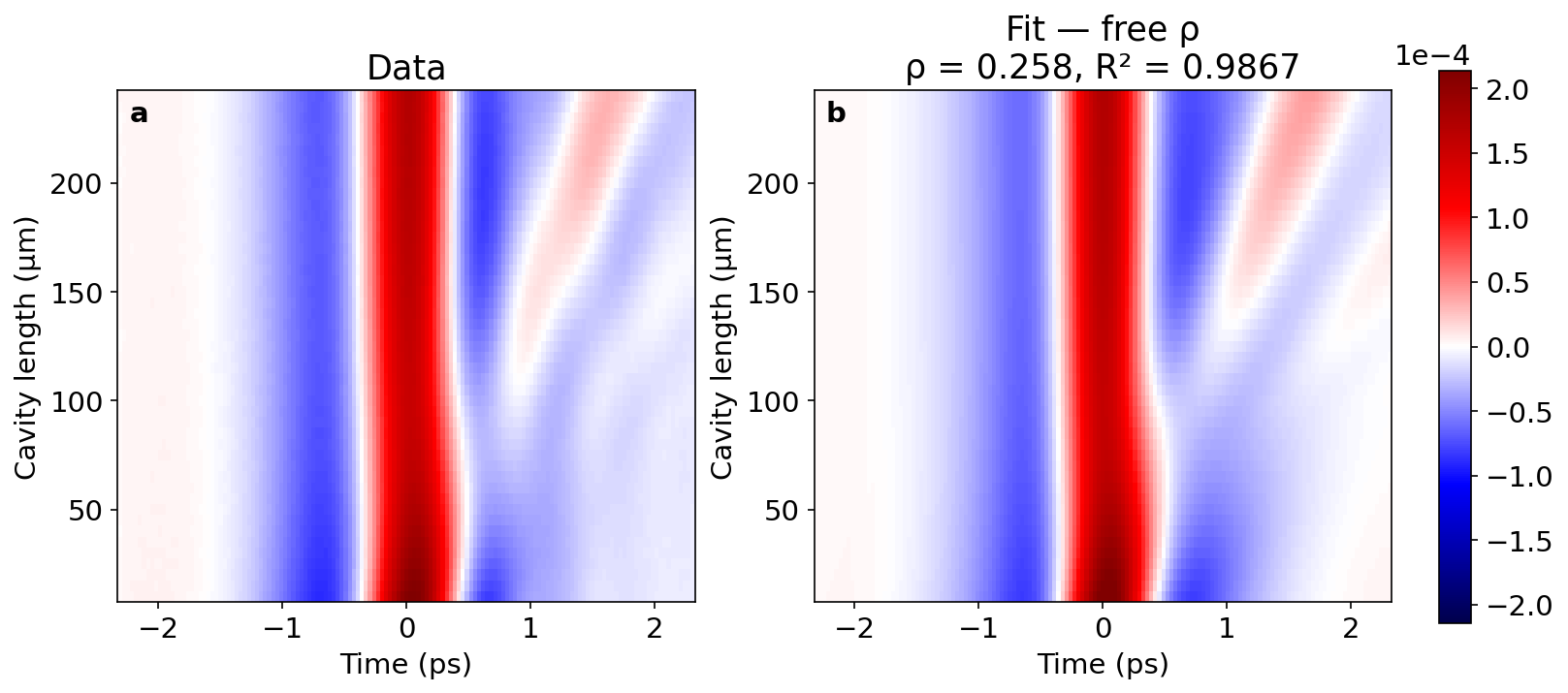}
\centering
\caption{Comparison between the experimental transmitted field from the cavity setup without the YBCO sample (\textbf{a}) and the fit using the expression in Eq.\ \ref{eq:td} with $\rho=r_0\tilde{r}_L$ left as a free parameter (\textbf{b}).}
\label{fig:lao_simul}
\end{figure}

We can use Eq.\ref{eq:td} to directly fit time-domain experimental data. We measured the THz transmission at room temperature of the cavity without the YBCO film by tuning the separation between the cavity mirror and the LAO substrate (Figure \ref{fig:lao_simul}a). To fit the data using Eq.\ref{eq:td}, we model the incoming field using a skew normal frequency distribution that reproduces the asymmetric shape of the incoming field spectrum. We introduce
\begin{equation}
f(\omega)=\frac{2}{\sigma}\phi\left(\frac{\omega}{\sigma}\right)\Phi\left(\alpha\frac{\omega}{\sigma}\right),
\end{equation}
where $f$ is obtained as the product of a gaussian $\phi(\omega)=1/\sqrt{2\pi}e^{-\omega^2/2}$ and an error function $\Phi(\omega)=1/2\left[1+\textrm{erf}(\omega/\sqrt{2})\right]$. We construct then 
\begin{equation}
E_i(\omega)=E_0e^{-i\omega \tau}\left[e^{i\phi}f(\omega-\omega_0)+e^{-i\phi}f(\omega+\omega_0)\right],
\end{equation}
where $\omega_0$ roughly coincides with center of the envelope, $\sigma$ is a proxy of its bandwidth and $\alpha$ introduces a skewness. The phase $\phi$ and the delay $\tau$ are introduced to control the phase and the arrival time of the carrier envelope in time domain. Indeed, we also notice that this shape presents a closed form for $E_i(t)$, reading
\begin{equation}
E_i(t)=\textrm{Re}\left[E_0\mathcal{E}(t-\tau)e^{i(\omega_0(t-\tau)+\phi)}\right],
\end{equation}
where the envelope $\mathcal{E}(t)$ can be expressed, introducing $\delta=\alpha/\sqrt{1+\alpha^2}$, as
\begin{equation}
\mathcal{E}(t)=e^{-\sigma^2t^2/2}\left(1+i\textrm{erfi}(\delta\sigma t/\sqrt{2})\right).
\end{equation}
Operatively, we leave $E_0,\phi,\alpha,\omega_0,t_0$ as free parameters in our fit, reabsorbing the factor $t_0\tilde{t}_L$ in the definition of $E_0$ and the factor $\bar{T}_d$ in the definition of $\tau$. 

In this way, we can fit the transmitted field $E_t(t)$ with the functional form in Eq.\ \ref{eq:td} for all the cavities at our disposal, obtaining an excellent agreement when $\rho=r_0\tilde{r}_L$ is left as a free parameter in the fit, as can be seen by comparing Fig.\ \ref{fig:lao_simul} a,b. %Fixing $\tilde{r}_L$ to the value obtained with $n_r=4.95$, we find $r_0=0.39$, which substantially smaller then what would be obtained using the conductivity of bulk gold. This can be attributed to an intrinsic, mesoscopic spatial granularity characterizing the thin metallic layer, which effectively reflects the THz pulse in an inhomogeneous way \cite{walther_Phys.Rev.B07}. 

%We notice, however, that there is a substantial decrease in the quality of the fit when $\rho$ is fixed at the expected value obtained for the parameters extracted from the control experiment, namely $d_0=15$ nm, $\sigma_0=40$ (m$\Omega$ cm)$^{-1}$, $n_l=2.1$ and $n_r=4.95$, as can be seen in Fig.\ \ref{fig:lao}c. The most striking difference is encoded in the height of the echo peaks with respect to the main transmitted pulse for longer cavities and in the enhancement of the main peak amplitude for shorter ones, which are both significantly lower than expected. The fact that an excellent fit, free from any systematic deviation as a function of cavity length, can be immediately recovered by allowing for the adjustment of the parameter $\rho$, leads us to the exclusion of a geometrical effect due to a misalignment of the cavity, which would worsen the fit for longer cavities. This motivates us to attribute the effect to an intrinsic, mesoscopic spatial granularity characterizing the metallic layer, which effectively reflects the THz pulse in an inhomogeneous way. 

%-----------------------------------------------------------------------
\section{Characterization of cavity elements}\label{sec:characterization}
In this section, we present the full characterization of all the optical elements in the cavity assembly. We measured independently the THz field transmitted by the optical windows of the cryostat chamber, the mirror's quartz substrate, the gold-layered mirror and the YBCO film's substrate. This step is crucial to disentangle extrinsic effects that are related to other optical elements, rather than the sample. In the case of the mirror and the sample's substrate, we measured the transmitted field for different temperatures.  

\begin{figure}
\centering
\includegraphics[width=1.0\linewidth]{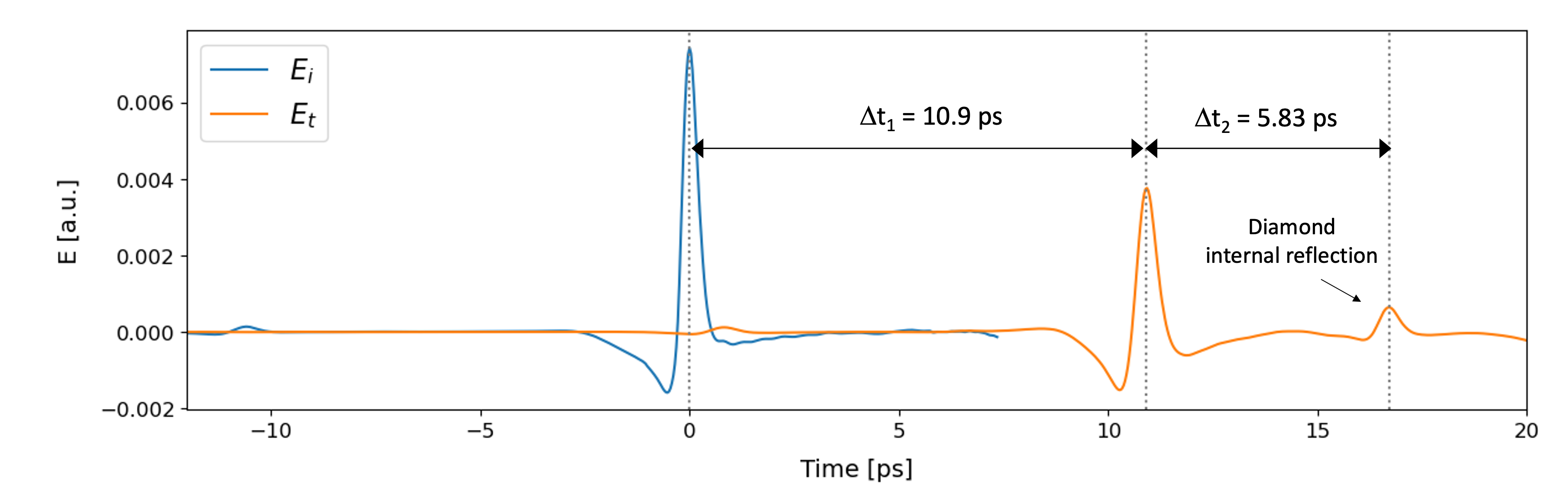}
\caption{\label{fig:chamber}Incoming (blue) and outgoing (orange) THz trace through the cryostat chamber. }
\end{figure}

\subsection{Cryostat windows}
In Figure \ref{fig:chamber}, we plot in blue the incoming THz field and in orange the field transmitted by the optical windows of the cryostat at room temperature. The transmitted field is delayed by approximately $\Delta t_1 = 10.9$ ps. This delay is due to propagation through $d\sim 370 \; \mu m$ of diamond ($n \sim 2.36$) and $d\sim 2.5$ mm of quartz ($n \sim 2.1$).

The second peak appearing in the transmitted field delayed by $\Delta t_2 = 5.83$ ps is due to the internal reflection within the thickness of the diamond window ($\Delta t_2 \sim 2dn/c$).

Many other internal reflections appear in the transmitted field at later times due to transmission through thicker elements. The diamond window is the optically thinner element in the cavity assembly and therefore all the THz traces analyzed and discussed in the following and in the main text will be measured only until a maximum time delay of 5 ps. This time-windowing procedure will allow us to neglect the contributions of thick dielectric components in the assembly. Importantly, both diamond and quartz have a refractive index that is flat across the spectral range of our THz pulse and is weakly absorbing due to the absence of optically-active phonons.

\subsection{Cavity mirror}\label{subsec:mirror}
\begin{figure}
\centering
\includegraphics[width=1.0\linewidth]{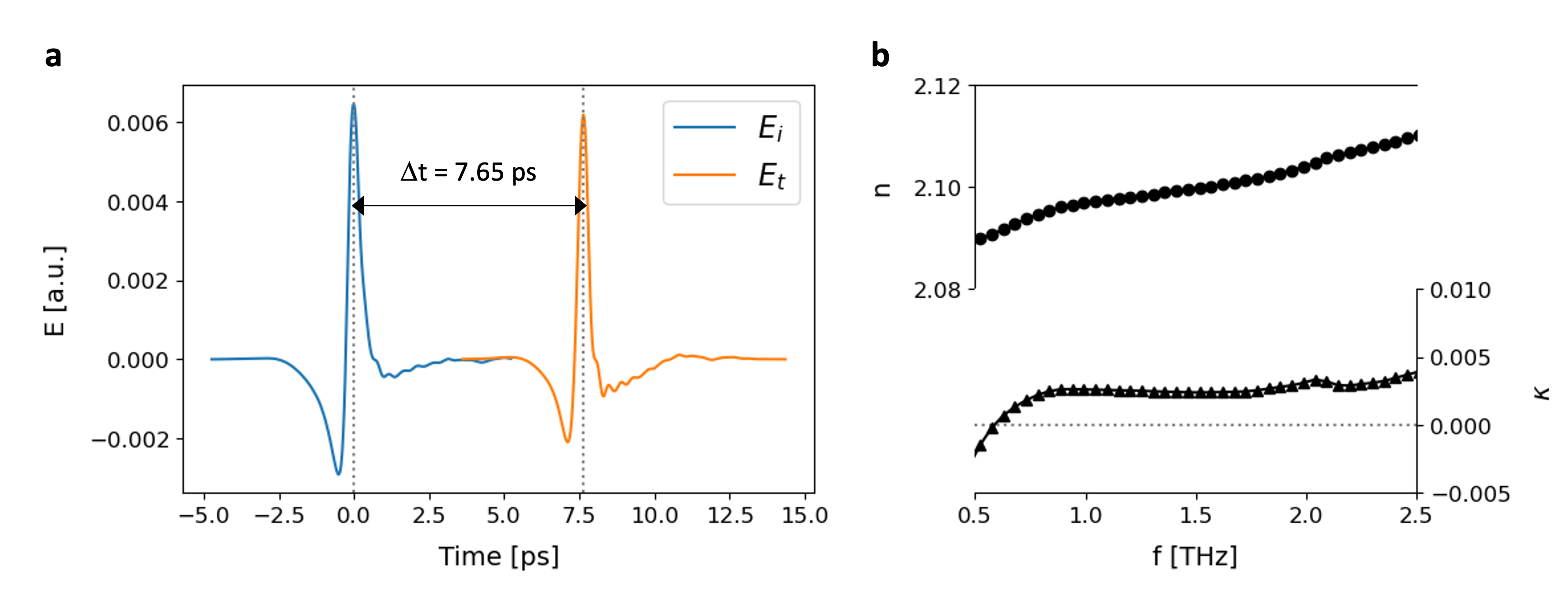}
\caption{\label{fig:zcut_quartz}\textbf{a,} Incoming (blue) and outgoing (orange) THz trace through the z-cut quartz substrate at room temperature. \textbf{b,} Extracted frequency-dependent index of refraction (circles) and extinction coefficient (triangles).}
\end{figure}

\begin{figure}
\centering
\includegraphics[width=1.0\linewidth]{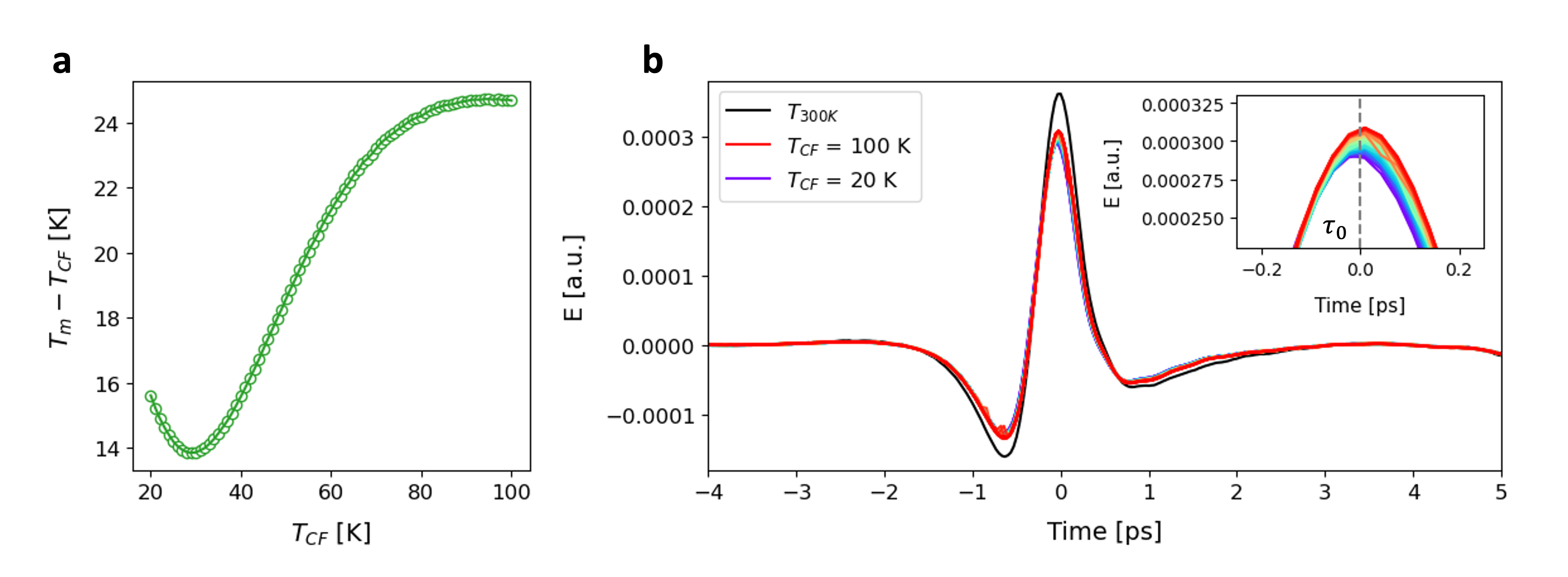}
\caption{\label{fig:mirror_tdep}\textbf{a,} Difference between the temperature measured at the mirror mount ($T_{m}$) and the one measured at the cold finger ($T_{CF}$) as a function of $T_{CF}$. \textbf{b,} THz trace transmitted by the cavity mirror (quartz substrate + Ti/Au layers) at room temperature (black) and at cryogenic temperatures (colored curves). The inset shows a zoomed-in region around the maximum of the field.}
\end{figure}

The characterization of the cavity mirror has been carried out into two steps. First, we measured the THz field transmitted by the z-cut quartz substrate before the deposition of the gold layer (Figure \ref{fig:zcut_quartz}). Then, we repeated the measurement after the deposition of a thin buffer layer of Ti (5 nm) and the gold layer (10 nm) at different temperatures (Figure \ref{fig:mirror_tdep}).

In Figure \ref{fig:zcut_quartz}a we plot in blue the incoming THz field and in orange the field transmitted by the 2mm-thick z-cut quartz substrate. The measurement was performed at room temperature and without the cryostat chamber. A measured delay of $\Delta t = 7.65$ ps is consistent with the tabulated refractive index of crystalline quartz in the THz range ($n \sim 2.1$). 

We used the thick-substrate approximation to extract the frequency-dependent refractive index $n$ and the extinction coefficient $\kappa$ of the substrate \cite{novelli2024terahertz}. The extracted optical functions are plotted in Figure \ref{fig:zcut_quartz}b. The obtained values are consistent with the ones reported in the literature \cite{quartz_tempdep}. As expected, the extinction coefficient is close to zero and the refractive index can be safely treated as a real quantity. Moreover, this dataset was acquired at room temperature, but no major changes are expected at lower temperatures in the frequency range of interest ($<$2.5 THz) \cite{quartz_tempdep}. %An optically-active phonon at $\sim3.9$~THz is responsible for a temperature-dependent absorption, but it lies well outside our spectral range.

A full temperature-dependent characterization of the mirror has been carried out after the deposition of the metallic layers. In this case, a temperature-dependent transmission is expected due to the presence of free charges. As mentioned in the Methods, the cavity mirror is held into a copper ring which is cooled down to cryogenic temperature through a copper braid. The temperature of the mirror is measured by a diode clamped onto the copper holder. The temperature measured at the mirror position $T_m$ is plotted against the temperature of the cold finger $T_{CF}$ in Figure \ref{fig:mirror_tdep}a. A difference of $\sim 20$ K is detected, due to finite resistance of the copper braid. This values were recorded after waiting for thermalization to occur.

In Figure \ref{fig:mirror_tdep}b, we plot in black the THz field transmitted by the mirror (mounted in the cryostat chamber) at room temperature. The colored curves have been instead recorded at different cryogenic temperatures of the colf finger, reported in the legend. As expected, a quench of transmission is measured as the temperature is lowered. A small shift in the central location of the THz peak ($\tau_0$) is also observed, as evidenced in the inset. \newline

\begin{figure}
\centering
\includegraphics[width=1.0\linewidth]{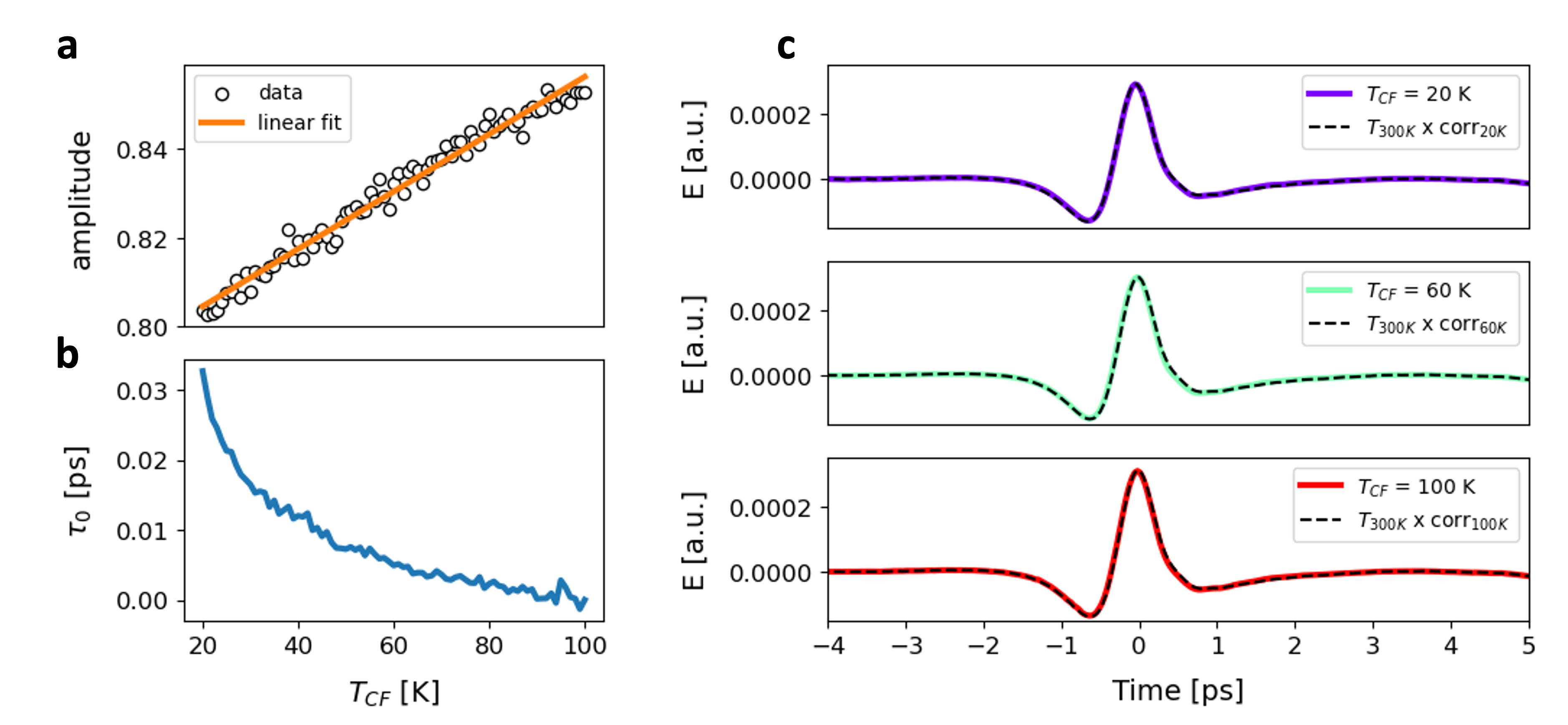}
\caption{\label{fig:mirror_tcorr}\textbf{a,} Temperature-dependent scaling factor of the amplitude of the THz field transmitted by the cavity mirror with respect to a reference measurement at room temperature. \textbf{b,} Temperature-dependent shift of the THz trace with respect to the reference measurement at room temperature. \textbf{c,} THz field transmitted at three representative temperatures (colored solid lines) and reference measurement at room temperature corrected according to the factors in (a) and (b) (black dashed lines).}
\end{figure}

In order to study the entity of these temperature-dependent effects, we performed a Gaussian-like fit of the THz peak to track down the change in amplitude and arrival time. In Figure \ref{fig:mirror_tcorr}a, we plot the amplitude of the THz transmission peak as a function of temperature. The datapoints have been normalized to the value at room temperature. The orange line is a linear fit to the data, which provides a temperature-dependent scaling factor of the transmission amplitude of the mirror if the THz transmission at room temperature is known. The temperature-dependent shift of the peak is instead plotted in Figure \ref{fig:mirror_tcorr}. The shift is not linear in temperature: the maximum value is measured at the lowest temperature and it barely exceeds 0.03 ps. This quantity is comparable with the temporal resolution of the electro-optic sampling detection, in which we use a step of $5 \;\mu m \; (= 33 \; fs)$. Moreover, the transmitted THz field shifts by less than $\sim~5~fs$ when cooled from room temperature down to $T_{CF} \simeq 60$K. It is then safe to assume that this shift is negligible in the temperature range of the YBCO superconducting transition. 

In Figure \ref{fig:mirror_tcorr}c, we consider three representative THz traces transmitted at temperature 20 K, 60 K and 100 K (colored solid curves). Superimposed to these curves is the THz field transmitted by the mirror at room temperature (dashed black line) corrected by the  scaling factor extrapolated in Figure \ref{fig:mirror_tcorr}a and the shift in Figure \ref{fig:mirror_tcorr}b at the given temperature. The agreement between the curves shows that the scaling factors correctly reproduce the temperature-dependent THz transmission.

\subsection{Sample's substrate}
\begin{figure}
\centering
\includegraphics[width=1.0\linewidth]{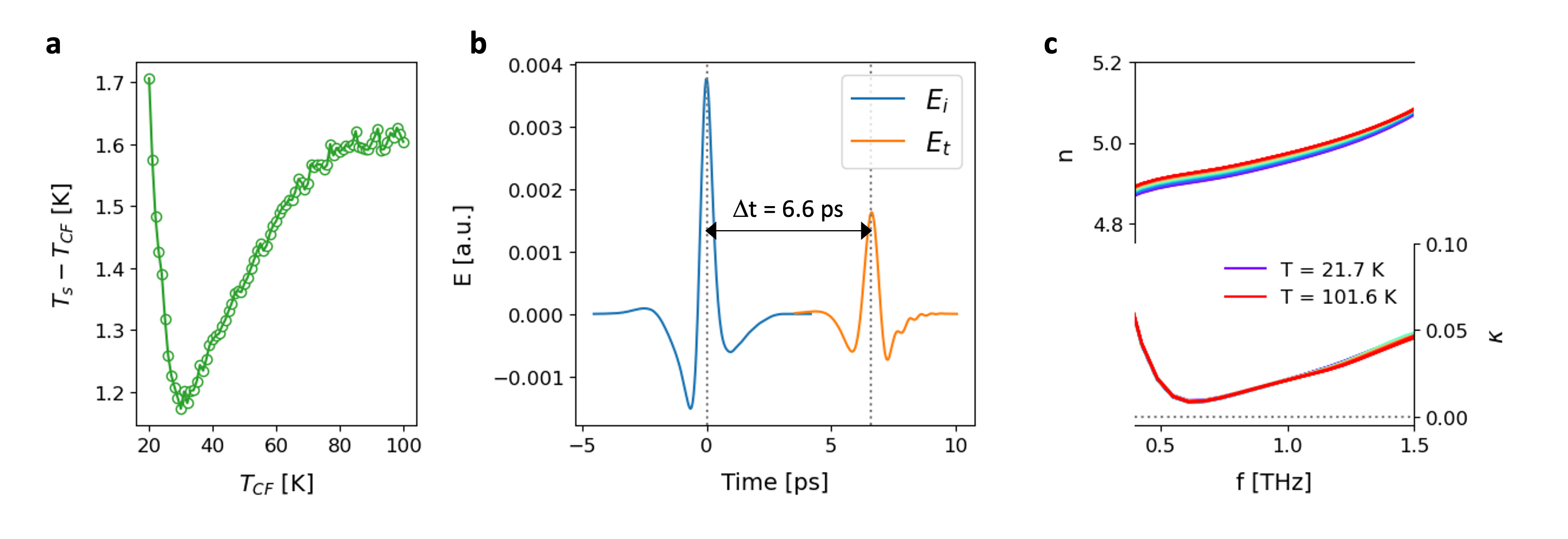}
\caption{\label{fig:lao_tdep}\textbf{a,} Difference between the temperature measured at the sample position ($T_{s}$) and the one measured at the cold finger ($T_{CF}$) as a function of $T_{CF}$. \textbf{b,} Incoming (blue) and outgoing (orange) THz trace through the LAO substrate measured at temperature $T_s$=100 K. \textbf{c,} Extracted frequency-dependent refractive index $n$ and extinction coefficient $\kappa$ in the temperature range of interest (20K - 100K).}
\end{figure}

The YBCO film was grown on a LaAlO$_3$ (LAO) substrate (500 $\mu m$) on top of a buffer layer of CeO$_2$ (10 nm). The optical characterization of the LAO+CeO$_2$ substrate was carried out on a twin sample. The substrate was glued with silver paint on a copper plate with a hole in the center. The mount is in direct thermal contact with the cold finger, and we measured a temperature difference of only $\sim 1.5$ K at the sample position with respect to the cold finger's temperature (Figure \ref{fig:lao_tdep}a). 

We plot the incoming field (blue) and transmitted one measured at 100 K (orange) in Figure \ref{fig:lao_tdep}b. The measured delay of $\Delta t = 6.6$ ps is compatible with the tabulated refractive index of LAO in the THz range ($n\sim 4.95$). The transmitted THz field features periodic oscillations at positive times. These oscillations are also visible in the THz field transmitted by the whole sample (YBCO film + LAO substrate), and should be uniquely attributed to the substrate's response. The frequency of these oscillations is $\sim2$~THz, and has been previously observed and attributed to crystal anisotropy rather than physical phonons \cite{zou_AIPAdvances12,lloyd-hughes_Opt.Lett.OL14}.

As for the quartz substrate, we use the thick-film approximation to extract the frequency-dependent refractive index $n$ and extinction coefficient $\kappa$ of the LaAlO$_3$ substrate.\footnote{The LaAlO$_3$ substrate includes a thin CeO$_2$ buffer layer, which is present both in the reference-substrate measurements and in the measurements performed with the YBCO film. Its contribution is therefore included into the extracted effective optical constants of the substrate. Since the extracted values of these effective $n$ and $\kappa$ agree with literature values for LaAlO$_3$, the optical contribution of the CeO$_2$ layer can be considered negligible.} The extracted optical constants are shown in Fig.~\ref{fig:lao_tdep}c for different temperatures and are in good agreement with literature values for LaAlO$_3$~\cite{lao_tempdep}. The extinction coefficient is close to zero in the spectral range of interest, and the refractive index shows no significant temperature dependence. In the following, we therefore describe the substrate using a real, temperature-independent refractive index $n_{\mathrm{LAO}}=4.95$.

%-----------------------------------------------------------------------
\section{Time-domain analysis}\label{sec:time_domain}

In this section, we describe complementary time-domain fitting procedures used to analyze the transmitted THz field. The analysis presented in the main text is based on the extraction of the complex optical conductivity of the YBCO film from frequency-domain data using the thin-film Tinkham formalism discussed in Sec.~\ref{sec:tinkham}. This approach is the most general one, since it does not impose a specific functional form on the optical conductivity and provides direct access to both $\sigma_1(\omega)$ and $\sigma_2(\omega)$.

On the other hand, the time-domain analysis provides a complementary and more immediate approach based on the direct fitting of the measured THz waveform. In particular, it allows us to track changes in the pulse amplitude and phase without relying on the full frequency-domain conductivity extraction. Although this approach does not provide the complete function $\sigma(\omega)$, it gives an intuitive measure of the superconducting response and serves as an independent consistency check of the results obtained from the Tinkham analysis.

An important limitation of the time-domain approach is set by the cavity round-trip time. In order to fit the directly transmitted THz pulse with a simple single-pulse parametrization, the first internally reflected cavity pulse must be temporally separated from the main waveform and removable by time-windowing. This condition is fulfilled only for sufficiently long cavities, approximately for cavity lengths $L > 200~\mu$m in our experiment. For shorter cavities, the first cavity reflection overlaps with the main THz waveform, so that the measured trace can no longer be described reliably by a single transmitted pulse. In this regime, a frequency-domain treatment is necessary, which is the reason why the Tinkham-based analysis is used for the short-cavity data discussed in the main text.

Within the long-cavity regime, however, the time-domain fits provide an additional and experimentally transparent way to verify the cavity-induced modification of the superconducting response. They do not replace the quantitative conductivity extraction, but they test whether the same physical trends are already visible directly in the measured electric-field waveform.\newline

We develop two time-domain fitting approaches. The first is based on a minimal parametrization of the transmitted THz pulse, in which the central frequency, temporal position, and envelope duration are fixed, while the amplitude and phase are treated as free parameters. By fitting the measured waveforms as a function of temperature and cavity length, we extract the evolution of these quantities and show in Sec.~\ref{subsec:phase_fit} that they encode clear signatures of the superconducting transition both outside and inside the cavity.

The second approach provides a more direct connection to microscopic quantities by constructing a time-domain analogue of the Tinkham formalism. Starting from the frequency-domain expression, we introduce a two-fluid model for the complex conductivity and derive the corresponding time-domain response. As discussed in Sec.~\ref{subsec:two_fluid}, this framework allows us to relate the fitted waveform to the superfluid density, thereby connecting the time-domain analysis to the physical observables of interest.

\subsection{Direct phase extraction}\label{subsec:phase_fit}

This first approach exploits the fact that, in the superconducting state, the imaginary part of the conductivity $\sigma_2(\omega)$ dominates the low-frequency response. In the London regime, $\sigma_2(\omega)\propto 1/\omega$, corresponding to an inductive response of the condensate. As a result, the transmitted THz field acquires a phase shift with respect to the incident field. In the ideal limit in which there is no residual dissipation from the real part of $\sigma$, the phase shift approaches $\pi/2$. This immediately suggests that the phase of the transmitted waveform provides a direct and sensitive probe of the superconducting response.

To quantify this effect, we analyze the transmitted THz waveform directly in the time domain using a minimal parametrization. The electric field is fitted with the function
\begin{equation}
    E_t(t)=E_0\,e^{-\frac{(t-\bar{t})^2}{2\tau^2}}\cos\left[\Omega(t-\bar{t})+\theta\right],
\end{equation}
which corresponds to a single-cycle pulse with Gaussian envelope. Its Fourier transform reads
\begin{equation}
    E_t(\omega)=E_0 e^{i\omega \bar{t}}\left(e^{i\theta}e^{-(\omega-\Omega)^2\tau^2/2}+e^{-i\theta}e^{-(\omega+\Omega)^2\tau^2/2}\right).
\end{equation}

The fitting procedure is performed as follows. For each cavity length, we first determine the parameters $\bar{t}$, $\tau$, and $\Omega$ by fitting the waveform in the normal state ($T>T_c$). These parameters are then kept fixed for all temperatures and cavity configurations, while only the amplitude $E_0$ and phase $\theta$ are treated as free parameters. This procedure isolates the physically relevant changes in the transmitted field associated with the superconducting transition.

To relate the fitted parameters to the optical response of the film, we assume that the experimental conditions are such that the Tinkham approximation derived in Sec.~\ref{sec:tinkham} applies. In this regime, the transmitted field can be written as
\begin{equation}
E_t(\omega)=\bar{T}e^{i\omega \bar{T}_d}t_0\frac{2}{1+n_r+Zd_L\sigma_L(\omega)}E_i(\omega),
\end{equation}
where $\bar{T}_d$ accounts for propagation through the dielectric stack and is effectively independent of the cavity length. Consequently, variations of the transmitted field with temperature and cavity length are directly encoded in the conductivity $\sigma_L(\omega)$.

In the normal state, we approximate $\sigma_L(\omega)\simeq\sigma_n$ as purely real, which yields
\begin{equation}
E_t(t,T>T_c)=\bar{T}t_0\frac{2}{1+n_r+Zd_L\sigma_n}E_i(t-\bar{T}_d).
\end{equation}
Assuming further that $Zd_L\sigma_n \gg 1+n_r$ (a condition well satisfied in the THz range), the transmitted field in the superconducting state can be approximated as
\begin{equation}
E_t(\omega,T,L)\simeq\frac{\sigma_n}{\sigma_L(\omega)}\,E_t(\omega,T>T_c).
\end{equation}

Within this framework, and assuming that $\sigma_L(\omega)$ varies weakly over the spectral width of the THz pulse, i.e.\ $\sigma_L(\omega\sim\pm\Omega)\simeq\sigma_L(\Omega)$, the fitted amplitude and phase can be directly related to the complex conductivity. In particular, the amplitude renormalization reads
\begin{equation}
\delta E=\frac{E_0(T<T_c)}{E_0(T>T_c)}=\left|\frac{\sigma_n}{\sigma_L(\Omega)}\right|,
\end{equation}
while the phase shift is given by
\begin{equation}
\delta\theta=\theta(T<T_c)-\theta(T>T_c)=\arg\left(\frac{\sigma_n}{\sigma_L(\Omega)}\right)=-\arctan\left(\frac{\sigma_L^{\prime\prime}(\Omega)}{\sigma_L^{\prime}(\Omega)}\right).
\end{equation}

Within the approximations outlined above, the evolution of $\delta E$ and $\delta\theta$ as a function of cavity length and temperature provides a direct and intuitive measure of changes in the complex conductivity. In particular, the phase shift $\delta\theta$ is primarily sensitive to the inductive response $\sigma_2$, and therefore captures the enhancement of the superconducting response induced by the cavity.

\begin{figure}
\centering
\includegraphics[width=1.0\linewidth]{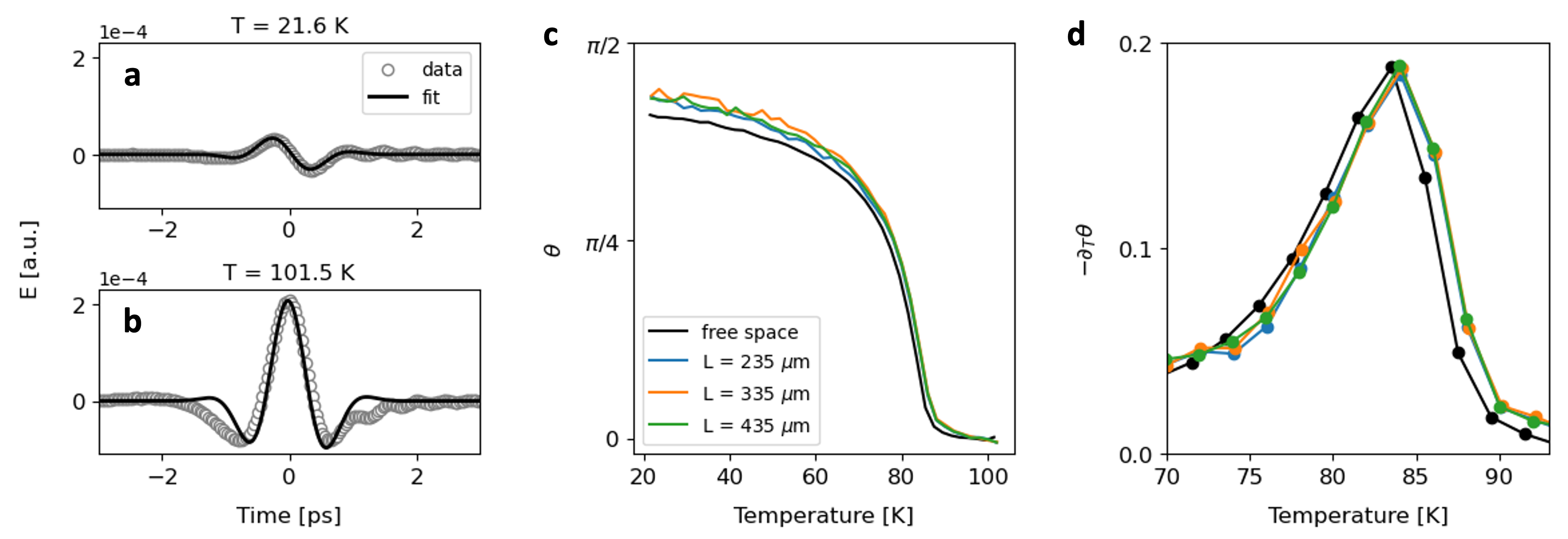}
\caption{\label{fig:phase_fit}
THz traces transmitted by the YBCO sample at \textbf{a,} $T = 21.6$ K and \textbf{b,} $T = 101.5$ K. The black solid line shows the fit using a single-cycle pulse with fixed central frequency (0.7 THz), temporal duration (0.51 ps), and envelope arrival time ($t=0$). 
\textbf{c,} Temperature dependence of the extracted phase for the bare sample (black) and for the sample in cavities of different lengths (colored curves). 
\textbf{d,} Temperature derivative of the phase curves shown in (c), highlighting the shift of the superconducting transition.}
\end{figure}

We first apply this fitting procedure to the THz traces transmitted by the bare YBCO film on its substrate (i.e., in free space) as a function of temperature. For each dataset, we determine the full set of fitting parameters $(E_0, \bar{t}, \tau, \Omega, \theta)$ in the normal state ($T>T_c$). We then fix $\bar{t}$, $\tau$, and $\Omega$ to these values and perform temperature-dependent fits leaving only the amplitude $E_0$ and phase $\theta$ as free parameters. Representative fits are shown in Fig.~\ref{fig:phase_fit}a,b for low temperature ($\sim 20$ K) and high temperature ($\sim 100$ K), demonstrating that this minimal parametrization accurately captures the measured waveforms across the full temperature range.

The extracted phase $\theta(T)$ is plotted in Fig.~\ref{fig:phase_fit}c. For clarity, the phase is referenced to zero in the normal state ($T>T_c$). As expected, upon entering the superconducting state, the phase exhibits a rapid increase, approaching the characteristic $\pi/2$ shift associated with the inductive response of the superfluid. The black curve corresponds to the YBCO film in free space, while the colored curves represent measurements performed in cavities of different lengths (235, 335, and 435~$\mu$m). Two key observations emerge: i) the magnitude of the phase shift is systematically enhanced in the cavity configurations, ii) the onset of the phase change occurs at higher temperature compared to the free-space case.

This behavior is further highlighted in Fig.~\ref{fig:phase_fit}d, where we plot the temperature derivative of the phase. In the cavity case, the peak associated with the superconducting transition is clearly shifted to higher temperature, providing a direct visualization of the cavity-induced modification of the superconducting response.

Overall, these results are fully consistent with the analysis based on the Tinkham formalism presented in the main text. We stress, however, that the present time-domain approach provides a qualitative mapping between the fitted phase shift and the complex conductivity, relying on simplifying assumptions such as the weak frequency dependence of $\sigma(\omega)$ over the pulse bandwidth. Despite these approximations, the method offers a direct and intuitive confirmation of the cavity-induced enhancement of the superconducting response.

\subsection{Two-fluid model}\label{subsec:two_fluid}

The direct phase extraction discussed in Sec.~\ref{subsec:phase_fit} provides an intuitive probe of the superconducting transition, but it does not directly yield a quantitative estimate of the superfluid response. To establish a closer connection between the time-domain waveform and the optical quantities discussed in the main text, we introduce a minimal two-fluid model for the complex conductivity of the YBCO film. This approach allows us to fit the transmitted THz electric field directly in the time domain while retaining parameters associated with the dissipative and inductive components of the superconducting response.

\begin{figure}
\centering
\includegraphics[width=0.9\linewidth]{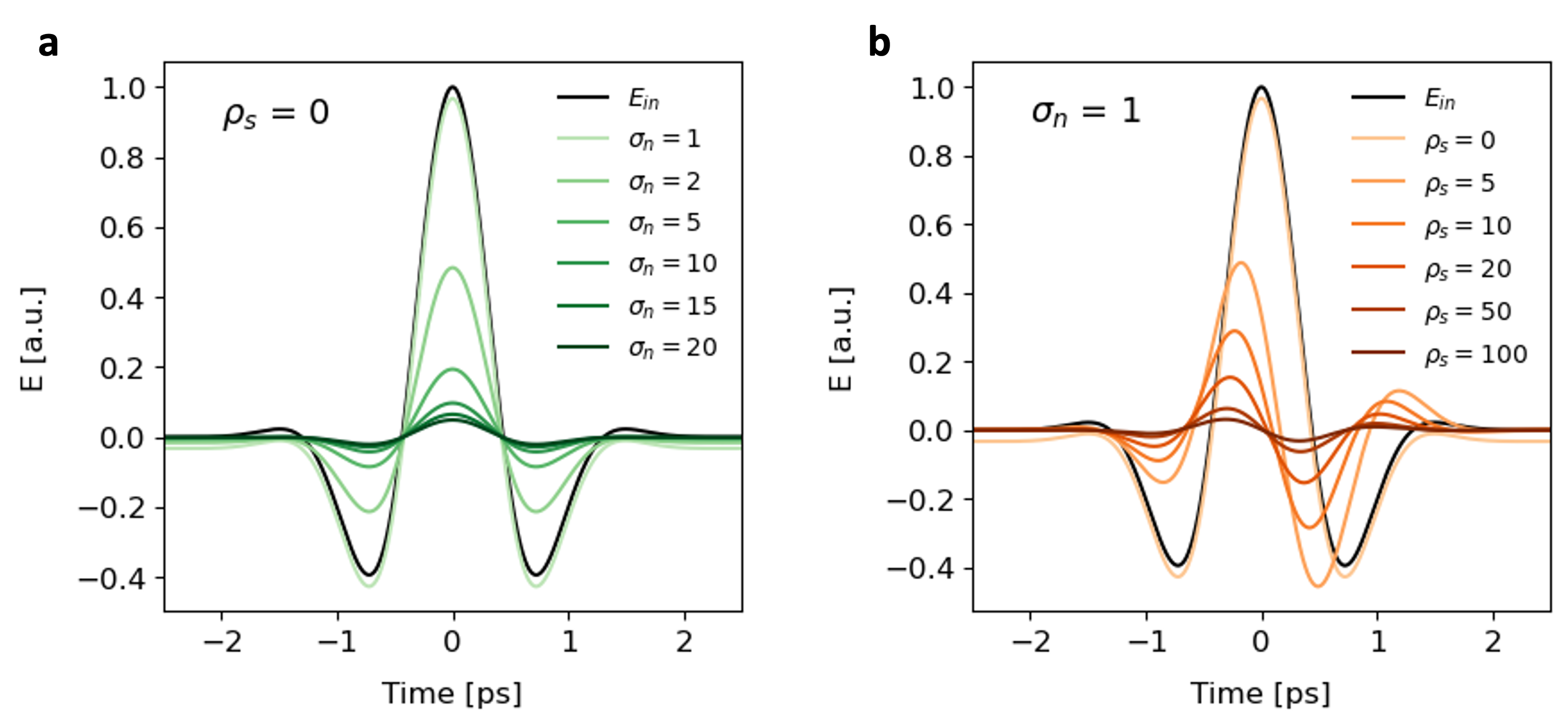}
\caption{\label{fig:twofluid_simul}
Simulated THz fields transmitted by the YBCO sample using the two-fluid model. A representative incoming field, $E_{\mathrm{in}}$, is shown in black. 
\textbf{a,} The superfluid contribution is set to zero, $\rho_s=0$, and the colored curves show the outgoing field for different values of the normal-fluid contribution $\sigma_n$. 
\textbf{b,} The normal-fluid contribution $\sigma_n$ is kept fixed, while the superfluid contribution $\rho_s$ is varied.}
\end{figure}

We start from the thin-film transmission formula in Eq. \ref{eq:tinkham}. In the long-cavity regime, where the first cavity reflection is temporally separated from the main transmitted pulse, the field transmitted through the YBCO film can be approximated as
\begin{equation}
E_t(\omega,T)=\frac{2}{1+n_r+Zd\,\sigma(\omega,T)}E_i(\omega),
\end{equation}
where $d$ is the YBCO film thickness, $n_r$ is the refractive index of the substrate and $Z$ is the impedance of free space. The conductivity is modeled using a phenomenological two-fluid form,
\begin{equation}\label{eq:twofluid}
\sigma(\omega,T)=\sigma_n(T)+i\frac{\rho_s(T)}{\omega}.
\end{equation}
Here $\sigma_n(T)$ describes the effective dissipative contribution in the THz range, while $\rho_s(T)$ parameterizes the inductive response of the condensate and is proportional to the superfluid density. Equivalently,
\begin{equation}
\rho_s(T)=\lim_{\omega\rightarrow 0}\omega\sigma_2(\omega,T).
\end{equation}

Before applying this model to the experimental data, we first illustrate its effect on the transmitted THz waveform. Figure~\ref{fig:twofluid_simul} shows simulated transmitted fields obtained by varying independently the two components of the two-fluid conductivity. When the superfluid contribution is set to zero, changing $\sigma_n$ mainly modifies the amplitude and shape of the transmitted pulse through dissipative attenuation. In contrast, when $\sigma_n$ is kept fixed and $\rho_s$ is increased, the waveform develops a pronounced phase shift and temporal reshaping, reflecting the inductive response of the superconducting condensate. This comparison shows that the phase evolution discussed in Sec.~\ref{subsec:phase_fit} is primarily sensitive to the emergence of the superfluid contribution.

We then apply the same model to the measured THz traces. The transmitted waveform is calculated by multiplying the reference field by the two-fluid transfer function in the frequency domain and transforming the result back to the time domain. For the YBCO in cavity, the reference field is corrected by the temperature-dependent scaling factor in Fig. \ref{fig:mirror_tcorr}a, which accounts for the temperature-dependent transmission of the mirror. The fit is performed within the temporal window containing the main transmitted pulse. As discussed above, this procedure is restricted to sufficiently long cavity lengths, for which the first cavity reflection does not overlap with the directly transmitted waveform.

\begin{figure}
\centering
\includegraphics[width=1.0\linewidth]{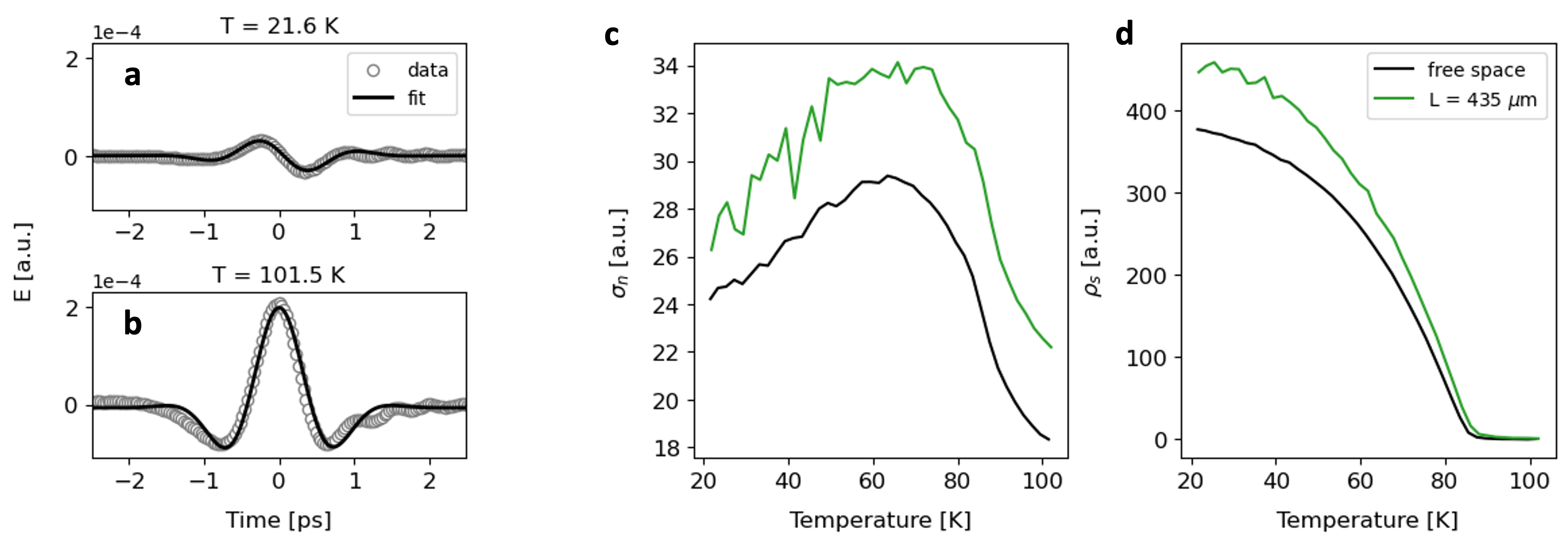}
\caption{\label{fig:twofluid_fit}
Fit of the THz traces transmitted by the YBCO sample at \textbf{a,} $T=21.6$ K and \textbf{b,} $T=101.5$ K using the two-fluid model transfer function. Temperature dependence of the fitted \textbf{c,} normal-fluid contribution $\sigma_n$ and \textbf{d,} superfluid contribution $\rho_s$ for the bare sample (black) and for the sample inside a cavity of length 435~$\mu$m.}
\end{figure}

Representative fits are shown in Fig.~\ref{fig:twofluid_fit}a,b for a low-temperature trace in the superconducting state and for a high-temperature trace above the transition. The model reproduces the measured waveforms in both regimes, capturing the phase shift and reshaping of the pulse that develop upon cooling below $T_c$. The corresponding fitted parameters are shown in Fig.~\ref{fig:twofluid_fit}c,d. The effective normal-fluid contribution $\sigma_n$ evolves smoothly with temperature, while the superfluid contribution $\rho_s$ emerges sharply below the superconducting transition. This behavior is observed both for the bare YBCO film and for the sample placed inside the cavity.

The resulting temperature dependence of $\rho_s$ is consistent with the trends obtained from the Tinkham analysis in the main text: in the cavity configuration, the superconducting response appears enhanced and develops at a higher temperature than in the free-space measurement.

We stress that the two-fluid fit should be interpreted as a consistency check rather than as a replacement for the full complex-conductivity analysis. The model assumes a simplified low-frequency form of the conductivity over the bandwidth of the THz pulse and therefore does not capture all spectral details of $\sigma(\omega)$. Nevertheless, within the long-cavity regime where the time-domain analysis is applicable, it provides an independent confirmation that the cavity-induced changes are already visible directly in the measured electric-field waveform.

%-----------------------------------------------------------------------
\section{Control experiments}\label{sec:control_exp}

In this section, we present a series of control experiments performed to test the robustness and reproducibility of the observed cavity-induced modifications of the superconducting response. These measurements are designed to address possible extrinsic contributions, including thermal lag, direction of the temperature scan, nonlinear THz excitation, imperfect thermalization, and changes in the optical path unrelated to the YBCO film. Together, they provide independent checks that the reported shift and enhancement of the superconducting response are not artifacts of the measurement protocol.

\subsection{Direction of the temperature scan across the superconducting transition}

\begin{figure}
\centering
\includegraphics[width=1.0\linewidth]{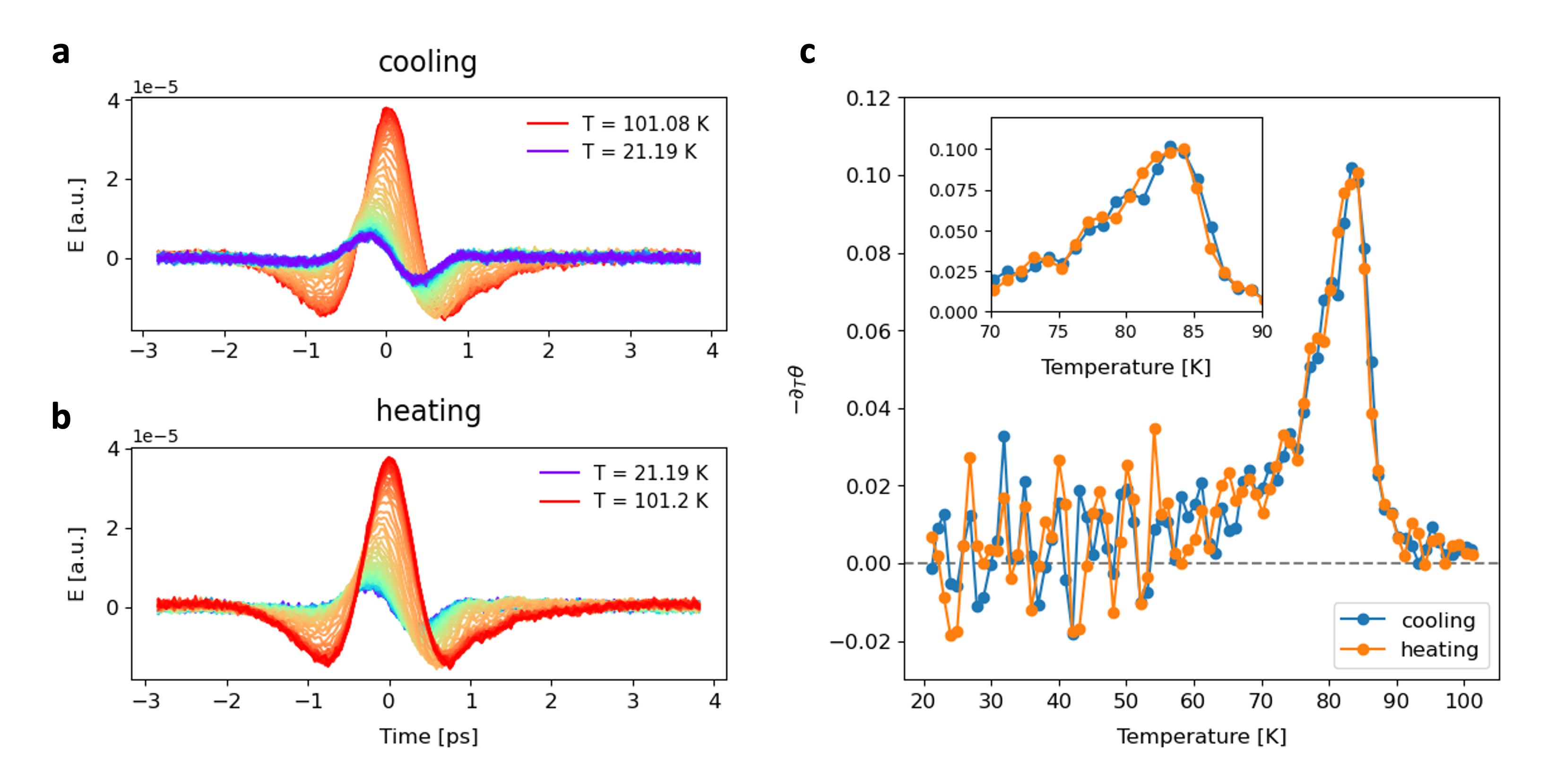}
\caption{\label{fig:ybco_hysteresis} THz traces transmitted by the YBCO sample in free space during \textbf{a,} cooling and \textbf{b,} heating across the superconducting transition. 
\textbf{c,} Temperature derivative of the THz phase extracted from the data in (a,b) using the procedure described in Sec.~\ref{subsec:phase_fit}. The inset shows a zoom around the transition temperature, demonstrating the overlap between cooling and heating scans.}
\end{figure}

We first verify that the superconducting transition extracted from the THz response is independent of the direction of the temperature scan. To this end, we measured the THz field transmitted by the YBCO sample in free space while sweeping the temperature both on cooling and on heating across the superconducting transition, as shown in Fig.~\ref{fig:ybco_hysteresis}a,b.

The sample temperature is controlled through a feedback loop referenced to the silicon diode mounted on the cryostat cold finger. At each temperature setpoint, the THz waveform is acquired only after thermalization of the cold finger and sample holder has been reached. This procedure is used to minimize possible temperature offsets arising from finite thermal resistance between the cold finger, sample mount, and YBCO film.

We then analyze the transmitted THz waveforms using the time-domain phase fitting procedure described in Sec.~\ref{subsec:phase_fit}. From the extracted phase $\theta(T)$, we compute the temperature derivative $d\theta/dT$, which provides a sensitive marker of the superconducting transition. The resulting curves for cooling and heating are shown in Fig.~\ref{fig:ybco_hysteresis}c. The two peaks overlap within the experimental uncertainty, as highlighted in the inset.

This control experiment confirms that the temperature feedback and thermalization protocol provide a reproducible temperature axis for the THz measurements. In particular, it rules out the possibility that apparent shifts of the superconducting transition arise from the direction of the temperature sweep or from delayed thermalization in the cryostat.

\subsection{Thermalization effects}

We performed an additional control experiment to verify that proper thermalization is achieved when the YBCO film is placed inside the cavity. This check is particularly important because the cavity assembly contains additional metallic and dielectric elements, including the movable mirror, which could in principle introduce thermal gradients or delayed thermalization.

For this purpose, we measured the THz field transmitted by the YBCO film for three different cavity lengths ($L=235$, 335, and 435~$\mu$m), while sweeping the temperature across the superconducting transition. At each temperature setpoint, the feedback loop was first allowed to reach thermal stability. We then acquired three consecutive THz scans: the first immediately after thermalization was reached ($t=0$), the second after 10 minutes, and the third after 20 minutes.

\begin{figure}
\centering
\includegraphics[width=1.0\linewidth]{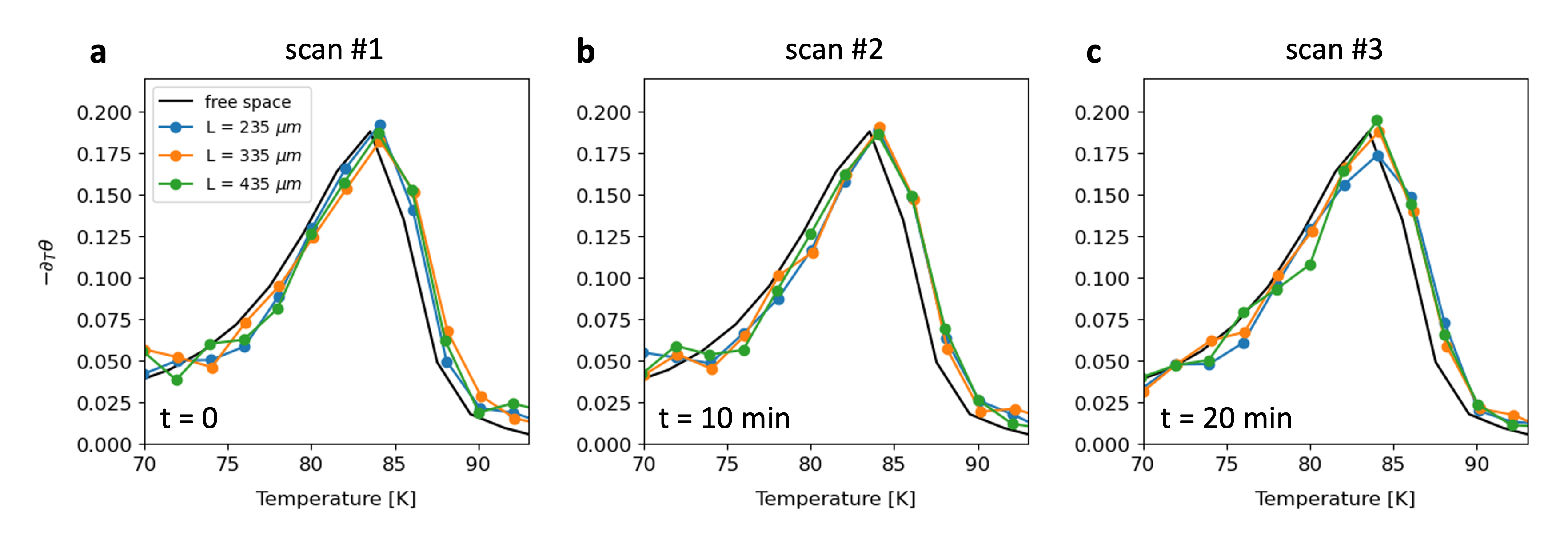}
\caption{\label{fig:thermalization}
Temperature derivative of the THz phase extracted for the YBCO film in free space (black) and inside cavities of different lengths (colored curves). Each panel corresponds to a different delay after thermalization at each temperature setpoint: \textbf{a,} $t=0$, \textbf{b,} $t=10$ minutes, and \textbf{c,} $t=20$ minutes.}
\end{figure}

For each of these three scans, we independently extracted the temperature-dependent phase using the procedure described in Sec.~\ref{subsec:phase_fit}. Figure~\ref{fig:thermalization} shows the corresponding temperature derivative of the phase shift. The same datasets are used in Fig.~\ref{fig:phase_fit}d, where the three scans are averaged. As a reference, the black curve shows the same quantity measured for the YBCO film in free space.

The position of the phase-derivative peak is unchanged between the scans acquired at $t=0$, 10 minutes, and 20 minutes. This confirms that the observed shift of the superconducting transition in the cavity is not caused by incomplete thermalization of the cavity assembly. Moreover, the measurements for the three cavity lengths were performed with different temperature-sweep directions: heating for $L=435~\mu$m, heating for $L=335~\mu$m, and cooling for $L=235~\mu$m. The systematic upward shift of the transition is independent of these different sweep protocols.

We also consider the possible effect of the cavity mirror, whose temperature is approximately 20~K higher than that of the sample holder (Figure \ref{fig:mirror_tdep}a). One might ask whether radiative heating from the mirror could locally increase the temperature of the YBCO film relative to the temperature read by the diode mounted on the sample holder. If this were the dominant effect, the actual film temperature would be higher than the recorded temperature. The superconducting transition would then appear at a lower measured temperature, because the film would reach its intrinsic $T_c$ while the diode still reads a smaller value. This is opposite to the experimentally observed shift, where the transition moves to higher temperature in the cavity. We therefore conclude that radiative heating from the mirror cannot account for the observed effect, and that any heat load from the mirror is efficiently dissipated through the copper sample mount.

\subsection{Cavity measurements without gold layer}\label{ssec:quartz}

We performed an additional control experiment to test whether the quartz substrate of the movable mirror, rather than the metallic mirror itself, could be responsible for the observed modification of the superconducting response. We replaced the Ti/Au-coated mirror with an identical bare quartz substrate, without any metallic layer. The quartz substrate was positioned at controlled distances from the YBCO film, $L=235$, 335, and 435~$\mu$m, and temperature-dependent THz transmission measurements were repeated using the same protocol as for the cavity measurements.

\begin{figure}
\centering
\includegraphics[width=1.0\linewidth]{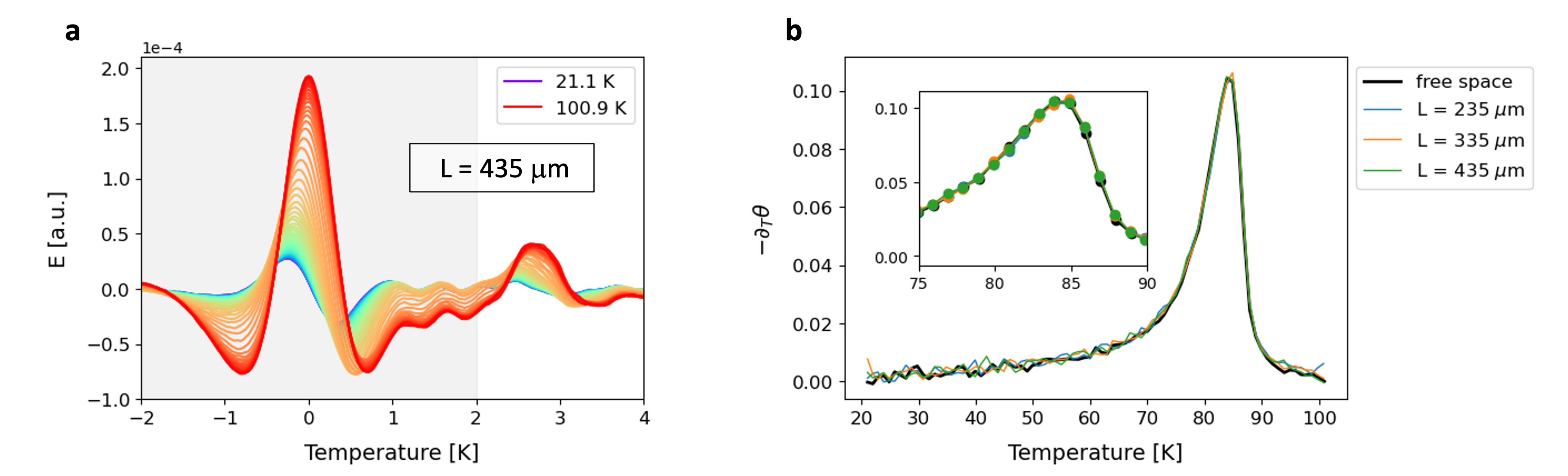}
\caption{\label{fig:quartz}
\textbf{a,} THz traces transmitted through the bare quartz substrate and the YBCO sample at different temperatures for a quartz-YBCO separation of $L=435~\mu$m. The grey shaded area indicates the temporal window used for the time-domain fit. 
\textbf{b,} Temperature derivative of the extracted phase for the data shown in (a) and for two additional quartz-YBCO separations (colored curves), compared with the corresponding free-space YBCO measurement (black curve). The inset shows a zoom around the superconducting transition.}
\end{figure}

Representative THz traces acquired for a quartz-YBCO separation of $L=435~\mu$m are shown in Fig.~\ref{fig:quartz}a. We analyze these data using the same time-domain phase fitting procedure described in Sec.~\ref{subsec:phase_fit}. Figure~\ref{fig:quartz}b shows the temperature derivative of the extracted phase for the three quartz-YBCO separations, compared with the corresponding free-space YBCO measurement shown in black. All curves peak at the same temperature. 

\begin{figure}
\centering
\includegraphics[width=0.7\linewidth]{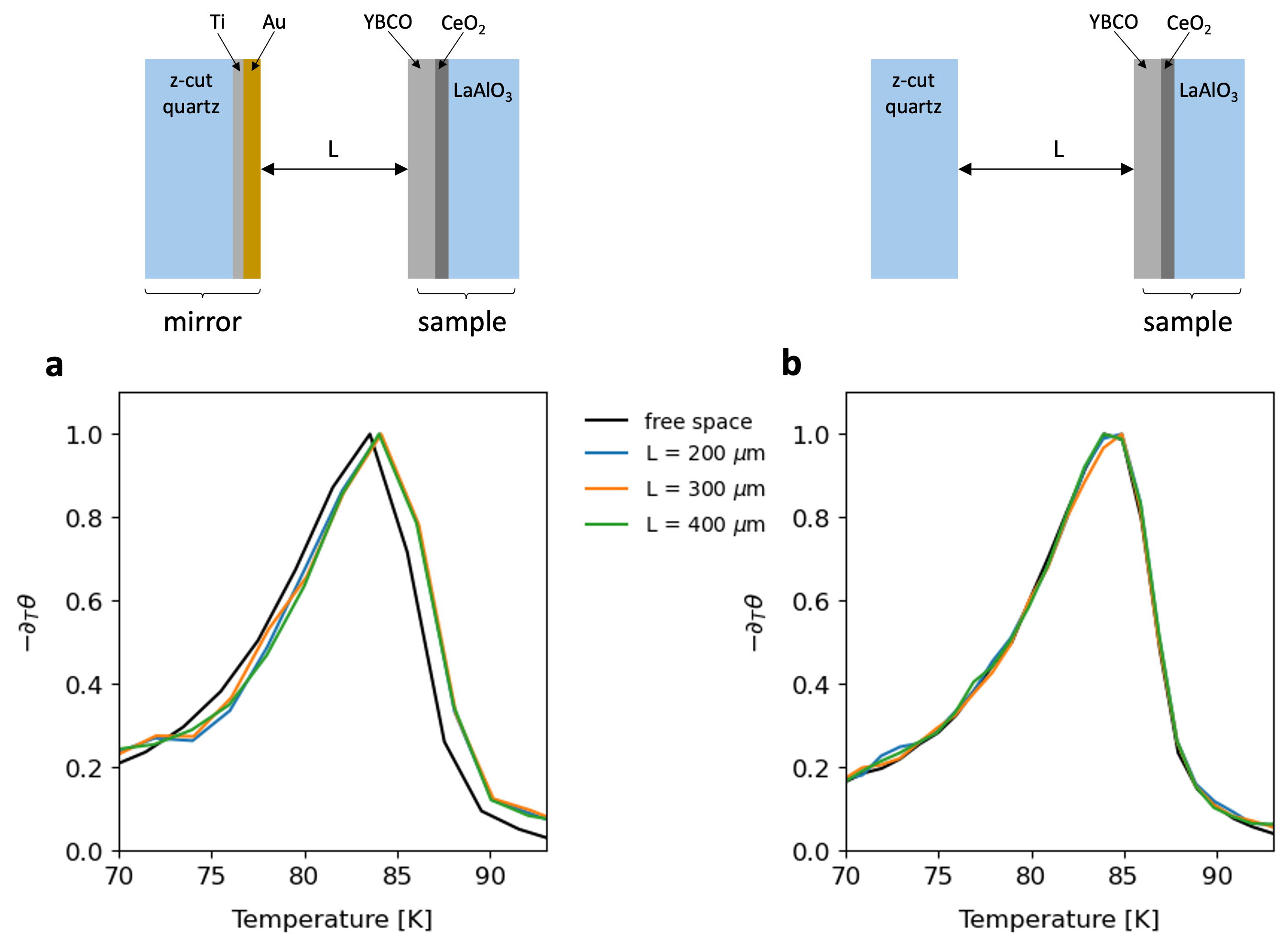}
\caption{\label{fig:mirr_quartz} Temperature derivative of the extracted phase for different cavity lengths in the presence (\textbf{a}) and in the absence (\textbf{b}) of the metallic Ti/Au layer.}
\end{figure}

This result demonstrates that the cavity-induced shift discussed in the main text is not caused by the presence of the quartz substrate, by a change in the dielectric environment alone, or by the modified optical path associated with inserting the mirror support. As shown in Figure \ref{fig:mirr_quartz}, the observed modification of the superconducting response appears only when the metallic Ti/Au layer is present, confirming that the relevant effect is associated with the electromagnetic boundary conditions imposed by the semi-reflecting cavity mirror.

\subsection{Terahertz intensity-dependent measurements}

Finally, we verify that the apparent transition temperature is not affected by the THz excitation strength. This control experiment is important because the incident THz field reaching the YBCO film is different in the free-space and cavity geometries. In the cavity configuration, the semi-reflecting mirror transmits only a fraction ($\sim 10\%$) of the incoming THz radiation, so that the field incident on the YBCO film is substantially reduced compared to the free-space measurement. If the stronger THz excitation in free space produced local heating of the film, the superconducting transition could appear artificially shifted to lower temperature, potentially mimicking the relative shift observed between the free-space and cavity configurations.

\begin{figure}
\centering
\includegraphics[width=1.0\linewidth]{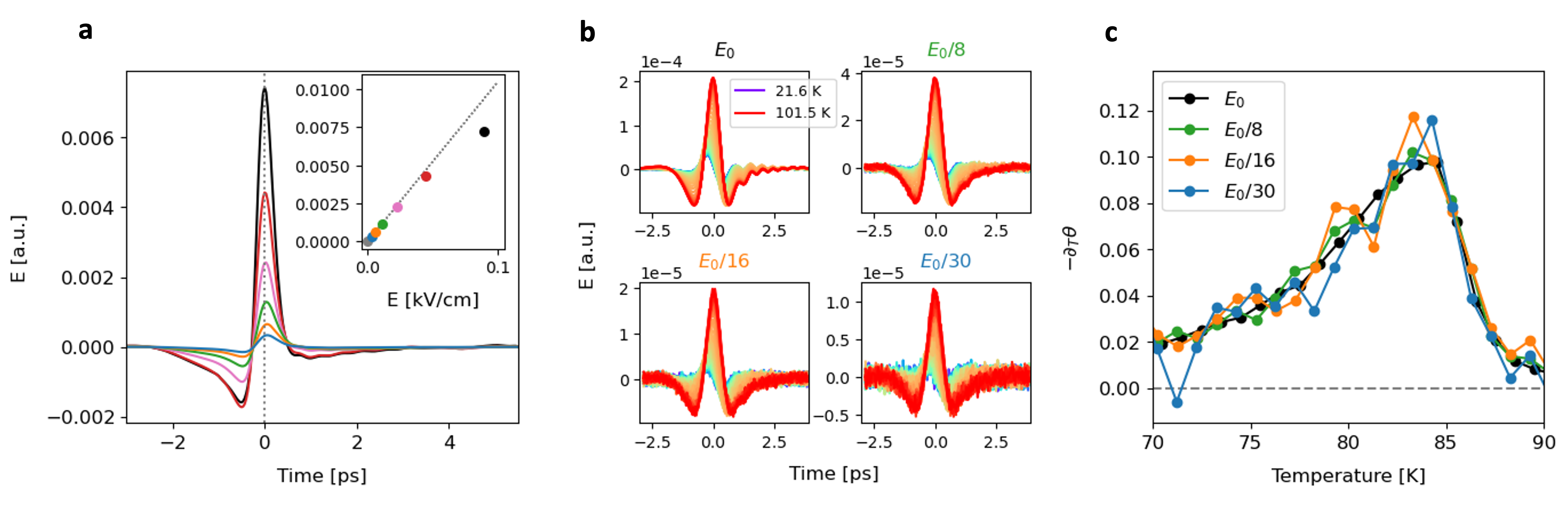}
\caption{\label{fig:thz_intensity}
\textbf{a,} THz fields generated by the photoconductive antenna for different bias voltages. The inset shows the THz peak field as a function of antenna bias voltage. 
\textbf{b,} THz traces transmitted by the YBCO sample at different temperatures for different incident THz field strengths. The maximum field amplitude $E_0$ corresponds to approximately $0.1$~kV/cm. 
\textbf{c,} Temperature derivatives of the phase extracted from the datasets in (b), showing that the transition marker is independent of the incident THz field strength.}
\end{figure}

To rule out this possibility, we performed temperature-dependent THz transmission measurements of the YBCO film in free space while systematically reducing the incoming THz field strength. The THz amplitude was varied by changing the bias voltage applied to the photoconductive antenna. Figure~\ref{fig:thz_intensity}a shows the incident THz fields measured without optical elements in the beam path for different antenna bias voltages. The inset shows the corresponding peak THz field as a function of field strength. The largest field amplitude, denoted by $E_0$, corresponds to approximately $0.1$~kV/cm and is the excitation strength used for the measurements discussed in the main text.

We then measured the THz field transmitted by the YBCO film as a function of temperature for four different incident field amplitudes, as shown in Fig.~\ref{fig:thz_intensity}b. The excitation amplitude $E_0/16$ is comparable to the field amplitude reaching the YBCO film after transmission through the cavity mirror. This range therefore directly tests whether the different THz excitation strengths in the free-space and cavity measurements can influence the extracted transition temperature.

The transmitted waveforms were analyzed using the same phase-extraction procedure described in Sec.~\ref{subsec:phase_fit}. Figure~\ref{fig:thz_intensity}c shows the temperature derivative of the extracted phase for the different THz field amplitudes. All curves peak at the same temperature. In particular, reducing the incident THz field by more than one order of magnitude does not shift the transition to higher temperature, as would be expected if local THz-induced heating were responsible for the lower apparent transition temperature in free space.

These measurements demonstrate that the THz excitation strength does not measurably affect the superconducting transition extracted from the phase response. We therefore conclude that the cavity-induced shift discussed in the main text cannot be attributed to a different thermal load from the THz pulse in the free-space and cavity configurations.

%\begin{mdframed}
%To conclude, 
%\end{mdframed}

\begin{boxE}
\textbf{Comment on thermal effects and comparison with 1T-TaS$_2$.}
We note that the present effect is distinct from the cavity-mediated thermal control reported in 1T-TaS$_2$ \cite{jarc2023cavity}. In that experiment, the cavity assembly was designed to minimize the thermal coupling to the cryogenic bath and maximize instead the radiative coupling to the photonic bath mediated by the cavity. Changing the cavity geometry resulted in a modified radiative balance of the sample and produced a shift of the transition through a change in the sample's electronic and vibrational modes temperatures. 

\

In the present experiment, the YBCO film is mounted in direct thermal contact with a copper holder that efficiently dissipates heat loads. In this configuration, radiative effects are minimized and the sample's temperature is locally monitored during the experiment. In YBCO, the cavity does not simply displace the transition curve, but modifies the superconducting electrodynamics below the transition, with the most pronounced effect appearing in the low-frequency inductive response $\sigma_2$. Moreover, the effect depends on the presence of the metallic Ti/Au layer: when only the quartz substrate of the mirror is brought close to the YBCO film, no comparable shift is observed. 

\

These features indicate that the observed response cannot be reduced to a simple effective cooling of the film, but requires the electromagnetic boundary condition imposed by the semi-reflecting mirror.
\end{boxE}

%-----------------------------------------------------------------------
\section{Quantitative extraction of complex optical conductivity}\label{sec:tinkham}

The central quantity used to characterize the superconducting response of the YBCO film is its complex optical conductivity,
\begin{equation}
\sigma(\omega,T)=\sigma_1(\omega,T)+i\sigma_2(\omega,T).
\end{equation}
In particular, the imaginary part $\sigma_2(\omega,T)$ provides direct access to the inductive response of the superconducting condensate and, in the low-frequency limit, to the superfluid density through $\rho_s(T)\propto \lim_{\omega\rightarrow0}\omega\sigma_2(\omega,T)$. A quantitative extraction of $\sigma(\omega,T)$ is therefore required in order to compare the superconducting response of the YBCO film in free space and inside the cavity.

The extraction is complicated by the fact that the measured transmitted field contains not only the intrinsic response of the YBCO film, but also propagation through the dielectric stack, transmission through the semi-reflecting metallic mirror, and multiple reflections inside the cavity. The propagation model derived in Sec.~\ref{ssec:cavity} provides the forward relation between the incident field and the measured transmitted field. In this section, we invert this relation to obtain the YBCO conductivity from the experimentally measured complex transmission.\newline

The procedure follows the same logic as the standard Tinkham analysis for optically thin conducting films, but generalized to the cavity geometry. We first derive a cavity-corrected Tinkham expression by normalizing the transmitted field measured in the presence of the YBCO film to a corresponding reference measurement in which the YBCO sheet contribution is absent. This normalization removes the common propagation factors and isolates the contribution of the conducting film. We then discuss two limiting regimes. For sufficiently long cavities, the first internal cavity reflection is temporally separated from the main THz pulse and can be removed by time-windowing; in this case, the usual Tinkham expression is recovered. For short cavities, the reflected field overlaps with the main transmitted waveform, and the cavity correction factor must be considered explicitly.

%Finally, we describe the correction required when the empty-cavity reference is measured at a fixed temperature, while the YBCO transmission is measured as a function of temperature. This accounts for the weak temperature dependence of the metallic mirror transmission and allows us to extract the temperature-dependent conductivity of the YBCO film consistently across different cavity lengths.

\subsection{Cavity-corrected Tinkham formula}

Starting from the cavity transmission derived in Eq.~\ref{eq:S_modtin}, we now show how the conductivity of the YBCO film can be extracted from the normalized transmission. We denote by $E_t(\omega)$ the field transmitted through the full cavity assembly in the presence of the YBCO film, and by $\tilde{E}_t(\omega)$ the corresponding reference field obtained when the YBCO sheet contribution is absent. Their ratio defines the complex transmission function
\begin{equation}
\mathcal{T}(\omega)\equiv\frac{E_t(\omega)}{\tilde{E}_t(\omega)}.
\end{equation}
Using Eq.~\ref{eq:S_modtin} for the two cases, one obtains
\begin{equation}
\mathcal{T}(\omega)=
\frac{t_L}{\tilde{t}_L}
\frac{1-r_0\tilde{r}_Le^{i2\omega L/c}}
     {1-r_0r_Le^{i2\omega L/c}},
\end{equation}
where
\begin{equation}
\tilde{t}_L=t_L(d_L=0)=\frac{2}{n_r+1},
\qquad
\tilde{r}_L=r_L(d_L=0)=\frac{n_r-1}{n_r+1}.
\end{equation}
Algebraic inversion gives the cavity-corrected conductivity
\begin{equation}
\label{eq:smodtin}
\sigma_L(\omega)=
\frac{n_r+1}{Zd_L}
\left(\frac{1}{\mathcal{T}(\omega)}-1\right)
\frac{1-r_0\tilde{r}_Le^{i2\omega L/c}}
     {1-r_0e^{i2\omega L/c}}
\equiv
\sigma_L^{\mathrm{Tin}}(\omega)F(\omega,L).
\end{equation}
Here, $\sigma_L^{\mathrm{Tin}}(\omega)$ is the conductivity that would be obtained by applying the standard Tinkham expression directly to the YBCO film on its substrate (Eq. \ref{eq:tinkham_first}), while $F(\omega,L)$ contains the additional correction associated with the finite cavity length and the semi-reflecting mirror.

\subsubsection{Long-cavity limit}

The expression above simplifies considerably when the cavity is sufficiently long so that the first internally reflected cavity pulse is separated in time from the directly transmitted THz pulse. In this regime, the time-domain window used for the Fourier transform contains only the first transmitted pulse, and the multiple-reflection terms do not contribute to the extracted spectrum.

The cavity factor in Eq.\ \ref{eq:cav} can always be rewritten as
\begin{equation}
\frac{t_0t_L(\omega)}{1-r_0r_L(\omega)e^{i2\omega L/c}}=t_0t_L(\omega)\sum_{m=0}^{+\infty}\left(r_0r_L(\omega)\right)^me^{i\omega2mT_{cav}}\equiv\sum_{m=0}^{+\infty}F_m(\omega)e^{i\omega2mT_{cav}},
\end{equation}
where $2T_{cav}=2L/c$ is the round-trip cavity time and $F_m(\omega)$ encodes powers of the transmission and reflection coefficients of the cavity mirrors. Unlike the non-superconducting case, where all coefficients are real, the $t_L$ and $r_L$ functions are here frequency-dependent and complex, due to the presence of the superconducting layer. This means that the Fourier transform of $E_t(t)$ is slightly more complicated:
\begin{align}
E_t(t)&=\sum_{m=0}^{+\infty}\int d\omega\, e^{-i\omega(t-T_d-T_{cav}-2mT_{cav})}F_m(\omega)E_i(\omega)\nonumber\\
&=\sum_{m=0}^{+\infty}\int dt^\prime F_m(t^\prime)E_i(t-T_d-T_{cav}-2mT_{cav}-t^\prime).
\end{align}
Due to the frequency dependence of $F_m(\omega)$, the transmitted field in real time is now the sum of time delayed copies of the incident field $E_i$ convoluted with the response kernel $F_m$ induced by the number $m$ of reflections inside the cavity mirror. 

There are two key observations to make here. Due to the form of $t_L(\omega)$ and $r_L(\omega)$ in Eq.\ \ref{eq:S_rtc} in terms of conductivity, the analytic behavior of $\sigma_L(\omega)$ in the upper-half plane is inherited by the function $F_m(\omega)$. This means that $F_m(t)=\theta(t)\bar{F}_m(t)$.

In addition, in the impulsive limit, $E_i(t)\sim E_i\delta(t)$, the transmitted field reduces to 
\begin{equation}
E_t(t)=E_i\sum_{m=0}^{+\infty}\theta(t-T_d-T_{cav}-2mT_{cav})\bar{F}_m(t).
\end{equation}
The two considerations above imply that if one cuts the transmitted signal in time domain before the arrival of the internally reflected fields, only the $m=0$ term will contribute to $E_t(t)$ and it is possible to approximate it as follows:
\begin{equation}
\label{eq:lc}
E_t(\omega)=\bar{T}e^{i\omega T_d}t_0t_L(\omega)e^{i\omega L/c}E_i(\omega).
\end{equation}
We stress that an analogous result can be obtained if instead of assuming an impulsive field, one retains a finite duration $\tau$ such that $\tau\ll2T_{cav}$, which is true only for long-enough cavity lengths. When $\tau\sim2T_{cav}$ interferences with the internal reflection occur and the above formula is no longer valid. 

Considering again the ratio between $E_t(\omega)$ and $\tilde{E}_t(\omega)$ in the long-cavity limit of Eq.\ \ref{eq:lc}, one immediately recovers Eq.\ \ref{eq:smodtin} with $F(\omega,L)=1$, which means that Tinkham's formula can be used directly to retrieve the conductivity.

\subsubsection{Short-cavity limit}

For short cavities, the cavity round-trip time becomes comparable to the duration of the THz waveform. The first internal cavity reflection then overlaps with the directly transmitted pulse, so that it cannot be removed by time-windowing. In this regime, the correction factor $F(\omega,L)$ in Eq.~\ref{eq:smodtin} must be assessed explicitly.

Generally, since $F(\omega,L)$ is complex, it will mix the real and imaginary parts of the conductivity extracted using the bare Tinkham formula, namely
\begin{align}
\textrm{Re}\sigma_L(\omega)&=\textrm{Re}\sigma_L^{Tin}(\omega)\textrm{Re}F(\omega,L)-\textrm{Im}\sigma_L^{Tin}(\omega)\textrm{Im}F(\omega,L),\nonumber\\
\textrm{Im}\sigma_L(\omega)&=\textrm{Im}\sigma_L^{Tin}(\omega)\textrm{Re}F(\omega,L)+\textrm{Re}\sigma_L^{Tin}(\omega)\textrm{Im}F(\omega,L).
\end{align}
We can then give an estimate on the variation of the conductivity retrieved through the application of the bare Tinkham formula $(F=1, \sigma_L^{Tin})$ and the one including the cavity correction $(\sigma_L)$
\begin{align}
\frac{\Delta\textrm{Re}\sigma_L}{\textrm{Re}\sigma_L^{Tin}}&=\textrm{Re}F(\omega,L)-1-\frac{\textrm{Im}\sigma_L^{Tin}(\omega)}{\textrm{Re}\sigma_L^{Tin}(\omega)}\textrm{Im}F(\omega,L),\nonumber\\
\frac{\Delta\textrm{Im}\sigma_L}{\textrm{Im}\sigma_L^{Tin}}&=\textrm{Re}F(\omega,L)-1+\frac{\textrm{Re}\sigma_L^{Tin}(\omega)}{\textrm{Im}\sigma_L^{Tin}(\omega)}\textrm{Im}F(\omega,L).
\end{align}

\begin{figure}
\includegraphics[width=0.75\textwidth]{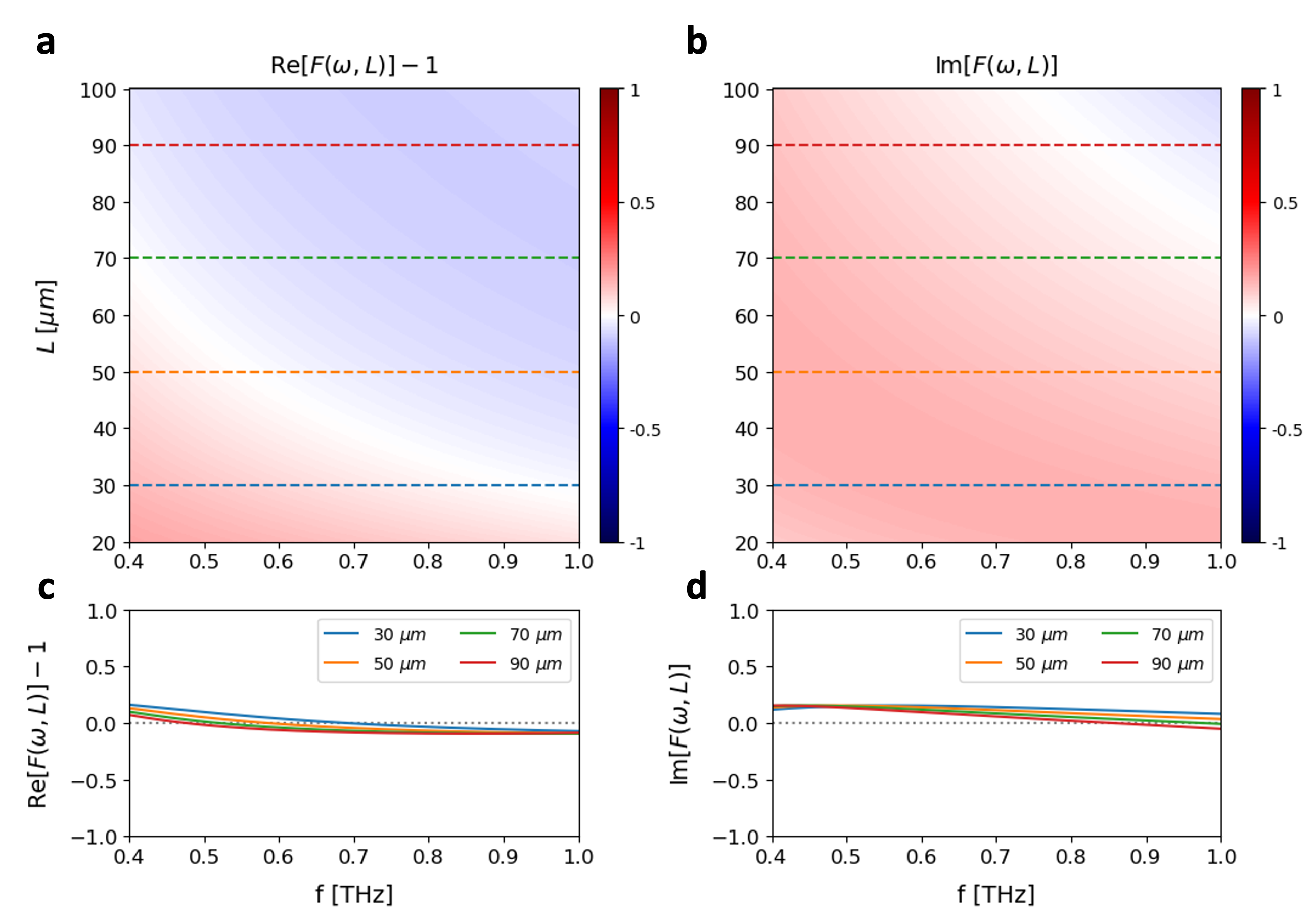}
\centering
\caption{Simulations of the real (\textbf{a}) and imaginary (\textbf{b}) parts of the function $F(\omega,L)$ for the frequencies and cavity lengths of interest. \textbf{c-d,} Horizontal cuts at fixed $L$. The simulations are performed considering the fitted value of $r_0=0.39$ for the gold-titanium mirror, derived from the fitted parameter $\rho$ in Figure\ \ref{fig:lao_simul}.}
\label{fig:F}
\end{figure}

To understand the behavior, we plot the the quantities $\textrm{Re}F(\omega,L)-1$ and $\textrm{Im}F(\omega,L)$ in Figure \ref{fig:F}. We focus on the range of cavity lengths $L$ experimentally investigated and we explore the frequency range where the incoming field spectrum is strong enough to allow for an optimal extraction of the conductivity. We set $r_0$ to the value extracted from the fit of the cavities in absence of YBCO, see Figure\ \ref{fig:lao_simul}. Focusing on the variation of the imaginary part of the conductivity, we observe that, in the first case, $\textrm{Re}F-1$ is actually positive for the shortest cavities and produces a $5-10\%$ of reduction in $\sigma_2$ in the longest cavities of the set. Moreover, $\textrm{Im}F$ provides an additional positive variation which would balance the aforementioned small reduction. 

In summary, setting $F=1$ also for the short cavities actually constitutes a conservative approximation to estimate the cavity-enhanced imaginary part of the conductivity, as discussed in Sec. \ref{subsec:extraction_sigma}. 

\subsection{Correction for a fixed-temperature empty-cavity reference}

The conductivity extraction discussed above is based on the normalization of the field transmitted through the YBCO film to a reference field measured through the same cavity geometry without the YBCO conducting layer. Since the dielectric elements of the setup have negligible temperature dependence in the relevant THz spectral range (see Sec.~\ref{sec:characterization}), a single empty-cavity reference acquired at a fixed temperature $\bar{T}$ is sufficient. The only temperature-dependent contribution to the reference measurement comes from the metallic Ti/Au mirror, which we characterized experimentally (see Sec. \ref{subsec:mirror}). The transmissivity and reflectivity of the metallic layer can be accounted for explicitly in the model through the coefficients $t_0(T)$ and $r_0(T)$.

From the measured temperature-induced change in transmission amplitude of approximately 15\% between $\bar{T}\sim300$ K and $T=100$ K, and using the thin-film Tinkham expression for the field transmission coefficient, we estimate a relative conductivity variation $\Delta\sigma_0/\sigma_0 \approx -\Delta t_0/t_0 \approx 15\%$. This approximation follows from the linearized form of the Tinkham relation in the experimentally relevant regime. Translating this conductivity variation into the \textit{extracted} cavity reflection coefficient $r_0$ yields only a weak relative variation of the cavity parameter, $\Delta r_0/r_0 \approx 1\%$. 

With this temperature-normalization procedure, the measured transmission ratio can be written as
\begin{equation}
\mathcal{T}(\omega,T)\equiv
\frac{E_t(\omega,T)}{\tilde{E}_t(\omega,\bar{T})}
=
\frac{t_L}{\tilde{t}_L}
\frac{t_0(T)}{t_0(\bar{T})}
\frac{1-r_0(\bar{T})\tilde{r}_Le^{i2\omega L/c}}
     {1-r_0(T)r_Le^{i2\omega L/c}}.
\end{equation}

that we want to invert to find $\sigma_L(\omega)$. Here, $E_t(\omega,T)$ is the field transmitted through the YBCO film in the cavity at temperature $T$, while $\tilde{E}_t(\omega,\bar{T})$ is the empty-cavity reference acquired at temperature $\bar{T}$ (room temperature in the experiment). The coefficients $\tilde{t}_L$ and $\tilde{r}_L$ correspond to the transmission and reflection coefficients of the right interface in the absence of the YBCO sheet contribution. This expression separates the intrinsic YBCO response, contained in $t_L$ and $r_L$, from the temperature-dependent response of the metallic mirror.

Since $\Delta r_0/r_0 \approx 1\%$, we can safely set $r_0(T)=r_0(\bar{T})\equiv r_0$. The previous Eq.\ \ref{eq:smodtin} is again rewritten in the form $\sigma_L(\omega)=\tilde{\sigma}_L^{Tin}(\omega)F(\omega)$, with $F(\omega)$ expressed in terms of the common $r_0$, while $\tilde{\sigma}_L^{Tin}(\omega)$ is now 

\begin{equation}\label{eq:sigma_temp_indep}
\tilde{\sigma}_L^{Tin}(\omega)=\frac{n_r+1}{Zd_L}\left(\frac{1}{\mathcal{\tilde{T}(}\omega)}-1\right).
\end{equation}
Here we redefined the corrected transmissivity ratio as:
\begin{equation}\label{eq:temp_corr}
\frac{1}{\mathcal{\tilde{T}}(\omega)}=\frac{1}{\mathcal{T}(\omega,T)}\frac{t_0(T)}{t_0(\bar{T})}\frac{1-r_0(\bar{T})\tilde{r}_Le^{i2\omega L/c}}{1-r_0(T)\tilde{r}_Le^{i2\omega L/c}}\approx\frac{1}{\mathcal{T}(\omega,T)}\frac{t_0(T)}{t_0(\bar{T})},
\end{equation}
where in the last step we have again set $r_0(T)=r_0(T^*)=r_0$. Eq. \ref{eq:sigma_temp_indep} would then read
\begin{equation}
\tilde{\sigma}_M^{Tin}(\omega)=\frac{n_r+1}{Zd_L}\left(\frac{1}{\mathcal{T}(\omega,T)}\frac{t_0(T)}{t_0(\bar{T})}-1\right), 
\end{equation}
with an intuitive renormalization of the ratio $\mathcal{T}$ by the different Ti/Au mirror transmissivity at the temperatures $T$ and $\bar{T}$ (i.e., the scaling factor extracted in Figure \ref{fig:mirror_tcorr}a). This scaling ensures that the extracted temperature dependence is assigned to the YBCO film rather than to the weak temperature dependence of the metallic mirror.

\subsection{Experimental extraction of temperature-dependent optical conductivity}\label{subsec:extraction_sigma}

We now describe the experimental procedure used to extract the temperature-dependent complex optical conductivity of the YBCO film. The aim is to construct, for each cavity length, a normalized complex transmission that isolates the response of the YBCO conducting layer from propagation through the remaining optical elements of the setup.

Operationally, the extraction proceeds as follows:
\begin{itemize}
    \item We first measure the temperature-dependent THz field transmitted by the YBCO sample inside the cavity for a selected set of cavity lengths.

    \item We then remove only the cavity mirror and measure the temperature-dependent THz field transmitted by the same YBCO sample in free space. Importantly, the sample mounting and thermal contacts are left unchanged, so that this measurement provides a direct free-space reference for the same film.

    \item We subsequently replace the YBCO sample by the reference substrate, consisting of LaAlO$_3$ with the CeO$_2$ buffer layer, and mount the cavity mirror again.

    \item For this reference-substrate configuration, we measure the empty-cavity transmission at room temperature for the same set of cavity lengths used in the YBCO measurements.

    \item The empty-cavity measurement provides the reference field $\tilde{E}_t(\omega,\bar{T})$ for each cavity length. As discussed in the previous subsection, this fixed-temperature reference is rescaled using the calculated temperature-dependent transmission of the Ti/Au mirror, shown in Fig.~\ref{fig:mirror_tcorr}a. This gives the corrected transmission ratio $\mathcal{\tilde{T}}(\omega,T)$ according to Eq.~\ref{eq:temp_corr}.

    \item Finally, the YBCO conductivity is extracted using the Tinkham expression introduced above (Eq.~\ref{eq:smodtin}).
\end{itemize}

In the following analysis, we take $F(\omega,L)=1$ unless explicitly stated otherwise. This approximation is justified in the two regimes relevant to our measurements. For sufficiently long cavities, $L\gtrsim400~\mu$m, the first internally reflected cavity pulse is temporally separated from the directly transmitted waveform and can be removed by time-windowing. In this limit, the long-cavity expression derived above applies directly. For the short-cavity datasets, $L\lesssim100~\mu$m, the internal reflection overlaps with the main THz waveform; however, the correction factor $F(\omega,L)$ can be evaluated explicitly in the relevant frequency and length range. As shown in Fig.~\ref{fig:F}, this correction produces only a small modification of the extracted conductivity. A direct comparison in Fig.~\ref{fig:comp} further shows that including $F(\omega,L)$ does not change the observed enhancement of the averaged $\sigma_2$ response discussed in the main text.

%For compactness, we characterize the temperature-dependent superconducting response using the frequency-integrated conductivities $\bar{\sigma}_1(T)$ and $\bar{\sigma}_2(T)$, obtained by integrating $\sigma_1(\omega,T)$ and $\sigma_2(\omega,T)$ over the spectral window from 0.4 to 1~THz.

\subsubsection{YBCO in free space}

\begin{figure}
\includegraphics[width=0.95\textwidth]{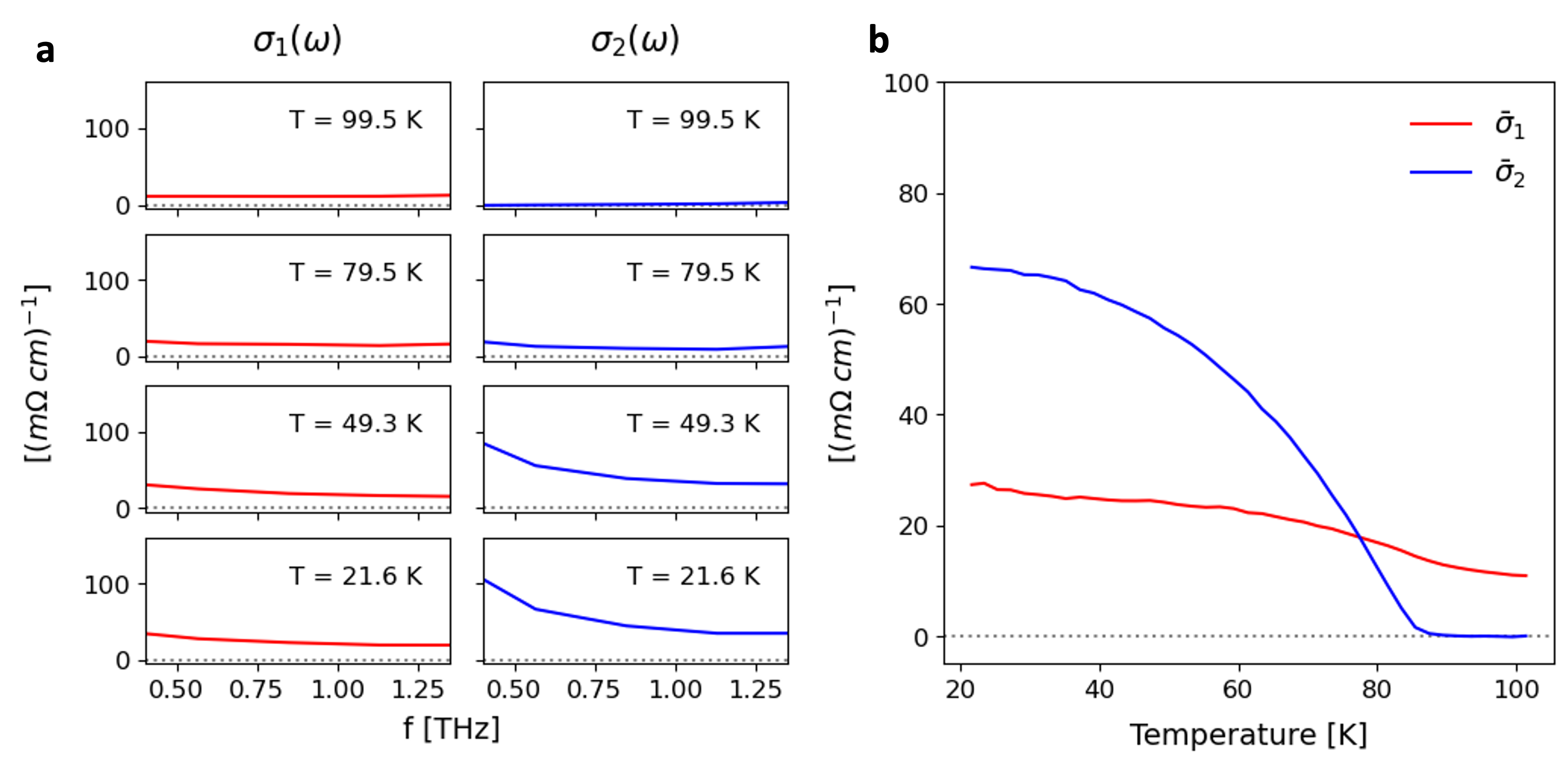}
\centering
\caption{\textbf{a,} Frequency-dependent real, $\sigma_1(\omega)$, and imaginary, $\sigma_2(\omega)$, parts of the YBCO conductivity in free space for selected temperatures. \textbf{b,} Frequency-averaged $\bar{\sigma}_1$ and $\bar{\sigma}_2$ as a function of temperature. The averaging range is 0.4--1~THz.}
\label{fig:sigma_y_free}
\end{figure}

We first validate the extraction procedure using the YBCO film measured in free space. In this case, the reference is the temperature-dependent THz field transmitted by the LaAlO$_3$+CeO$_2$ substrate stack, shown in Fig.~\ref{fig:lao_tdep}. The extracted conductivity is shown in Fig.~\ref{fig:sigma_y_free}. The real and imaginary parts, $\sigma_1(\omega,T)$ and $\sigma_2(\omega,T)$, show the expected evolution across the superconducting transition. In particular, the low-frequency $\sigma_2(\omega,T)$ develops the characteristic $1/\omega$ inductive response below $T_c$, while the temperature dependence of the frequency-averaged imaginary conductivity $\bar{\sigma}_2(T)$ provides a clear marker of the superconducting transition.

The extracted conductivity is in good quantitative agreement with previous THz measurements on YBCO films \cite{ybco_pumpprobe}, confirming that the normalization and thin-film extraction procedure reproduce the expected superconducting electrodynamics in the absence of the cavity.

\subsubsection{YBCO in the long-cavity regime}

We next apply the same extraction procedure to the YBCO film inside the cavity. As a representative long-cavity case, we consider a cavity length $L=435~\mu$m. For this length, the first cavity reflection is temporally separated from the main transmitted THz pulse and can be removed by time-windowing. The analysis is therefore performed in the long-cavity limit, for which $F(\omega,L)=1$.

\begin{figure}
\includegraphics[width=0.95\textwidth]{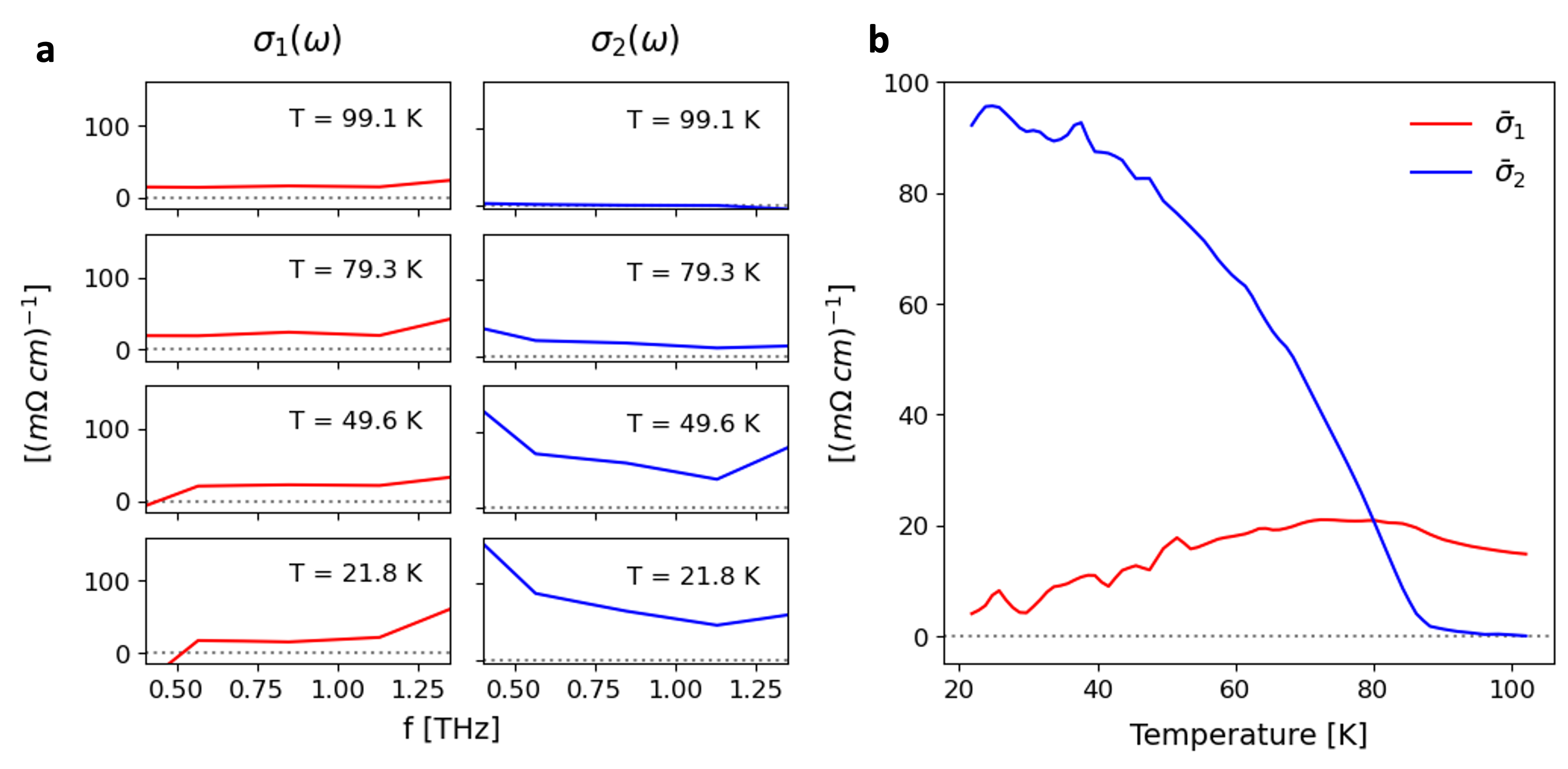}
\centering
\caption{\textbf{a,} Frequency-dependent real, $\sigma_1(\omega)$, and imaginary, $\sigma_2(\omega)$, parts of the YBCO conductivity inside the cavity for $L=435~\mu$m at selected temperatures. \textbf{b,} Frequency-averaged $\bar{\sigma}_1$ and $\bar{\sigma}_2$ as a function of temperature. The averaging range is 0.4--1~THz.}
\label{fig:sigma_y_cavity}
\end{figure}

The extracted complex conductivity is shown in Fig.~\ref{fig:sigma_y_cavity}. As in the free-space case, the onset of superconductivity is marked by the emergence of a strong low-frequency $\sigma_2(\omega,T)$ response. Comparing the averaged imaginary conductivity $\bar{\sigma}_2(T)$ with the corresponding free-space result reveals the enhancement of the superconducting response reported in the main manuscript. This comparison is the basis of the long-cavity data shown in Figure~2 of the main text.

\subsubsection{YBCO in the short-cavity regime}

We then extend the conductivity extraction to the short-cavity regime, with cavity lengths in the range $20$--$95~\mu$m. In this regime, the first cavity reflection overlaps with the main THz waveform and cannot be removed by time-windowing. It is therefore essential to ensure that the YBCO cavity measurement and the empty-cavity reference are acquired at the same mirror--sample separation.

To do so, we first define a common reference position at a sufficiently long cavity length, where the first cavity echo is clearly distinguishable from the main transmitted pulse. This procedure is performed for both the YBCO cavity configuration and the empty-cavity reference configuration. Starting from this matched position, the piezoelectric nanopositioners are then displaced by the same calibrated amount, ensuring that the final short-cavity length is the same in the sample and reference measurements. This protocol is repeated for all cavity lengths in the short-cavity set.

Using the corrected transmission ratio defined above, we extract the frequency-dependent complex optical conductivity for each short-cavity length. We then average the imaginary part over the same 0.4--1~THz spectral window to obtain $\bar{\sigma}_2(T)$. These data correspond to the short-cavity results shown in Figure~3c of the main text.

\begin{figure}
\includegraphics[width=0.75\textwidth]{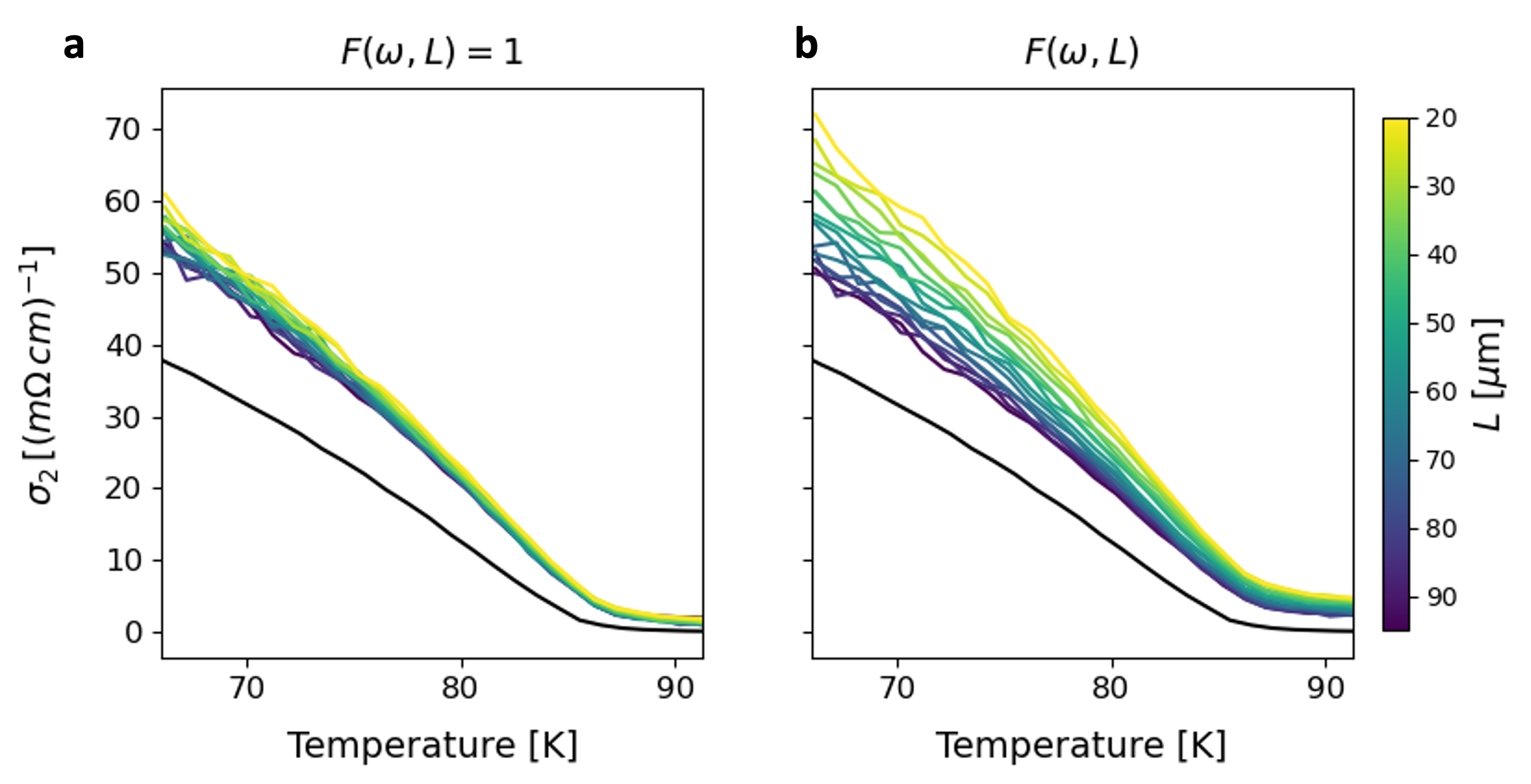}
\centering
\caption{Comparison of $\bar{\sigma}_2$ extracted for different cavity lengths using \textbf{a,} the bare Tinkham expression, $F=1$, and \textbf{b,} the calculated correction factor $F(\omega,L)$ from Fig.~\ref{fig:F}. Importantly, the analysis discussed in Figure 3c in the main text considers $F=1$ and should be thus regarded as a conservative estimate of the cavity-enhanced effect.}
\label{fig:comp}
\end{figure}

For the main analysis, we keep $F(\omega,L)=1$. To verify that the observed enhancement is not introduced by this approximation, we compare the averaged $\bar{\sigma}_2$ obtained using the bare Tinkham expression with the result obtained after applying the calculated correction factor $F(\omega,L)$. The comparison is shown in Fig.~\ref{fig:comp}. Including the correction changes the absolute magnitude only weakly and does not alter the qualitative temperature dependence or the enhancement of $\bar{\sigma}_2$ in the short-cavity regime. Thus, the extracted enhancement of the superconducting response is not a consequence of residual cavity-propagation effects. On the contrary, using $F(\omega,L)=1$ provides a conservative estimate of the cavity-induced modification.

\subsection{Error analysis}

The uncertainty on the extracted optical conductivity was estimated by propagating the experimental noise of the measured THz electric fields through the conductivity extraction. The statistical contribution originates from the time-domain signal recorded by the lockin amplifier. We measured a standard deviation $\sigma_t$ for the recorded THz traces, corresponding to the noise level of the lockin signal \footnote{The lockin noise level is $\sigma_t=4\times10^{-7}~V$. The amplitude of the measured THz peak with no optical elements in transmission is $6 \times 10^{-3}~V$, resulting in a signal-to-noise ratio of $\sim2\times10^4$.}.

The same preprocessing steps used for the conductivity extraction were applied to estimate this contribution: the sample and reference traces were placed on a common delay axis, time-windowed, and Fourier transformed. The resulting frequency-domain uncertainty ($\sigma_f = \sigma_t \sqrt{N/2}$, where $N$ is the number of points in the time window used for the Fourier transform) was then propagated to the complex transmission $\mathcal{T}(\omega,T)=\frac{E_{t}(\omega,T)}{\tilde{E}_{t}(\omega,T)}$ expressed in terms of its amplitude $A$ and phase $\phi$. 
%The uncertainty in these two qunatities was evaluated as follows:
%\begin{align}
%   \delta A (\omega)=& \\
%    \delta \phi (\omega) =& \sqrt{\left(\frac{\delta E_{\mathcal{sam}}}{E_{\mathcal{sam}}(\omega)}\right)^2 + \left(\frac{\delta E_{\mathcal{ref}}}{E_{\mathcal{ref}}(\omega)}\right)^2} \simeq \frac{\delta E_{\mathcal{sam}}}{E_{\mathcal{sam}}(\omega)} = \frac{\sigma_f}{E_{\mathcal{sam}}(\omega)}
%\end{align}

Using the Tinkham formula, the statistical uncertainty of the imaginary conductivity $\sigma_2(\omega)$ was obtained by standard error propagation from the uncertainties in the transmission amplitude and phase.

In addition to this statistical contribution, we include a systematic uncertainty associated with the prefactor of the Tinkham formula. The dominant contribution comes from the uncertainty in the YBCO film thickness, which is approximately $10\%$. Since the film thickness enters as an overall multiplicative factor in the conductivity extraction, this uncertainty affects the absolute scale of $\sigma_2$ but not its relative temperature dependence.

The total uncertainty on $\sigma_2(\omega)$ was obtained by adding the statistical and systematic contributions in quadrature. This procedure provides the uncertainty used for the extracted imaginary conductivity and for the frequency-averaged quantity discussed in the main text.

%-----------------------------------------------------------------------
\section{Theoretical model}\label{sec:model}

Here we present the theoretical framework underlying the model discussed in the main text, which predicts a cavity-enhanced superfluid phase stiffness. The goal is to capture, at a phenomenological level, how cavity coupling can enhance the superconducting response by influencing phase fluctuations. We propose a mechanism that is relevant for superconductors where the two-dimensional physics of phase fluctuations is important. %, like the layered YBCO film used in the present experiment. 
The comparison with the experimental results should therefore be understood as a qualitative interpretative framework.

\

A further guidance in the choice of the model is offered by the fact that underdoped cuprates with particularly low superfluid stiffness show a pseudogap regime above the critical temperature $T_c$, commonly associated with robust pairing correlations but reduced phase stiffness compared to mean-field expectations. Spectroscopic probes \cite{corson1999vanishing,PhysRevLett.106.047006,yuan2024dynamical} observe a gap-like suppression of low-energy excitations up to a higher temperature $T^*$, while diamagnetism \cite{PhysRevLett.95.247002,PhysRevB.81.054510} and Nernst \cite{xu2000vortex,wang2006nernst,daou2010broken,PhysRevB.97.064502,hu2024vortex} signals indicate strong superconducting fluctuations over an extended range above $T_c$. 
Within the widely used phase-only model for quasi-two-dimensional systems, the regime near $T_c$ is characterized by dominant fluctuations of the order parameter phase, while the pairing amplitude remains rigid \cite{halperin1979resistive,emery1995importance,PhysRevLett.83.612, LOKTEV20011, PhysRevB.66.094515,PhysRevLett.94.217001, PhysRevLett.99.117004,zhou2025universal}. In this description, the long-wavelength physics can be captured by an effective two-dimensional XY model, which in the strictly 2D case exhibits a Berezinskii-Kosterlitz-Thouless (BKT) transition driven by vortex-antivortex unbinding, thereby setting the superconducting transition temperature $T_c$ \cite{berezinskii1971destruction,kosterlitz1973ordering,RevModPhys.59.1001}. %BKT physics should be robust against weak interlayer coupling and inhomogeneities in real materials, the latter however smearing the transition \cite{PhysRevB.80.214506,PhysRevB.93.024519,kracht2026crossover}.

\

We consider the experimental situation in which superconducting YBCO film is measured by a THz pulse in the long-wavelength limit, and therefore probed in the London regime. In this limit, the imaginary part of the complex optical conductivity is $\sigma_2(\omega) \propto \rho_s/\omega$ (Eq. \ref{eq:twofluid}), allowing us to relate the effective phase stiffness $J$ obtained from our model to the superfluid density $\rho_s$ and in turn to $\sigma_2$. Our model predicts that the cavity environment enhances $J$ 
%and thus $\sigma_2$ 
with respect to free space, which in turn leads to a tiny increase of the BKT transition temperature $T_c$. This behavior is consistent with the experimental observations inferred from the frequency dependence of $\sigma_2(\omega)$. 

\subsection{Cavity-increased phase stiffness}\label{ssec:model_J}

\textit{
In this section we derive an effective theory for the superconducting phase dressed by electromagnetic fluctuations. The idea is to use the phenomenological coarse-grained theory for the superconducting order parameter phase in 2D, minimally coupled to the electromagnetic field. The latter can then be integrated out to obtain a phase-only description with cavity-dressed plasmon mode. The cavity modifies the electromagnetic component of the plasmon and increases the energetic cost of the collective mode. By raising this energy, the cavity suppresses thermally excited phase fluctuations, thereby reducing their disordering effect on the superconducting phase and effectively enhancing the total phase stiffness.
}

\subsubsection{London model for a phase-fluctuating superconductor}

As was mentioned, the 2D spin XY model can be used to model the phase-fluctuations in cuprates, as it is intrinsically nonlinear and thus naturally contains both spin-wave and vortex phase fluctuations. The angle of the spin on the lattice site $\theta_i$ can be associated with the phase of the order parameter, which couples to the electromagnetic (EM) gauge field via standard Peierls substitution, giving rise to the following Hamiltonian ${H_{XY} = - J_0 \sum_{\langle i,j\rangle}\cos(\theta_i-\theta_j - e^* A_{ij})}$. Here $J_0$ is the phase stiffness, and $A_{ij}=\int_{{\bf r}_i}^{{\bf r}_j} {\bf A}d{\bf r}$ with effective charge $e^*=2e$ in superconductors. Hereafter we use the natural units $\hbar = c = 1$. For the long-wavelength physics one can use the coarse-grained approximation of the XY model, $H_{XY} \approx  J_0 \int d{\bf r} (\nabla\theta - e^* {\bf A})^2$, which describes the smooth phase fluctuations and can also account for vortices if the latter are introduced explicitly (see Sec. 6.2). This coarse-grained XY model is equivalent to the London theory of superconductivity \cite{tinkham2004introduction}, where the physics is determined purely by the phase of the order parameter \cite{PhysRevB.69.184510}. In this minimal model, the strength of the light-matter coupling is determined solely by the phase stiffness $J_0$ and the effective charge $e^*$. These parameters also set the fundamental electrodynamic scale of the superconductor, namely the London penetration depth $\lambda_L^{-2}\propto J_0 (e^*)^2$.

\

\begin{figure}[h] 
\includegraphics[width=0.5\textwidth]{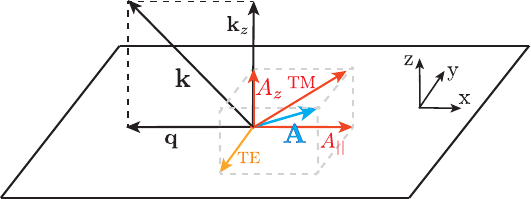}
\centering
\caption{\small{ Projection of the transverse electromagnetic vector potential {\bf A} with the wavevector ${\bf k}$ onto the plane of the 2D superconductor. }}
\label{sketch_theory}
\end{figure}

Here we specifically consider a two-dimensional superconductor with a lateral size $S$ lying in the $z=z_0$ plane. The corresponding Euclidean action is given by
\begin{gather}\label{S}
    S[\theta,\phi,{\bf A}] = \int d\tau \int d{\bf r} \left[  \frac{\varkappa}{2}\Big( \partial_\tau \theta+ e^*\phi \Big)^2+ \frac{J_0}{2}\Big( \nabla_{||} \theta- e^*{\bf A} \Big)^2 
    \right]\delta\left(z-z_0\right) 
    \\ \notag
    + \frac{1}{8\pi}\Big[  (\nabla \phi)^2 + (\partial_\tau {\bf A})^2 + (\nabla\times{\bf A})^2
    \Big],
\end{gather}
where $\theta$ is the phase of the order parameter $\Delta=\Delta_0e^{i\theta}$ with constant gap amplitude $\Delta_0$; $\phi$ and ${\bf A}({\bf r},\tau)$ are respectively the 3D EM scalar and vector potentials; $J_0=\rho_s^{2D}/m$ is the bare phase stiffness connected to the two-dimensional superfluid density $\rho_s^{2D}=\rho_s^{3D}d$ via the superconductor's thickness $d$; $\tau$ is imaginary time; and $T$ is the temperature. The term proportional to charge compressibility $\varkappa$ accounts for the phase dynamics, with its time-derivative coupled to the scalar potential. The term proportional to the bare phase stiffness $J_0$ corresponds to the coarse-grained XY-model term already discussed above. Finally, the last term between square brackets is the free EM field contribution. We use the Coulomb gauge $\nabla \cdot {\bf A}=0$ and set the dielectric permittivity and magnetic permeability as $\varepsilon=\mu=1$. The full 3D momentum ${\bf k}=({\bf q},k_z\hat{\bf z})$ has the in-plane component ${\bf q}$ (see Fig. \ref{sketch_theory}). For the in-plane Fourier transform to momentum and Matsubara representation $f({\bf r},\tau) = \sqrt{\frac{T}{S}}\sum_{q} f(q,z) e^{i{\bf q}{\bf r}_{||}+\Omega_m \tau}$ we use the short notation $q=({\bf q},\Omega_m)$.

\

Physically, the effect of the EM fluctuations arises from the coupling of the superconducting phase $\theta$ to the in-plane component of the vector potential. Although the EM field is transverse with respect to the full 3D wavevector $({\bf q},k_z\hat{\bf z})$, it can have a finite projection onto the superconducting plane along the in-plane momentum ${\bf q}$ and we split the vector potential with respect to it as ${\bf A}={\bf A}_{||}+{\bf A}_\bot$. From the viewpoint of a 2D film, ${\bf A}_{||}$ acts as an in-plane longitudinal vector-potential component and therefore couples to the phase gradient $\nabla\theta$.

Because the static longitudinal vector potential is a pure gauge, it can be absorbed into the phase through the substitution $\nabla\theta \to \nabla\theta - e^* T \int d\tau' {\bf A}_{||}({\bf r},\tau')$. This transformation removes its coupling to the phase from the action \eqref{S}, leaving only the coupling to finite-frequency component of the longitudinal vector potential, i.e. to electric field fluctuations. These fluctuations can be effectively screened in cavities by modifying the available EM modes, as considered below. We note that this differs essentially from the magnetostatic Halperin-Lubensky-Ma mechanism in 3D superconductors, where the fluctuations of the transverse vector potential ${\bf A}({\bf r})$  dress the amplitude of the superconducting order parameter and can change the nature of the phase transition \cite{PhysRevLett.32.292}. In the present London phase-only theory in 2D, however, the amplitude $\Delta_0$ is fixed, so a similar type of gauge-field feedback appears as nonlocal renormalization of the phase kernel. 

\subsubsection{Cavity-confined electromagnetic fluctuations}

First, we consider how the cavity changes the EM field. The scalar potential is instantaneous in the Coulomb gauge, so its spatial structure in the cavity is determined by the Poisson equation. We split the action \eqref{S} into a part $S[\theta,\phi]$ describing the effect of electrostatics onto the order-parameter phase, and an electrodynamics part $S[\theta,{\bf A}]$, containing the vector potential whose longitudinal in-plane component couples to the phase:
\[
S[\theta,\phi,{\bf A}]=S[\theta,\phi]+S[\theta,{\bf A}].
\]
The phase-electrostatics part reads:
\begin{gather}\label{S_phi}
    S[\theta,\phi] =  \frac{1}{2T} \sum_q \varkappa \Omega_m^2 |\theta(q)|^2 + 2\varkappa e^* i\Omega_m \theta(-q)  \phi({\bf q},z_0) +  \int dz dz' \phi(-{\bf q},z) U^{-1}({\bf q},z,z') \phi({\bf q},z'),
\end{gather}
where $U^{-1}({\bf q},z,z') = \frac{1}{4\pi} (-\partial_z^2+{\bf q}^2)\delta(z-z')+\varkappa (e^*)^2\delta(z-z_0)\delta(z'-z_0)$ is the electrostatic kernel at the position of the superconducting film $z=z_0$. By means of the bare Green function, which satisfies $(-\partial_z^2+{\bf q}^2)U_0(q,z,z')=4\pi \delta(z-z')$,  we can compute the electrostatic potential with the help of the Dyson equation $U^{-1}({\bf q},z,z')=U_0^{-1}({\bf q},z,z')+\varkappa (e^*)^2\delta(z-z_0)\delta(z'-z_0)$.

The electrostatic response is sensitive to setup-specific details, including grounding conditions, contacts, the substrate, and the surrounding electromagnetic environment outside the cavity. In particular, imposing ideal grounded-metal boundary conditions on both mirrors would describe a specific electrostatic environment and would generally screen the scalar potential. However, the experimental setup differs substantially from this idealization, as the superconducting film itself forms one of the mirrors, and the opposite mirror is not well grounded. We therefore do not model mirror-induced electrostatic screening explicitly. Instead, we treat the mirrors as defining the transverse cavity modes, while approximating the Coulomb interaction by its free-space form. Equivalently, the bare Green function is taken to satisfy the free-space  boundary condition $U_0(q,z,z')\to 0$ as $|z-z'|\to\infty$, which gives
\begin{gather}\label{U_0}
    U_0({\bf q})=2\pi/|{\bf q}|,
\end{gather}
and the corresponding full kernel $U({\bf q})=U_0({\bf q})/(1+\varkappa (e^*)^2U_0({\bf q}))$.

\

The phase-electrodynamics part contains the vector potential which is a dynamical (fluctuating) field whose quantized excitations correspond to photons. This term accounts for the coupling of the superconducting phase to the dynamical EM field and therefore incorporates retardation and photon fluctuation effects beyond the instantaneous electrostatic interaction. The corresponding part of the phase action $S[\theta, {\bf A}]$ can be written without specifying a particular cavity geometry. The relevant part of the EM field which couples to the phase gradient is the longitudinal in-plane component of the vector potential at the position of the superconducting layer $z=z_0$. Using Fourier transform ${\bf A}_{||}({\bf r}_{||},z_0,\tau) = \sqrt{\frac{T}{S}}\sum_{q} {\bf A}_{||}(q) e^{i{\bf q}{\bf r}_{||}+\Omega_m \tau}$, one can generally expand the corresponding component in EM normal modes as 
\begin{gather}
{\bf A}_{||}(q)=\hat{\bf q}\sum_\ell u_\ell({\bf q})A_\ell(q).
\end{gather}
Here the mode functions $u_\ell({\bf q})$ and the corresponding photon propagators $D_\ell(q)$ are determined by the EM normal modes of the cavity.  After we implement the additional transformation, introduced above to absorb ${\bf A}_{||}({\bf q},\Omega_m=0)$ into the phase of the order parameter, the action takes the form
\begin{gather}\label{S_A}
    S[\theta,{\bf A}] =   \frac{1}{2T} \sum_q  J_0 {\bf q}^2 |\theta(q)|^2 - 2 J_0 e^* i{\bf q} \theta(-q)  {\bf A}_{||}(q) (1-\delta_{m,0})
    \\ \notag
    + J_0 (e^*)^2 {\bf A}_{||}(-q){\bf A}_{||}(q) +  \frac{1}{2T} \sum_q\sum_\ell A_\ell(-q) D^{-1}_\ell(q) A_\ell(q). 
\end{gather}
The phase couples to the total longitudinal in-plane field. Therefore the photon propagator entering the phase dynamics is the mode-weighted sum $D(q)=\sum_\ell u^2_\ell({\bf q}) D_\ell(q)$.

We now specialize this general form to a Fabry-Pérot cavity oriented along the $z$-axis, with mirror separation $L$. We assume the cavity mirrors to have the same lateral area $S$ as the superconductor. Solving Maxwell's equations with the corresponding boundary conditions gives standing-wave modes with quantized momenta along the cavity axis $k_z\to k_\ell=\pi \ell/L$. For a superconducting film located in the middle of the cavity ($z_0=L/2$), only the TM polarization has a longitudinal in-plane component, which gives gives $u_\ell({\bf q})=\sqrt\frac{2}{L}\sin(\frac{\pi \ell}{2}) \frac{k_\ell}{\sqrt{{\bf q}^2+k_\ell^2}}$. At the same time the TE polarization, together with the transverse in-plane part of the TM polarization, decouples from the phase sector considered here (as they constitute the perpendicular component ${\bf A}_\bot$) and will not be kept explicitly below. For the Fabry-Pérot cavity, the corresponding photon propagator of the $\ell$-th mode reads simply as $D_\ell(q)=\frac{4\pi}{\Omega_m^2+{\bf q}^2+k_\ell^2}$. Thus, the full photon propagator in our model has the following form:
\begin{gather}\label{D_full}
    D(q,L) = \frac{2\pi}{\Omega_m^2} \left[ \sqrt{{\bf q}^2+ \Omega_m^2} \tanh\frac{\sqrt{{\bf q}^2+ \Omega_m^2}L}{2} - |{\bf q}| \tanh\frac{|{\bf q}|L}{2} \right].
\end{gather}

\subsubsection{Phase fluctuations dressed by the electromagnetic field}

Since the full-phase action is quadratic in the scalar (Eq. \eqref{S_phi}) and vector (Eq. \eqref{S_A}) potentials, the EM field can be integrated out exactly. Performing the Gaussian integration and using the photon propagator \eqref{D_full} and electrostatic kernel \eqref{U_0}, we obtain the following effective action for order-parameter phase dressed by cavity-confined EM fields: 
\begin{gather}\label{S_eff}
    S[\theta] =  \frac{1}{2T} \sum_q  J_0\left[ \frac{\Omega_m^2}{J_0/\varkappa+\frac{1}{2\pi\lambda_P}U_0({\bf q})}  +  \frac{{\bf q}^2 }{1+\frac{1}{2\pi\lambda_P}D(q,L)(1-\delta_{m,0})} \right]|\theta(q)|^2.
\end{gather}
Here we introduced the Pearl length $\lambda_P = 1/2\pi J_0 (e^*)^2 = m/2\pi (e^*)^2n_s^{2D}$, which characterizes the screening of the transversal magnetic field in two-dimensional superconductors, and is connected to the London penetration depth and the film thickness via the relation $\lambda_P = 2\lambda_L^2/d$ \cite{10.1063/1.1754056}. The inverse Pearl length $1/\lambda_P$ quantifies the strength of light-matter coupling in our model. The action \eqref{S_eff} shows that coupling a two-dimensional superconductor to a three-dimensional cavity-confined EM field makes the superconducting phase dynamics explicitly dependent on the cavity length.

\begin{boxE}
\textbf{Connection between model geometry and experimental cavity.}
In the experimental setup, the superconductor itself forms one of the cavity boundaries, rather than being placed at the center of a Fabry-Pérot cavity. It therefore experience reduced EM fluctuations compared with the theoretical model considered here, where the fundamental standing wave has an antinode at the superconductor position. Nevertheless, the electric field at the film remains finite because the relevant field-penetration length, either the skin depth $\delta_s$ or the London penetration depth $\lambda_L$, exceeds the film thickness $d\ll$min($\delta_s,\lambda_L$).

\

Consequently, the cavity-induced  modification of the photon propagator in Eq. \eqref{D_full} is quantitatively reduced, reflecting the weaker coupling between quantized cavity modes and the superconducting phase when the superconductor stays at the cavity boundary. Nevertheless, the central physical idea that the cavity modifies phase dynamics remains qualitatively valid. For clarity and simplicity, we therefore focus on the idealized case of $z_0=L/2$. 
\end{boxE}

\

One can verify that the phase-only action \eqref{S_eff} preserves charge continuity, as required by the gauge invariance of the original action \eqref{S} \cite{PhysRevB.69.184510}. After analytic continuation, the saddle-point equation of the full action gives the continuity equation $\partial_t\rho + \nabla \cdot {\bf j}_s=0$, with the gauge-invariant charge $\rho = \varkappa e^*(\partial_t\theta+e^*\phi)$ and supercurrent ${\bf j}_s = J_0 e^*(\nabla\theta-e^*{\bf A})$. On the other hand, since the action is Gaussian in the EM fields and since the parameters $J_0$ and $\varkappa$ are independent of $\theta$, integrating out $\phi$ and ${\bf A}$ is equivalent to substituting their saddle-point values for a given phase configuration. These saddle-point fields follow from $\delta S/\delta(\phi,{\bf A})=0$, and are given by $\phi_{cl}[q,\theta(q)] =  i\Omega_m \theta(q) \frac{\varkappa e^*U_0({\bf q})}{1+\varkappa (e^*)^2U_0({\bf q})}$ and ${{\bf A}_{cl}[q,\theta(q)] = i {\bf q} \theta(q) \frac{J_0 e^*D(q)}{1+J_0 (e^*)^2D(q)}}$.
Substituting them into the continuity equation $\Omega_m \rho(q) + {\bf q}\cdot {\bf j}_s(q)=0$, and removing the $\Omega_m=0$ component of ${\bf A}_\text{cl}$, yields 
\begin{gather}
     ie^* \left[ \frac{\varkappa \Omega_m^2}{1+\varkappa (e^*)^2U_0({\bf q})} + \frac{J_0 {\bf q}^2}{1+J_0 (e^*)^2D( q)(1-\delta_{m,0})} \right]\theta(q)=0.
\end{gather}
This is precisely the saddle-point equation obtained from the phase-only action \eqref{S_eff}. This, integrating out the EM does not violate charge conservation, and the resulting phase-only description preserves the continuity equation at the saddle point $\theta=\theta_{cl}$ \cite{PhysRevB.60.564}.

\subsubsection{Phase plasmon screened by cavity}

In this subsection, we discuss how the interplay between the cavity-confined EM fluctuations and superconducting phase modifies the phase dynamics and the associated plasmon mode. Performing analytical continuation, we obtain from the action \eqref{S_eff} the dispersion of the phase collective mode with 
\begin{gather}\label{w_q}
\omega^2\left(1+\frac{1}{2\pi\lambda_P} D^R({\bf q},\omega, L)\right) =  {\bf q}^2 \left(v_s^2 + \frac{1}{2\pi\lambda_P} U_0({\bf q})\right),
\end{gather}
where $v_s = \sqrt{J_0/\varkappa}$ is the sound velocity and retarded photon propagator is a real function:
\begin{gather}
D^R({\bf q},\omega, L) = \frac{2\pi}{\omega^2}
\begin{cases}
         -\sqrt{{\bf q}^2 - \omega^2} \tanh\frac{\sqrt{{\bf q}^2 - \omega^2}L}{2} + |{\bf q}| \tanh\frac{|{\bf q}|L}{2}  , \qquad \omega \leq |{\bf q}| \\
          \sqrt{\omega^2 - {\bf q}^2} \tan\frac{\sqrt{\omega^2 - {\bf q}^2}L}{2} + |{\bf q}| \tanh\frac{|{\bf q}|L}{2} , \qquad \omega > |{\bf q}| 
    \end{cases}.
\end{gather}
\begin{figure}[] 
\includegraphics[width=0.65\textwidth]{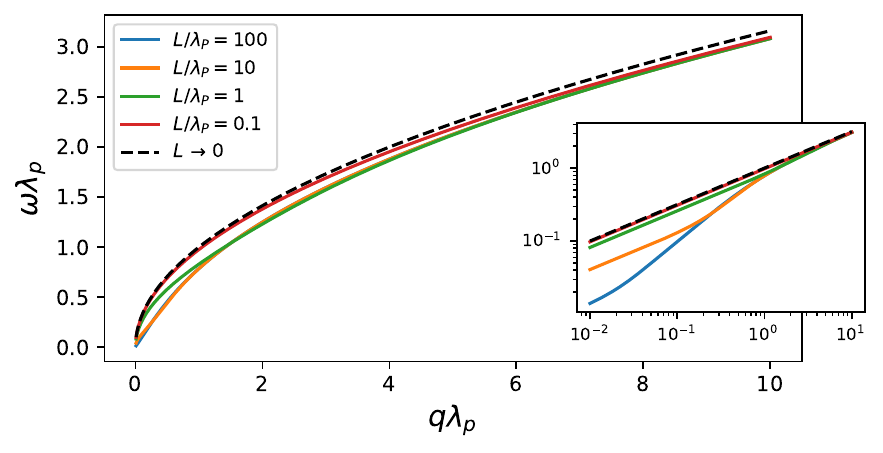}
\centering
\caption{\small{  Spectrum of the plasmon in two-dimensional superconductor for different cavity sizes $L$. Black dashed line shows the dispersion without retardation effects, given by Eq.\eqref{no_D_R}, which corresponds to the $L\to0$ limit. The inset shows the same plot on a log-log scale. Here we set $v_s=10^{-5}$ (with respect to $c=1$).  }}
\label{plasmon}
\end{figure}
The cavity-screening of this plasmon mode can be understood as follows:

\

(i) For large $L$, the cavity effect is negligible and we restore the standard dispersion relation for a 2D superconductor 
\begin{gather}
    \omega^2-v_s^2{\bf q}^2 - \lambda_P^{-1}\sqrt{-\omega^2+{\bf q}^2}=0.
\end{gather}
This equation admits one propagating solution, which corresponds to the longitudinal plasmon mode with the following asymptotes:
\begin{gather}
    \omega \approx 
    \begin{cases}
        |{\bf q}| , \qquad |{\bf q}|\ll \frac{1}{2\lambda_P\sqrt{1-v_s^2}} \\
        \frac{\sqrt{1-v_s^2}}{2\lambda_Pv_s } +v_s |{\bf q}| , \qquad |{\bf q}|\gg \frac{\sqrt{1-v_s^2}}{\lambda_P v_s^2} 
    \end{cases}
\end{gather}
Dynamical EM fluctuations, including retardation effects, reshape the phase-mode dispersion. At small wave vectors, the plasmon is governed by the EM sector and becomes photon-like, with a linear dispersion. This is in contrast to the case with only instantaneous Coulomb interactions, where $\omega\propto \sqrt{|{\bf q}|}$ at low momenta and the mode is not well resolved in the long-wavelength limit \cite{PhysRevResearch.2.023413}. In the London model, the strength of the coupling between the phase and EM field is controlled by $1/\lambda_P$, and formally for $\lambda_P\to\infty$ the plasmon reduces to a phase-like mode with $\omega = v_s |{\bf q}|$.

\

(ii) With decreasing $L$, the cavity creates a gap $\Omega_0=\pi/L$ in the photon spectrum, thus reducing the effect of the photons on the phase plasmon and deforming the dispersion $\omega(q)$ starting from small momentum ${\bf q}$:
\begin{gather}
    \omega \approx 
    \begin{cases}
        \sqrt\frac{|{\bf q}|/\lambda_P}{1+L/2\lambda_P} , \qquad |{\bf q}|\lesssim \frac{1}{L} \\
        \frac{\sqrt{1-v_s^2}}{2\lambda_Pv_s } + v_s |{\bf q}| , \qquad |{\bf q}|\gg \frac{\sqrt{1-v_s^2}}{\lambda_P v_s^2}
    \end{cases}.
\end{gather}

(iii) At very small cavity size $L$, the photons are strongly gapped and actually "leave" the phase mode, since $D^R(q,L\to 0)\to \pi L$. In this regime the phase plasmon restores the ordinary dispersion for the 2D case 
\begin{gather}\label{no_D_R}
    \omega = \sqrt{v_s^2{\bf q}^2+|{\bf q}|/\lambda_p}.
\end{gather}

This means that the cavity environment reshapes the two-dimensional superconducting plasmon through the dynamical EM contribution, pushing the mode to higher energies, as shown in Fig. \ref{plasmon}. In other words, the cavity modifies the energy cost of  low-energy phase excitations. This in turn can have a profound impact on the thermodynamic phase stiffness $J_0$, since such excitations generally suppress the stiffness, as discussed below.

\subsubsection{Cavity-induced enhancement of the phase stiffness}

Here we study how the EM fluctuations affect the phase stiffness $J_0$ through the low-energy plasmon excitations. Capturing this effect requires going beyond the Gaussian approximation and retaining nonlinear terms in the action action \eqref{S}. At the Gaussian level, the collective modes are independent - the stiffness fixes their dispersion, but the fluctuations do not renormalize $J_0$ itself. The feedback appears only once interaction between the modes is included, encoded here by the nonlinear fluctuation terms.

Assume phenomenologically that the phase field has a slow background twist and fluctuations $\theta=\theta_0+\delta\theta$. Quartic terms, such as $S_\text{int}\propto\int(\nabla \theta)^4$, then generate couplings of the form $\propto(\nabla\theta_0)^2(\nabla\delta\theta)^2$. Averaging over the fluctuation field produces a correction to the bare phase stiffness $J_0$, which is the coefficient of $(\nabla\theta_0)^2$. This fluctuation contribution is governed by the thermal occupation of the collective modes, schematically encoded in $\langle (\nabla\delta\theta)^2\rangle$. Thus, nonlinearities provide a mechanism by which thermally populated sound (in neutral) or plasmon (in charged systems) modes feed back on the Gaussian stiffness. 

Physically, these thermally excited collective modes carry part of the momentum response as a normal component and therefore reduce the superfluid stiffness, in the same sense as the standard Landau/Bogoliubov reduction of the superfluid density \cite{PhysRev.60.356,RevModPhys.76.599}. More generally, this is the hydrodynamic version of the renormalization-group idea that nonlinear fluctuations dress the parameters of the Gaussian effective action \cite{popov1987functional}.

\

\begin{figure}[h] 
\includegraphics[width=0.55\textwidth]{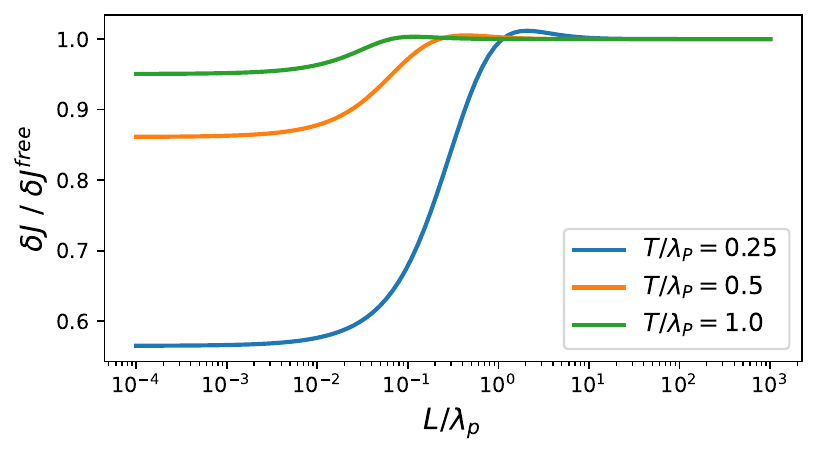}
\centering
\caption{\small{ Fluctuation correction to the phase stiffness $\delta J$ from Eq. \eqref{d_J} normalized by its free-space limit vs cavity length at different temperatures for fixed $v_s = 10^{-5}$. The absolute value of the negative correction is reduced for small cavities. }}
\label{J_L_q}
\end{figure}

In case of a charged phase-fluctuating superconductor, the corresponding nonlinear correction of the form $J_0 \to J_0+\delta J$ has been studied by Benfatto et. al. in Ref. \cite{PhysRevB.69.184510}. We make use of their result and, to keep this discussion concise, do not explicitly write the nonlinear terms in our action \eqref{S} or perform the corresponding diagrammatic expansion in the presence of the EM fluctuations. Instead, we phenomenologically apply their theory to our case and focus on physical interpretation. The result of Ref. \cite{PhysRevB.69.184510} gives a correction of the following form:
\begin{gather}\label{d_J}
    %\delta J = -\frac{1}{4m^2} \frac{1}{T} \int \frac{d{\bf q}}{(2\pi)^2} \ {\bf q}^2 n_B(\omega_{\bf q})\big[1+n_B(\omega_{\bf q})\big],
    \delta J = - \frac{\gamma}{T} \int \frac{d{\bf q}}{(2\pi)^2} \ {\bf q}^2 n_B(\omega_{\bf q})\big[1+n_B(\omega_{\bf q})\big],
\end{gather}
where $n_B(x)=(e^{x/T}-1)^{-1}$ is the Bose distribution; and $\gamma$ denotes the strength of the nonlinearity. Direct comparison with the continuum-limit calculation of Ref. \cite{PhysRevB.69.184510} gives $\gamma=1/4m^2$. We note, however, note that this value depends sensitively on the underlying microscopic model, and in our phenomenological theory $\gamma$ is therefore treated as a free parameter. In our case the collective mode is the plasmon with dispersion $\omega_{\bf q}=\omega({\bf q},L)$ extracted from Eq. \eqref{w_q}, which is controlled by the cavity. By reducing the cavity size, we increase the energetic cost of these excitations by shifting plasmon to higher energies (see Fig. \ref{plasmon}), thereby reducing their negative contribution to the stiffness. The cavity dependence of $\delta J(L)$ is shown in Fig. \ref{J_L_q}: for large cavities, it is essentially unchanged relative to the free space value $\delta J^\text{free}=\delta J(L\to \infty)$, while for small cavities it increases and eventually saturates. Since the nonlocal plasmon mode is deformed inhomogeneously for different ${\bf q}$ (see Fig. \ref{plasmon}), the subsequent momentum integration in Eq. \eqref{d_J} produces a non-monotonic function with a small bump.

We also should restore the temperature dependence of the bare stiffness $J_0$ due to the thermal suppression of the superfluid density, and we model it phenomenologically as $J_0(T)=J_0(1-T/T^*)$, where $T^*$ is the critical temperature of the pseudogap phase. Correspondingly, we introduce the temperature-dependent Pearl length as $\lambda_P(T)=\lambda_P\cdot(J_0/J_0(T))$ used also in Eq. \eqref{w_q}. 
With these changes the total phase stiffness takes the following form: 
\begin{gather}\label{J_eff_T}
    J(L, T) = J_0 \left[1 - \frac{\tilde\gamma}{T \lambda_p(T)} \int dx \ x^3 \ n_B(\omega_x)\big[1+n_B(\omega_x)\big] \right], 
\end{gather}
where $\tilde\gamma$ denoted the dimensionless strength of the nonlinearity. 
The ratio $J/J^\text{free}$ as a function of the cavity length $L$ is plotted in Fig. 4(c) of the main text for the following parameters: $\lambda_P=70 \mu$m; $T=0.1\lambda_P^{-1}=0.5T^*$; $\tilde\gamma=0.8$. The plasmon dispersion $\omega_x$, where $x=q\lambda_P$, is extracted numerically from Eq. \eqref{w_q}.

\

The proposed mechanism leads to an effective increase of the phase stiffness and, consequently of the superfluid density. This effect originates essentially from the cavity-induced suppression of thermal EM fluctuations in the superconducting film. At the same time, the cavity can enhance vacuum EM fluctuations, which  couple directly to the electrons. As was shown in Ref. \cite{plastovets2026cavity}, such cavity-enhanced vacuum fluctuations actually suppress the superfluid stiffness within a coarse-grained Ginzburg-Landau theory near the critical point. This contribution, however, is predicted to be particularly small for high-$T_c$ superconductors, and its additional correction to $J_0$ is therefore neglected in Eq. \eqref{J_eff_T}.

\subsection{Cavity-increased BKT transition temperature}

\textit{Here we show how the enhancement of the phase stiffness can shift the BKT transition to higher temperatures. For this purpose, we employ the standard analysis: we incorporate vortex fluctuations into our coarse-grained theory and analyze how the vortex-antivortex pairs renormalize the temperature dependence of the cavity-dressed phase stiffness $J(L,T)$. The role of the cavity then becomes transparent: an increased value of $J(L,T)$ at the transition implies a higher critical temperature, consistent with the universal BKT relation $J(L,T_\text{BKT})\propto T_\text{BKT}(L)$. The physical picture behind this is also clear. Within the London theory, we have shown that the long-range effective phase stiffness is enhanced below the critical temperature. This directly affects finite-temperature vortex excitations: Since the energy cost of phase gradients increases with $J$, long-wavelength circulating currents become more expensive, thereby enhancing the binding of vortex-antivortex pairs. As a result, their unbinding into free vortices is suppressed up to higher temperatures compared to free space. This provides a natural explanation for the increase of the BKT transition temperature $T_\text{BKT}^\text{free}<T_\text{BKT}^\text{cav}\ll T^*$, as observed experimentally.}

\subsubsection{Vortex excitations}

Unlike the full nonlinear XY model, the Gaussian phase-only action $S[\theta]$ does not explicitly include vortex configurations with windings $\theta \to \theta+2\pi n$. Nevertheless, their contribution can be incorporated by separating phase fluctuations into smooth and singular parts $\theta=\theta_{SW}+\theta_v$ \cite{halperin1979resistive,beasley1979possibility,RevModPhys.59.1001,PhysRevLett.98.117008,PhysRevLett.99.207002,benfatto2013berezinskii}. Here a vortex excitation satisfies $\nabla^2\theta_v = 2\pi \sum_i n_i\delta({\bf r}-{\bf r}_i)$ with ${\bf r}_i$ being the vortex/antivortex position ($n_i=\pm 1$). When substituting the full phase into the bare phase action $S_\theta \approx \frac{J_0(T)}{2} \int d\tau \int d{\bf r} \big[ \nabla \theta({\bf r}) \big]^2$, the boundary conditions (in the absence of real long-range interactions) can be used to eliminate the mixed $\theta_{SW}\theta_v$ term, so we neglect it in our case as well. In momentum space a vortex excitation corresponds to the phase profile $\theta_v({\bf k})=-2\pi g({\bf k})/{\bf k}^2$, where $g({\bf k})=\sum_i n_i e^{i{\bf k}{\bf r}_i} $. Substituting this phase profile into $S_\theta$ we get the vortex part of the action $S_{v} = \frac{2\pi^2 J_0(T)}{T}\sum_{i,j}n_in_jV(r_{ij})$, where $V(r)=\frac{1}{2\pi}\ln(L/r)$ is the interaction potential between vortices in the system. For simplicity, we assume that the cavity does not change the form of the vortex interaction potential, and its effect on the BKT physics in therefore included only through the renormalized stiffness $J(L,T)$.

\subsubsection{Shift of the transition temperature}

The vortex-induced BKT transition can be described using a renormalization group (RG) analysis \cite{kosterlitz1973ordering,RevModPhys.59.1001}. With vortex excitations included as above, the action $S_\theta$ can systematically account for vortex-induced renormalization of the phase stiffness, in addition to the cavity-induced effects, and reproduce the Coulomb gas model of charged vortices, thus leading to an RG flow for the phase coupling $K=J/T$ and the vortex fugacity $y_0=e^{-\mu/T}$, where $\mu$ is the vortex core energy \cite{altland2006condensed}. The standard RG equations read 
\begin{gather} \label{RG}
    \frac{d K^{-1}}{dl}=4\pi^3y_0^2, \quad \quad
    \frac{d y_0}{dl}=(2-\pi K )y_0,
\end{gather}
where the distance scaling is $r=a e^l$, and $a$ is the UV truncation of the London theory, which corresponds to the zero-temperature coherence length $a=\xi_0$. These equations are complemented by the following initial conditions for the RG flow: $K(l=0)=J(T,L)/T$, which account for the cavity-modified phase stiffness without vortices $J(T,L) $ defined in Eq. \eqref{J_eff_T}, 
and $y_0(l=0)=\exp(-\mu(T)/T)$, where $\mu(T)=J_0(T)\alpha_v$ is the vortex core energy with microscopically defined parameter $\alpha_v$, which is assumed to be unchanged by the cavity. We solve the equations numerically using the Runge-Kutta scheme. The solution $J(l\to \infty,T,L)$ gives the temperature dependence of the phase stiffness for different cavity sizes, which is shown in Fig. 4(b) in the main text. The temperature where the stiffness jumps to zero corresponds to the BKT transition temperature, which is growing with decreasing cavity size. For the equations \eqref{RG} the standard relation holds $J(T_\text{BKT})/T_\text{BKT}=2/\pi$, which means the higher transition temperatures correspond to the higher stiffness. For the plot in Fig. 4(b) we use the same parameters as in the Fig. 4(c), and additionally $\xi_0=2$nm; $\alpha_v=15$.

%-----------------------------------------------------------------------
\section{Cavity alignment}\label{sec:cav_align}

The minimum attainable cavity length is limited by the parallelism between the YBCO sample and the movable cavity mirror. Accurate alignment is therefore essential both to reach short cavity lengths and to avoid spurious thermal or mechanical contact between the two elements.

The alignment is performed at room temperature using a visible laser beam collinear with the THz propagation axis. We first adjust the azimuthal and polar orientation of the sample holder so that the sample surface is normal to the beam. This is done by monitoring the back-reflected beam in the far field, approximately 2~m away from the sample position. We then adjust the orientation of the cavity mirror using the three piezoelectric actuators until the back-reflection from the mirror overlaps with that from the sample. This procedure ensures that the two surfaces are parallel within the precision required for the short-cavity measurements.

\begin{figure}
\includegraphics[width=0.75\textwidth]{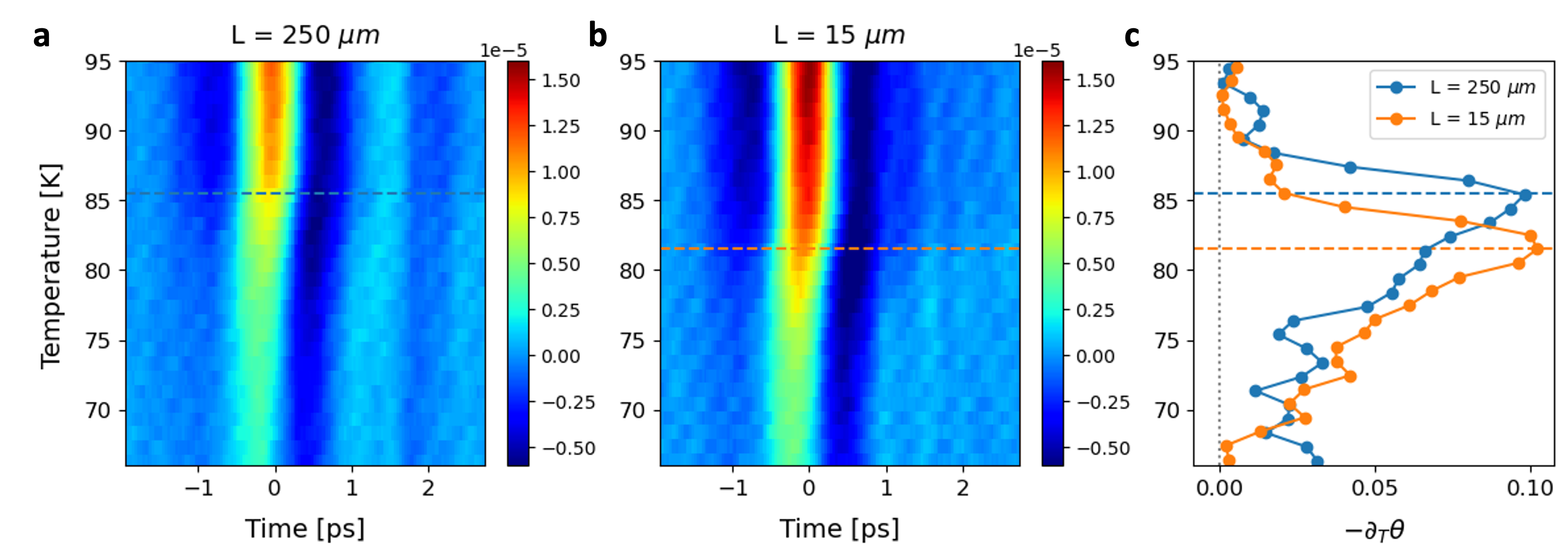}
\centering
\caption{\label{fig:map_touching}
Temperature-dependent THz transmission of YBCO in a cavity with length \textbf{a,} $L=250~\mu$m and \textbf{b,} $L=15~\mu$m. 
\textbf{c,} Temperature derivative of the THz phase, showing a downward shift of the apparent superconducting transition when the mirror is brought too close to the sample.}
\end{figure}

Using this alignment protocol, we reproducibly reached minimum cavity lengths of approximately $20~\mu$m. When attempting to move the mirror closer, we observed a reduction of the apparent superconducting transition temperature by about 4~K, as shown in Fig.~\ref{fig:map_touching}. A similar downward shift was also observed when the mirror was kept at a nominally safe distance but deliberately misaligned. These observations indicate that misalignment or excessive approach can produce local thermal contact or enhanced radiative/near-field heating of the sample by the mirror, whose temperature is approximately 20~K higher than that of the sample holder. In this case, the YBCO film is locally warmer than the temperature recorded by the sample thermometer, causing the superconducting transition to appear at a lower measured temperature.

This effect is opposite to the upward shift of the superconducting response reported in the main text. It therefore cannot account for the observed cavity-induced enhancement. All datasets discussed in the manuscript were acquired at cavity lengths and alignments for which this spurious downward shift was absent.

%-----------------------------------------------------------------------
\section{Temperature dependence of the cavity length}\label{sec:cav_length}

We verified that the effective cavity length does not measurably change with temperature due to thermal contraction of the cavity assembly. For this purpose, we performed cavity-length scans at different temperatures using an identical set of piezo-actuator positions. The experimental protocol was as follows. First, the cavity was aligned at room temperature. We then performed a cavity scan by moving the three piezo actuators by the same displacement $\Delta L$, so that the mirror remained parallel to the sample while the cavity length was varied. The same set of actuator positions was subsequently used at different temperatures, allowing us to test whether the optical cavity length changed upon cooling.

\begin{figure}
\includegraphics[width=0.8\textwidth]{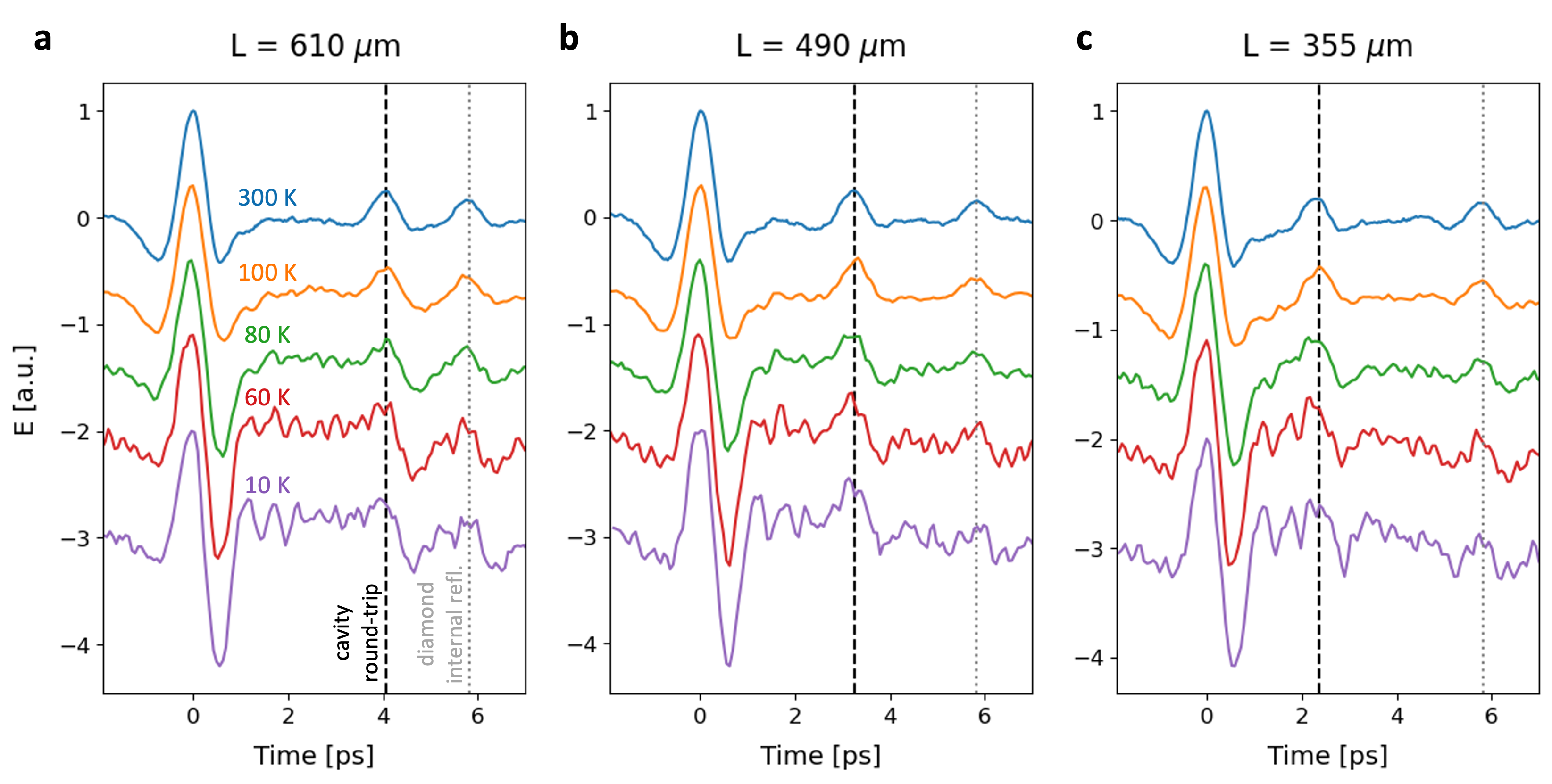}
\centering
\caption{\label{fig:temp_cav_l}
THz field transmitted by the YBCO sample in the cavity at different temperatures, from room temperature to 10~K, for different nominal cavity lengths: \textbf{a,} $L=610~\mu$m, \textbf{b,} $L=490~\mu$m, and \textbf{c,} $L=355~\mu$m. The black dashed line indicates the position of the cavity round-trip peak, while the grey dotted line marks the internal reflection from the diamond window.}
\end{figure}

Figure~\ref{fig:temp_cav_l} shows the THz field transmitted by the YBCO sample in the cavity at five temperatures, from room temperature down to 10~K. The curves have been normalized to their peak and arbitrarily shifted for clarity. The three panels correspond to three selected nominal cavity lengths. Importantly, for each panel the piezo-actuator positions are identical for all temperatures. The grey dotted line marks a peak at $\sim 5.83$~ps, corresponding to an internal reflection inside the cryostat's diamond window. This feature is independent of both temperature and cavity length, as also discussed in Fig.~\ref{fig:chamber}. The black dashed line marks the cavity round-trip peak, whose position changes when the mirror position is changed.

For a fixed actuator position, the relative delay between the main transmitted THz pulse and the cavity round-trip peak remains unchanged over the full temperature range. This demonstrates that the effective cavity length is stable upon cooling within our experimental resolution. We therefore conclude that thermal contraction of the cavity assembly does not introduce a measurable temperature-dependent change of the cavity length in the analysis.

%-----------------------------------------------------------------------
\section{Determination of the critical temperature}\label{sec:find_tc}

\begin{figure}
\includegraphics[width=0.8\textwidth]{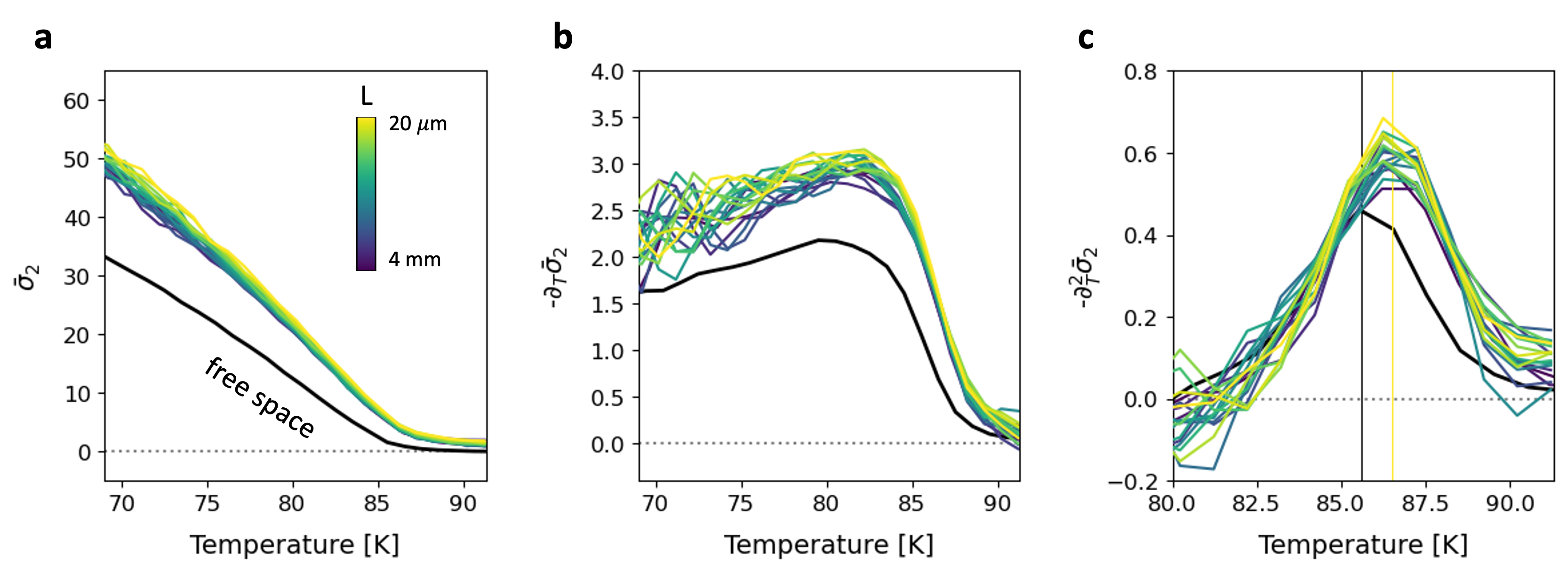}
\centering
\caption{\label{fig:crit_temp}
\textbf{a,} Temperature dependence of the frequency-averaged imaginary conductivity, $\bar{\sigma}_2$, for different cavity lengths compared to YBCO in free space (black curve). 
\textbf{b,} First temperature derivative of the curves shown in (a). 
\textbf{c,} Second temperature derivative of the curves shown in (a). The vertical lines identify the maxima of $-\partial_T^2\bar{\sigma}_2$ for YBCO in free space (black) and for the shortest cavity, $L=20~\mu$m (yellow).}
\end{figure}

To compare the transition temperature of YBCO with and without the cavity, we start from the temperature-dependent frequency-averaged imaginary conductivity, $\bar{\sigma}_2(T)$, shown in Figure~3c of the main text and reproduced here in Fig.~\ref{fig:crit_temp}a.

Rather than assigning the transition temperature from an arbitrary threshold value of $\bar{\sigma}_2(T)$, we use its temperature derivatives. The first derivative, shown in Figure~\ref{fig:crit_temp}b, identifies the temperature range over which the response changes most rapidly. We then define the transition temperature as the maximum of $-\partial_T^2\bar{\sigma}_2(T)$, shown in Figure~\ref{fig:crit_temp}c. This criterion provides a reproducible measure of the onset of the rapid increase in $\bar{\sigma}_2(T)$ and is applied identically to the free-space and cavity measurements.

The vertical lines in Figure~\ref{fig:crit_temp}c indicate the transition temperatures extracted for YBCO in free space and for the shortest cavity length, $L=20~\mu$m. The cavity dataset shows a clear shift of this transition marker to higher temperature, consistent with the trend discussed in the main text and with the results of the time-domain fitting (Sec. \ref{subsec:phase_fit}).

% =====================================================
% Supplementary references
% =====================================================

\clearpage
\renewcommand{\refname}{Supplementary References}
\putbib[supplementary_references]
\end{bibunit}

\end{document}